\documentclass[12pt,preprint]{aastex}
\newcommand\rme{{\rm e}}

\newcommand\scr{S_{\rm tot}}
\newcommand\acr{\alpha_{\rm tot}}

\shortauthors{Bryans et al.}
\shorttitle{Updated Ionization Balance}

\begin{document}

\title{Collisional Ionization Equilibrium for Optically Thin Plasmas. I.
Updated Recombination Rate Coefficients for Bare through Sodium-like Ions}

\author{P. Bryans\altaffilmark{1}, 
        N. R. Badnell\altaffilmark{2}, 
	T. W. Gorczyca\altaffilmark{3},
        J. M. Laming\altaffilmark{4}, 
	W. Mitthumsiri\altaffilmark{1}, 
	\\and D. W. Savin\altaffilmark{1}}
\altaffiltext{1}{Columbia Astrophysics Laboratory, Columbia University,  New York,
NY 10027}
%\email{bryans@astro.columbia.edu}
\altaffiltext{2}{Department of Physics,
University of Strathclyde, Glasgow, G4 0NG, United Kingdom}
%\email{badnell@phys.strath.ac.uk}
\altaffiltext{3}{Department of Physics, Western Michigan University, Kalamazoo, MI 49008}
%\email{thomas.gorczyca@wmich.edu}

\altaffiltext{4}{E. O. Hulburt Center for Space Research, US Naval Research
Laboratory,
Code 7674L, Washington, DC 20375}
%\email{jlaming@ssd5.nrl.navy.mil}

%\email{savin@astro.columbia.edu}

\begin{abstract}

Reliably interpreting spectra from electron-ionized cosmic plasmas requires
accurate ionization balance calculations for the plasma in question.  However,
much of the atomic data needed for these calculations have not been generated
using modern theoretical methods and are often highly suspect. This translates
directly into the reliability of the collisional ionization equilibrium (CIE)
calculations.  We make use of state-of-the-art calculations of dielectronic
recombination (DR) rate coefficients for the hydrogenic through Na-like ions of
all elements from He up to and including Zn. Where measurements exist, these
published theoretical DR data agree with recent laboratory work to within
typically 35\% or better at the temperatures relevant for CIE. We also make use
of state-of-the-art radiative recombination (RR) rate coefficient calculations
for the bare through Na-like ions of all elements from H through to Zn. Here we
present improved CIE calculations  for temperatures from $10^4$ to $10^9$~K
using our data and the recommended electron impact ionization data of
\citet{Mazz98a} for elements up to and including Ni and Mazzotta (private
communication) for Cu and Zn. DR and RR data for ionization stages that have not
been updated are also taken from these two additional sources. We compare our
calculated fractional ionic abundances using these data with those presented by
Mazzotta et al.\ for all elements from H to Ni. The differences in peak
fractional abundance are up to  60\%. We also compare with the fractional ionic
abundances for Mg, Si, S, Ar, Ca, Fe, and Ni derived from the modern DR
calculations of  \citet{Gu03a,Gu04a} for the H-like through Na-like ions, and
the RR calculations of \citet{Gu03b} for the bare through F-like ions. These
results are in better agreement with our work, with differences in peak
fractional abundance of less than 10\%.

\end{abstract}


\section{Introduction}
\label{sec:Intro}

Electron ionized plasmas (also called collisionally ionized plasmas)
are formed in a diverse variety of objects in the universe.  These
range from stellar coronae and supernova remnants through to the interstellar
medium and gas in galaxies or in clusters of galaxies.  The physical
properties of these sources can be determined using spectral
observations coupled with theoretical models.  This allows one to
infer electron and ion temperatures, densities, emission measure
distributions, and ion and elemental abundances.  But, reliably
determining these properties requires accurate fractional abundance
calculations for the different ionization stages of the various
elements in the plasma (i.e., the ionization balance of the gas).

Since many of the observed sources are not in local thermodynamic
equilibrium, in order to determine the ionization balance of the
plasma one needs to know the rate coefficients for all the relevant
recombination and ionization processes.  Often, the observed systems
are optically-thin, low-density, dust-free, and in steady-state or
quasi-steady-state.  Under these conditions the effects of any
radiation field can be ignored, density effects are insignificant,
three-body collisions are unimportant,
and the ionization balance of the gas is time-independent.  This is
commonly called collisional ionization equilibrium (CIE) or sometimes
coronal equilibrium.  

Because CIE occurs in a wide range of cosmic sources, accurate calculations have
long been an issue of concern.  One of the continuing challenges of theoretical
and experimental atomic physics is to provide reliable data for all the relevant
collision processes. In CIE, recombination is due to dielectronic recombination
(DR) and radiative recombination (RR).  At the temperature of peak formation in
CIE, DR dominates over RR for most ions. Ionization is a result of electron
impact ionization (EII).  At temperatures low enough for both atoms and ions to
exist, charge transfer (CT) can be both an important recombination and
ionization process.  Considering all the ions and levels that need to be taken
into account, it is clear that vast quantities of data are needed.  Generating
them to the accuracy required pushes theoretical and experimental methods to the
edge of what is currently achievable and often beyond.  For this reason progress
has been slow.

Over the years a number of different groups have evaluated the available atomic
data and produced CIE calculations.  Some of the most commonly cited results are
those of \citet{Shul82a}, \citet{Arna85a}, \citet{Land91a}, \citet{Arna92a}, and
\citet{Mazz98a}. \citet{Masa97} investigated the astrophysical implications for
several different CIE models on the fractional abundance of Fe. More recently, 
\citet{Gu03a} carried out CIE calculations using his DR and RR rate coefficients
and compared to the  CIE results of \citet{Mazz98a} for Mg, Si, S, Ar, Ca, Fe,
and Ni. Our efforts here are just another step in what promises to be a long
line of studies aimed at providing the astrophysics community with the most
reliable CIE calculations currently possible.

The work here is motivated in specific by recent advances in our understanding
of DR.  With the development in the late 1980s of electron beam ion traps
(EBITs) and heavy-ion storage rings combined with electron coolers, it has
become possible to carry out detailed DR measurements for a wide range of
systems \citep{Mull97a,Schi99a,Beie03a}.  These results, in turn, have been used
to test various state-of-the-art theoretical methods for calculating DR.  Using
these benchmarked methods, over the last few years a number of groups have
systematically calculated DR for K- and L-shell ions and M-shell Na-like ions of various
elements \citep[][ --- see also Table~\ref{tab:dr}]{Gu03a,Gu04a,Badn03a,Badn06a}.
These groups have also recently calculated state-of-the-art RR rate
coefficients for K- and L-shell ions and M-shell Na-like ions for a number of elements
\citep{Gu03b, Badn06b, Badn06d}. 

Using these new DR and RR results we have computed the
CIE fraction abundances for the various ionization stages of all elements from H
up to and including Zn.  We present results for plasma temperatures from 10$^4$
to 10$^9$~K.
The rest of this paper is organized as follows: In Sec.~\ref{sec:DR} we review
recent developments in our understanding of DR. Section~\ref{sec:RR} discusses
recent improvements in the theoretical calculation of RR rate coefficients.  
In Sec.~\ref{sec:unified} we discuss recent published unified DR+RR
rate coefficients.
Section~\ref{sec:EII} briefly discusses the status of the EII rate coefficients.
Updating these data will be the subject of a future paper.  In Sec.~\ref{sec:CT}
we give a short overview of the importance of CT in electron ionized plasmas.   
We will incorporate CT into future calculations.  Section~\ref{sec:calc}
outlines the equations relating ionization fractions to the rate coefficients
and describes how we solve these equations. In Sec.~\ref{sec:Results} we present
our new CIE calculations and compare these results to the ionization balance
results of \citet{Mazz98a} and to CIE calculations based on the data of
\citet{Gu03a,Gu03b,Gu04a}. Section~\ref{sec:Discussion} discusses the results of
our calculations and, in particular, how they differ from previous studies.
Concluding remarks are given in Sec.~\ref{sec:Conclusions}.

\section{Dielectronic Recombination (DR)}
\label{sec:DR}

Dielectronic recombination (DR) is a two-step recombination process that begins
when an electron collisionally excites a core electron of an ion and is
simultaneously captured. The core electron excitation can be labeled $nl_j\to
n^\prime l^\prime_{j^\prime}$, where $n$ is the principal quantum number, $l$
the orbital angular momentum, and $j$ the total angular momentum.  We label the
change in principal quantum number as $\Delta n=n^\prime-n$. The energy of this
intermediate system lies in the continuum and the complex may autoionize. The DR
process is complete when the system emits a photon, reducing the total energy of
the recombined system to below its ionization threshold.  Conservation of energy
requires that for DR to go forward  $E_k=\Delta E-E_b$. Here $E_k$ is the
kinetic energy of the incident electron, $\Delta E$ the excitation energy of the
initially bound electron in the  presence of the captured electron, and $E_b$
the binding energy released when the incident electron is captured onto the
excited ion. Because $\Delta E$ and $E_b$ are quantized, DR is a resonant
process.

\citet{Badn03a} have calculated the DR rate coefficients using the 
semi-relativistic {\sc
autostructure} code \citep{Badn86a} for the H- through Na-like isoelectronic
sequences of all elements from He through to Zn (see Table~\ref{tab:dr}; we
use the convention here of identifying the recombination process by the initial
charge state of the ion). These new DR data have been collected together and are
available online \citet{Badn06a}. In addition, some of the original data have
been refitted so as to extend the validity of the fits to lower temperatures.
\citet{Gu03a} has calculated DR rate coefficients using the relativistic 
{\sc fac} code for
the H- through Ne-like isoelectronic sequences of Mg, Si, S, Ar, Ca, Fe and Ni
and for the Na-like sequence for Mg through Zn \citep{Gu04a}. Both the
calculations of \citet{Gu03a,Gu04a} and \citet{Badn03a} were performed in the
independent processes, isolated resonance approximation \citep{Seat76}, using a
distorted-wave representation.  
For the ions considered here, 
the low collision energies of the important DR resonances
means that a fully relativistic
treatment is not necessary.
The methodology of the {\sc autostructure} and {\sc fac} calculations is
basically the same, with the differences coming mostly from the atomic
structure.

In the temperature range where the fractional CIE abundance of an ion is greater
than 0.01 (what we call here the CIE formation zone), we find the DR data  of
\citet{Badn06a} and \citet{Gu03a,Gu04a}  to be in good agreement except for
Ne-like ions. The agreement is to within better than $\sim35\%$ for Li-like Mg
and  $\sim25\%$ for other Mg ions, $\sim25\%$ for Si and S ions, $\sim20\%$ for
Ar and Ca ions, and $\sim15\%$ for Fe and Ni ions.  

For Ne-like ions the agreement is significantly poorer \citep{Gu03a,Zat04b,
Fu06}. Differences are seen of up to $\sim140\%$ for Mg$^{2+}$ and Si$^{4+}$,
$\sim55\%$ for S$^{6+}$,  $\sim60\%$ for Ar$^{8+}$, $\sim35\%$ for Ca$^{10+}$,
and $\sim15\%$ for Fe$^{16+}$ and Ni$^{18+}$.  However, these differences occur
at temperatures below the peak in the DR rate coefficient, where recombination
is dominated by RR and DR is unimportant. A comparison of peak DR rate 
coefficients reveals differences of $\sim30\%$ for Mg$^{2+}$, and $\sim 10\%$
for the remaining six ions calculated by \citet{Gu03a}.

In CIE, for K-shell ions of the elements considered here DR proceeds
via $\Delta n \ge 1$ core excitations independent of the atomic number
$Z$ of the system.  These have been well studied experimentally using
EBITs and storage rings.  State-of-the-art DR theory such as that of
\citet{Badn06a} and \citet{Gu03a}  reproduce the experimental results
with agreement on the order of $\sim 20\%$ \citep{Mull95a,Savi02c}.

For Li- to Na-like L- and M-shell ions of the elements considered here, 
at low $Z$ DR proceeds primarily via $\Delta n=0$
core excitations (except for Ne-like ions which have no $\Delta n=0$
channels).  For intermediate $Z$, DR proceeds via a mix of
$\Delta n=0$ and 1 core excitation.  At higher $Z$, DR
proceeds primarily via $\Delta n=1$ excitations.  DR for all these ions is
not as well understood as for K-shell systems.

Storage ring measurements of L-shell have been reviewed most recently by
\citet{Schi99a} and \citet{Savi06a}.  Experimental work on M-shell Na-like systems is
given in \citet{Link95}, \citet{Mull99a}, and \citet{Fogl03a}.  For $\Delta n=0$
DR, quite a number of laboratory measurements exist for Li- and Be-like
ions.  Significantly less work exists for B-, C-, N-, O-, F-, and Na-like ions.
For C-, N-, and O-like ions, storage ring measurements exist for only a single
ion in each sequence. For the B-, F-, and Na-like ions, they exist for only 2 ions in
each sequence.  The situation for $\Delta n=1$ DR is even spottier.  Results
have been published for some Li-, Be-, O-, F-, and Na-like ions.  
For the O-, F-, and Na-like ions, measurements exist for only one ion in each sequence.
We are unaware
of any published $\Delta n=1$ measurements for ions in the the B-, C-, N-, or Ne-like
isoelectronic sequences.  Clearly additional benchmark laboratory work is called
for.

To summarize the comparison between state-of-the-art theory and experiment for
the above L- and M-shell isoelectronic sequences, for $\Delta n=0$ DR at
collision energies $\gtrsim$~1--3~eV and for $\Delta n=1$ DR, agreement with the
stronger DR resonances is typically better than 35\%. Problems arise, however,
with $\Delta n=0$ DR for collision energies below $\lesssim$~1--3~eV where
modern theory has difficulty reliably calculating DR resonance energies.  These
differences translate directly into an uncertainty in the DR rate coefficient
for $T_\rme \lesssim$~10,000--35,000~K.  There is no clear $Z$ dependence
scaling for this energy or temperature limit.  For example, Be-like C$^{2+}$
\citep{Fogl05a}, B-like Ar$^{13+}$ \citep{DeWi96a}, and C-like Fe$^{20+}$
\citep{Savi03a} all show discrepancies between theory and experiment for
energies below $\sim 3$~eV.   On the other hand, O-like Fe$^{18+}$, F-like
Fe$^{17+}$, and Na-like Ni$^{17+}$ all show good agreement between theoretical
and experimental resonance energies down to 0.1~eV
\citep{Savi97a,Savi99a,Fogl03a}. The 1--3~eV limit given above is more a
function of the Rydberg level into which the incident electron is captured.  For
high levels, correlation effects are unimportant and theory can reliably
calculate the DR resonance energies and strengths and hence reliable DR rate
coefficients.  But for low levels, this is not the case.

The reliability of DR rate coefficients for $T_\rme \lesssim$~35,000~K
must be evaluated on a case-by-case basis.  Theoretical calculations can
be used as an evaluation guide by determining which Rydberg levels are
important below 3~eV.  But laboratory benchmark measurements are also
needed.  Fortunately in CIE, only singly- and doubly-charged ions form
in significant abundances at these temperatures (based on our fractional
abundance calculations below).  So any theoretical uncertainties will
affect mostly DR data only for these ions.  In a future work we will
investigate theoretically for which ions this is most likely to be an issue.

Below 25,000~K, CT with atomic H is also important, as is discussed in
Sec.~\ref{sec:CT}.  Since we do not include CT in our CIE calculations here, we
have also chosen not to include what little experimentally-derived DR data for 
singly- and doubly-charged ions exist.  Both CT and any published experimental
DR results will be included in future work.

For ionization stages not included in the state-of-the-art calculations of
\citet{Badn06a} and \citet{Gu03a,Gu04a}, we use the DR rate coefficients
recommended by \citet{Mazz98a} for elements up to and including Ni and Mazzotta
(private communication) for Cu and Zn.  These older data come from a variety of
sources and are typically less reliable than more modern results. The sources
of DR data used in our CIE calculations are listed in
Tables~\ref{tab:data_summary} and \ref{tab:data_summary_gu}.

\section{Radiative Recombination (RR)}
\label{sec:RR}

Radiative recombination (RR) is a one-step recombination process that occurs
when a free electron is captured by an ion.
Energy and momentum are conserved in the process
by the simultaneous emission of a photon. Quantum mechanically, DR and RR are
indistinguishable processes that interfere with each other. \citet{Pin92} have
shown that this interference is a very small effect and can safely be neglected
in most cases. This gives the independent processes approximation whereby DR and
RR can be considered separately. At high temperatures RR is unimportant
in comparison to DR so
relativistic effects of the colliding electron need not be considered.

\citet{Gu03b} has calculated RR rate coefficients for ions of Mg, Si, S, Ar, Ca,
Fe and Ni for bare through  F-like ions using {\sc fac}. \citet{Badn06d}  has
calculated RR rate coefficients for all elements from H through to Zn for the
bare through Na-like isoelectronic sequences using {\sc autostructure}. These
{\sc autostructure} data are available online \citep{Badn06b}.

The {\sc autostructure} results agree with those of
\citet{Ver96} to better than 5\% in the CIE zone.
The results of \citet{Gu03b} and \citet{Badn06b}  agree to within $\sim10\%$ in
the CIE formation zone, except for the H-like ions of Ar, Ca, Fe and Ni. For
these ions, the {\sc fac} rate coefficients of \citet{Gu03b} are systematically
smaller than the {\sc autostructure} data of  \citet{Badn06b}. 
The {\sc fac} results show negligible differences with the {\sc
autostructure} data at the low temperature limit of the CIE formation zone, but
these differences rise to $\sim20\%$ at the high temperature limit. For these
four H-like ions in this temperature range, RR dominates over DR. It should be
noted, however, that in CIE the total electron-ion recombination rate is
generally dominated by DR rather than RR. So uncertainties in RR data typically
have less of an effect than those of DR data for fractional abundance
calculations of most ions in CIE. 

For ionization stages not included in the calculations of \citet{Badn06b} and
\citet{Gu03b}, we use the RR rate coefficients recommended by \citet{Mazz98a}
for elements up to and  including Ni and Mazzotta (private communication) for Cu
and Zn. The sources of RR data used in our CIE calculations are listed in
Tables~\ref{tab:data_summary} and \ref{tab:data_summary_gu}.

\section{Unified DR+RR Calculations}
\label{sec:unified}

\citet{Naha1} have presented theoretical unified electron-ion recombination rate
coefficients (i.e., DR+RR). The currently available data from these works
include all ionization stages of  C \citep{Naha1,Naha4}, N \citep{Naha1,Naha11},
and O \citep{Naha2,Naha7}; the bare, H-, and He-like ions of F \citep{Naha11},
Ne \citep{Naha10}, Fe \citep{Naha5}, and Ni \citep{Naha9}; and B-like Ar
\citep{Naha8}. We have compared these results with the summed DR+RR results from
the {\sc autostructure} calculations \citep{Badn06a,Badn06b}. 

In the CIE formation zone, agreement is within 20\% for C$^{q+}$ ($q=1$, 2, 4,
and 5), N$^{q+}$  ($q=3$, 4, and 6), O$^{q+}$ ($q=1$, 2, 4, and 5), F$^{q+}$
($q=7$ and 8), Ne$^{q+}$ ($q=8$ and 9), Ar$^{13+}$, and  Ni$^{q+}$ ($q=26$ and
27). It is within 30\% for  N$^{q+}$ ($q=2$ and 5), O$^{q+}$ ($q=3$ and 7), and
Fe$^{25+}$. For the bare ions, where there is no DR contribution to the
recombination rate coefficient, agreement is to within a few percent. The only
ions that have differences greater than $\sim30\%$ are C$^{3+}$,  N$^{1+}$, and
Fe$^{24+}$. 

For the C$^{3+}$ ion the difference is $\sim35\%$. However,  the results of
\citet{Naha1} were calculated using LS-coupling for this ion and are in good
agreement (better than 18\%) with the LS-coupling  results
of {\sc autostructure} (unpublished). The CIE peak DR rate coefficient for
C$^{3+}$ is enhanced by 30\% when using intermediate coupling. Using such a
coupling scheme, \citet{Prad01} have carried out  Breit-Pauli $R$-matrix
recombination calculations for C$^{3+}$ so as to compare with experiment but, to
our knowledge, no Maxwellian rate coefficient has been made publicly available.
We expect their data would be in better agreement with the summed DR+RR
{\sc autostructure} results.

The disagreement in the N$^{1+}$ rate coefficient is largest at the low
temperature end of the CIE range (up to 60\% at $10^4$~K) where the
$2s^22p^{4\,\,2}D$ DR resonance dominates. 
The source of this difference may lie in the energy used for this resonance.
The {\sc autostructure} results are
in close agreement with those of \citet{Nuss83}. Both use the observed position
of this resonance. 
The difference is unlikely to be due to fine structure as DR via fine structure
core excitations does not become important for this ion
until below $10^3$~K, well outside of the CIE range.

For Fe$^{24+}$ the calculations of \citet{Naha5} track the RR calculations of
\citet{Badn06b} closely; but above $10^7$~K, where the DR contribution to the
total recombination rate coefficient becomes important, we find the
\citet{Naha5} results to be around 45\% larger than the summed (DR+RR) {\sc
autostructure} results. The source of this difference is unclear, but
\citet{Gorc97} have shown that DR resonance interference for Fe$^{24+}$
is negligible so this is unlikely to be the cause. The {\sc autostructure} RR
rate coefficients differ by no more than 5\% from those of \citet{Ver96} over
the entire CIE temperature range while the DR rate coefficients differ by no
more than 10\% from those of \citet{Gu03a} over this range.

\section{Electron Impact Ionization}
\label{sec:EII}

Electron impact ionization (EII) can occur through either direct ionization or
indirect processes such as excitation-autoionization (EA) and
resonant-excitation double autoionization (REDA). Direct ionization is a
non-resonant process. Direct outer-shell ionization typically changes the charge
of the initial atom or ion by one. Direct inner-shell ionization produces a hole
in the shell and a free electron. As the ion stabilizes to fill the hole, up to
six Auger electrons can be emitted \citep{Kaas93,Gorc03}. EA occurs when an
incident electron collisionally excites an ion to a state that then decays by
autoionization rather than radiative decay.  REDA begins when the incident
electron is captured by an ion and simultaneously excites a bound electron of
the ion. REDA is complete when this recombined system autoionizes by emission of
two electrons. Thus the initial ion has moved one higher in charge state. For
ions with certain electron configurations, 
such as those with one or two valence electrons,
EA and, to a lesser extent, REDA can
significantly enhance ionization cross-sections compared to the direct
ionization contribution \citep[e.g.,][]{Link95}.

\begin{figure}
%  \vspace{2.in}
  \includegraphics{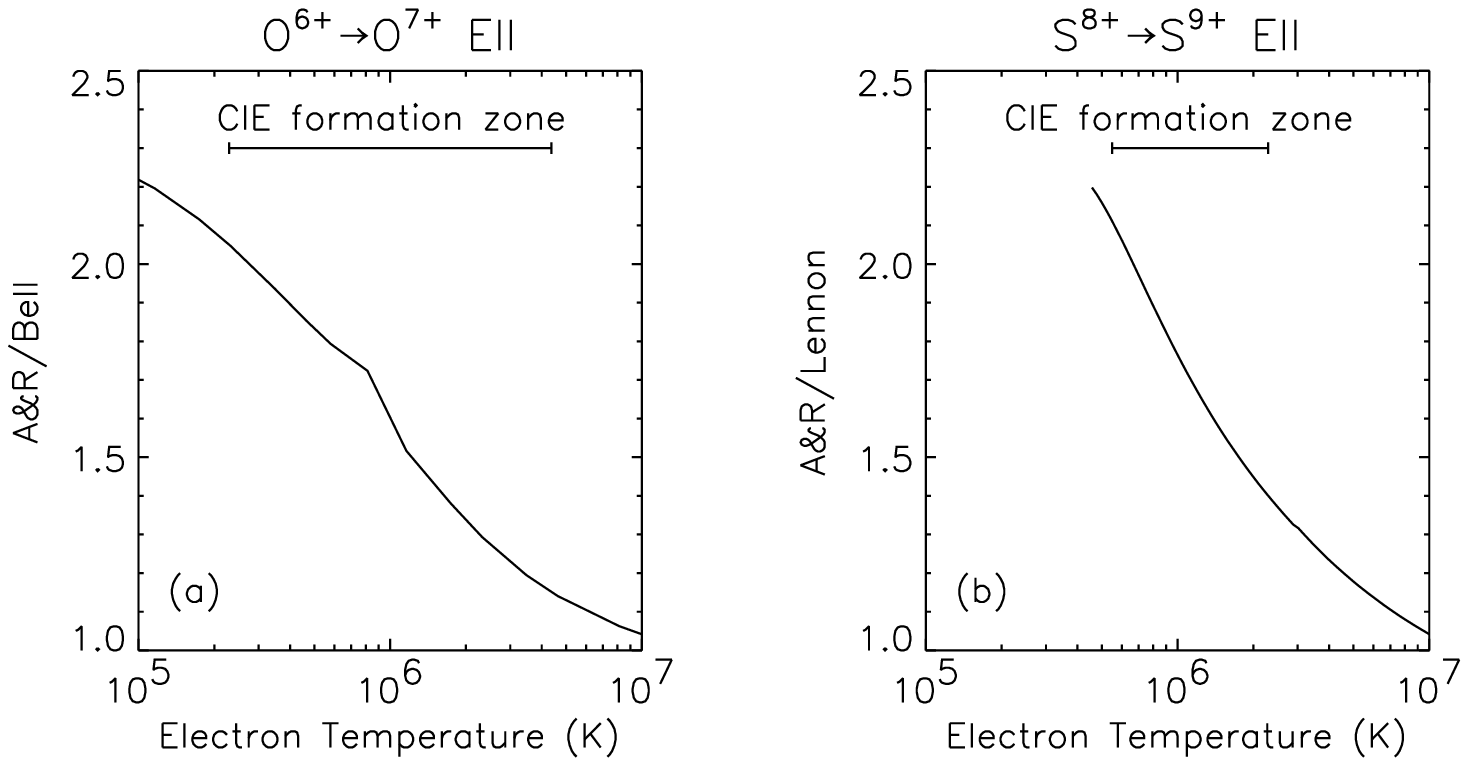}
  \caption[]{The ratio of the recommended EII rate coefficients of
             Arnaud \& Rothenflug (1985; A\&R) relative to the recommended 
	     data of (a)
	     \protect\citet{Bell83} for He-like O$^{6+}$ and (b) \protect\citet{Lenn88} for
	     O-like S$^{8+}$. The horizontal bars show the temperature range
	     over which these ions form in CIE.}
  \label{fig:eii}
\end{figure}

The most recent set of CIE calculations \citep{Mazz98a} used the recommended
data of \citet{Arna85a} and \citet{Arna92a}. These EII data are derived from a
combination of laboratory measurements and theoretical calculations. Other
workers have derived recommended rate coefficients using essentially the same
measurements and calculations \citep{Bell83,Pind87,Lenn88}. All of these
recommended EII rate coefficients have been compared by \citet{Kato91}. Taking
into account known typographical errors in the recommended EII data, Kato et
al.\ found differences between the various recommended data of up to a factor of
2--3 for many ions (see, e.g., Fig~\ref{fig:eii}). These differences are not in
the fits to the data but in the derived recommended data. This is somewhat
surprising considering that the recommended rate coefficients are basically all
derived from the same experimental and theoretical data.

In the present paper we  use the EII rate coefficients for all ionization
states of H through Ni from \citet{Mazz98a}.  This means that any and all
subsequent differences between our new fractional abundances and those of
\citet{Mazz98a} can be attributed to the changes in the recombination rate
coefficients used (barring any computational or round-off errors). For Cu and
Zn we use the recommended rate coefficients of Mazzotta (private communication)
based on extrapolation of the fitting parameters from other elements. The
sources of EII data used in our CIE calculations are listed in
Tables~\ref{tab:data_summary} and \ref{tab:data_summary_gu}. 

\citet{Mm02} have
also published EII rate coefficients for Cu and Zn, but they differ from the
Mazzotta (private communication) data only for the lowest 3 ionization stages
of Cu and the lowest 4 ionization stages of Zn. For these 7 ions, however, the
data of Mazzitelli \& Mattioli do not offer any significant improvement on the
Mazzotta data as  the Mazzitelli \& Mattioli data are taken from sources that
predate the Mazzotta work \citep{Lotz68a,Higg89a}. Since the intention of this
paper is to investigate the effects of updated recombination rate coefficients,
and for consistency with the comparisons for other elements, we use the
Mazzotta EII rate coefficients for Cu and Zn.

A fully relativistic treatment of electron impact ionization is often required
for highly ionized species, which require much higher incident electron energies
to ionize.  For example, the ionization of near fully-stripped U is highly
dependent on the relativistic treatment \citep{Pind88}, particularly for
$s$-orbitals. On the other hand, \citet{Loch05} found a semi-relativistic
treatment of W$^{9+}$ at 5~keV to be in close agreement with experiment. The
same study showed that a fully relativistic treatment was not required until
W$^{64+}$.  For all lower tungsten ion stages, a semi-relativistic treatment
produced good results. Since we do not consider elements above Zn in this work, 
theoretical calculations using a fully
relativistic treatment are not needed for the present CIE modeling.
A semi-relativistic approach
should be able to produce accurate results for the higher charge states of the
ions in this work.

It is not clear, however, that reliable semi-relativistic calculations exist among
the currently used recommended EII rate coefficients. We have
already mentioned the problems noted by \citet{Kato91}. Additionally, much of
the data are based on experiments with unknown metastable fractions. The
resulting rate coefficients represent some average over a distribution of ground
state and metastable populations. This is often an acceptable approximation for
magnetically confined fusion plasmas, which can have high metastable ion
content. But this is generally unsuitable for astrophysical plasmas of the type
considered here where the ions are in their ground state.  Lastly, the
recommended EII data currently used by the astrophysics community has not
undergone any significant revision or laboratory benchmarking since around 1990.
It is clear that an updating of the EII database is sorely needed.

\section{Charge Transfer}
\label{sec:CT}

Charge transfer (CT), also known as charge exchange or electron capture, is  the
reaction whereby an electron is captured from a donor atom. For plasmas of
cosmic abundances, this is typically atomic hydrogen but, in some instances, can
also be atomic helium.

\begin{figure}
  %\centering
  \centering
  \includegraphics[angle=90,scale=.9]{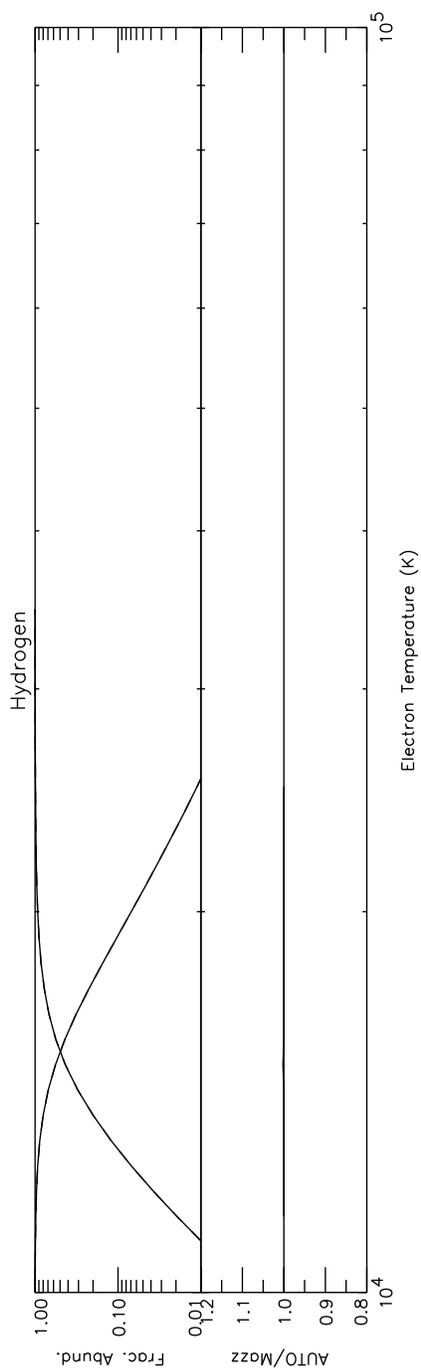}
  \caption{{\small{Ionization fractional abundance versus electron temperature for 
             all ionization stages of H. 
	     The upper graph shows our
	     results using the {\sc autostructure} RR  rate coefficients of 
	     Badnell (2006b; {\it solid curves}) and the abundances 
	     calculated by  
	     Mazzotta et al.\ (1998; {\it dashed curves}).
	     Here these curves lie on top of one another and cannot be
	     distinguished. The lower graph shows 
	     the ratio of the calculated abundances.
	     Comparison is made only for fractional abundances greater than
	     $10^{-2}$.
	     We label the results using the data of \protect\citet{Badn06b} 
	     as `AUTO' and \protect\citet{Mazz98a} as `Mazz'.}}}
  \label{fig:bad_maz_h}
\end{figure}

The importance of CT with H can be readily demonstrated \citep[e.g.,][]{King96}.
In CIE, CT is most important for near-neutral systems, up to 4 times ionized
\citep{Arna85a}. Using the data of \citet{Badn06a,Badn06b} and 
\citet{Gu03a,Gu03b,Gu04a} for these ions, a typical rate coefficient for 
DR+RR is on the order of
$\alpha_{\rm DR+RR} \approx 10^{-11}\;{\rm cm}^3\;{\rm s}^{-1}$. A large CT rate
coefficient is on the order of $\alpha_{\rm CT} \approx 10^{-9}\;{\rm
cm}^3\;{\rm s}^{-1}$ \citep{King96}. 
Using these values one finds that, for a given ion, the
ratio of the CT to DR+RR rates is given by 
\begin{equation} 
  \frac{R_{\rm CT}}{R_{\rm DR+RR}} = \frac{\alpha_{\rm CT}n_{\rm H^0}}
  {\alpha_{\rm DR+RR}n_\rme}  \approx 10^2\frac{n_{\rm H^0}}{n_\rme} \approx
  10^2\frac{n_{\rm H^0}}{n_{\rm H^+}} 
\end{equation}  
where $n_{\rm H^0}$ is the neutral H density, $n_\rme$ the electron density, and
$n_{\rm H^+}$ the H$^+$ density. The last approximation makes use of
$n_\rme\approx n_{\rm H^+}$ for plasmas with cosmic abundances. As one can see,
the CT rate will be equal to or greater than the DR+RR rate provided that
$n_{\rm H^0}/n_{\rm H^+} \gtrsim 0.01$. This inequality holds for electron
temperatures $T_\rme\lesssim$~25,000~K (see Fig.~\ref{fig:bad_maz_h} and
Table~\ref{tab:H}).

\citet{Arna85a} have investigated the effects of CT on CIE and found the process
to be important for a number of ions of astrophysical abundance. Those they list
for CT with H are He$^{q+}$ ($q=$1--2); C$^{q+}$, N$^{q+}$, O$^{q+}$ and S$^{q+}$
($q=$1--4); and  Ne$^{q+}$, Mg$^{q+}$, Si$^{q+}$ and Ar$^{q+}$  ($q=$2--4). For CT
with He, they list C$^{q+}$, N$^{q+}$, O$^{q+}$, Ne$^{q+}$ and Ar$^{q+}$
($q=$2--4); and Mg$^{q+}$, Si$^{q+}$ and S$^{q+}$ ($q=$3--4). The reverse
reaction, CT ionization with H$^+$, was also found to be comparable to EII for
O$^{0+}$, Si$^{0+}$, S$^{0+}$, Mg$^{1+}$ and Si$^{1+}$. CT ionization with
He$^{1+}$ was found to be important for C$^{1+}$, N$^{1+}$, Si$^{1+}$,
Si$^{2+}$, S$^{1+}$, S$^{2+}$ and Ar$^{1+}$.   Many of the ions listed in this
paragraph form at temperatures above 25,000~K. This points out the crudeness of
our above back-of-the-envelope estimate. Clearly a more detailed study is needed
using data more modern than that used by \citet{Arna85a}.

We will incorporate CT into our
results in a future work. Until then, our CIE results at low temperatures should be
used with caution, particularly for temperatures $\lesssim$~25,000~K, where neutral
H is abundant.

\section{Collisional Ionization Equilibrium Calculation}
\label{sec:calc}

We work in the coronal approximation where each ionization stage is represented
by its ground population only, i.e., metastable populations are assumed zero. We
neglect the effects of any radiation field, three-body processes, charge
transfer, and electron density effects.

The total abundance of element X is given by,
\begin{equation}
\label{eqn:tot}
  N_{\rm tot} = \sum_{i=0}^{Z} N^i ,
\end{equation}
where $N^q$ is the population of ion X$^{q+}$, $q$ is the charge, and $Z$ is the
atomic number of X. The fractional abundance of charge state $q$ is then given
by,
\begin{equation}
  f^q = {N^q \over  N_{\rm tot}} .
\end{equation}
This leads naturally to the normalization,
\begin{equation}
\label{eqn:normal}
  \sum_{i=0}^{Z} f^i = 1 .
\end{equation}

For a given system, the nearby charge stages are linked by the total
recombination and ionization coefficients. For the present calculations, the
total recombination coefficient is the sum of the DR and RR rate coefficients.
The total ionization rate coefficient is simply the EII rate coefficient. For
the more general CIE case, one would need also to account for CT recombination and
ionization.  For recombination from stage $q$ to $q-1$, we write the total
recombination rate coefficient as $\acr^{q\to q-1}$ and, for ionization from
stage $q$ to stage $q+1$, we write the total ionization rate coefficient as
$\scr^{q\to q+1}$. Here we only consider changes in charge state of $\Delta
q=\pm1$, as has commonly be done in the past for CIE calculations. Changes of 
$\Delta q>\pm1$ will be considered in future work.

In coronal
equilibrium, the populations are unchanging in time and can be written in terms
of $\acr$, $\scr$ and electron density, $n_\rme$, in matrix
form as,
\begin{equation}
  \label{eqn:matrix}
  N_{\rm tot}\frac{\rm d}{{\rm d}t}
  \left[\begin{array}{c}
    f^0 \\ f^1 \\ \vdots
  \end{array}\right]
  =N_{\rm tot}n_\rme \left[
  \begin{array}{ccc}
    -\scr^{0\to1} & \acr^{1\to0} & 0 \\
    \scr^{0\to1} & -\acr^{1\to0}-\scr^{1\to2} & \acr^{2\to1} \\ 
    0 & \scr^{1\to2} & \ddots
  \end{array}
  \right]\left[
  \begin{array}{c}
    f^0 \\ f^1 \\ \vdots
  \end{array}
  \right] = \left[
  \begin{array}{c}
    0 \\ 0 \\ \vdots
  \end{array}\right].
\end{equation}
We thus have a tridiagonal system in which the solution to all the ionization
stage populations is given in terms of any one population. The system is
tridiagonal since we consider only changes in charge state of $\Delta q=\pm1$.
Coupling equations~(\ref{eqn:normal}) and (\ref{eqn:matrix}) 
gives $Z+2$ equations
with $Z+1$ unknowns. The set of equations are then degenerate, so we 
divide equation~\ref{eqn:matrix} by $N_{\rm tot}n_\rme$ and then arbitrarily
replace the  first row of equation~(\ref{eqn:matrix}) with
equation~(\ref{eqn:normal}). This set of equations is then solved using the
ionization and recombination rate coefficients detailed previously. Our results
are presented in terms of the calculated fractional abundances $f$.

\section{Results}
\label{sec:Results}

\begin{figure}
  \centering
  \includegraphics[angle=90]{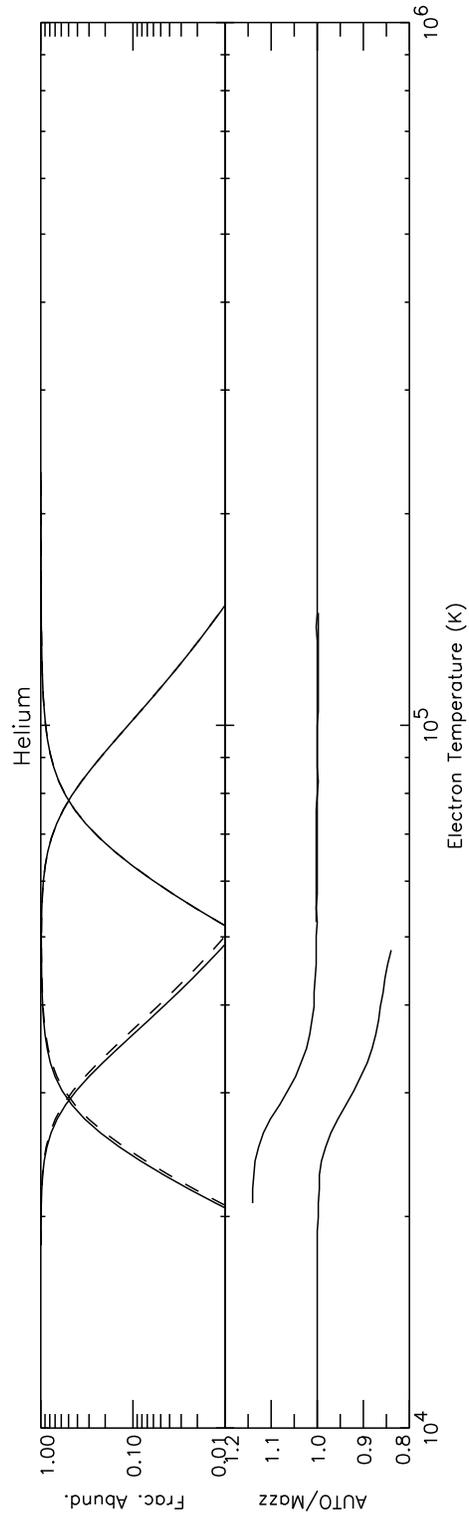}
  \caption[]{Same as Fig.~\protect\ref{fig:bad_maz_h} but for He and using the
             {\sc autostructure} DR data of \protect\citet{Badn06a}.}
  \label{fig:bad_maz_he}
\end{figure}

\begin{figure}
  \centering
  \includegraphics[angle=90]{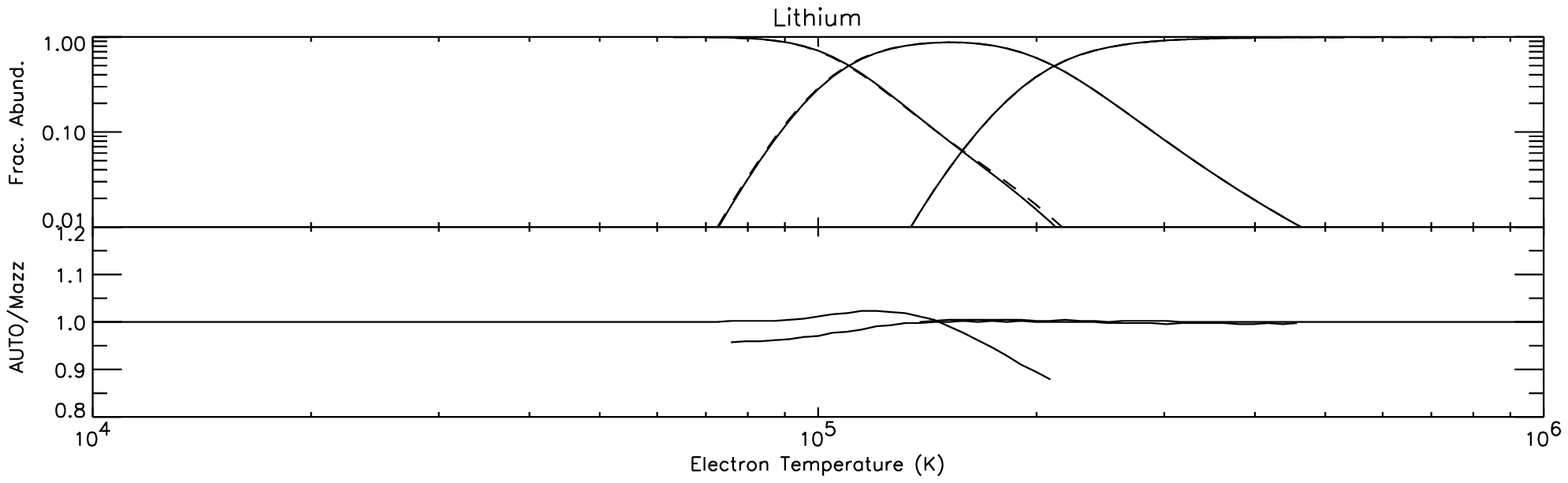}
  \caption[]{Same as Fig.~\protect\ref{fig:bad_maz_he} but for Li. The lowest ionization
             stage shown is He-like, the neutral ion forming at temperatures
	     below $10^4$~K.}
  \label{fig:bad_maz_li}
\end{figure}

\begin{figure}
  \centering
  \includegraphics[angle=90]{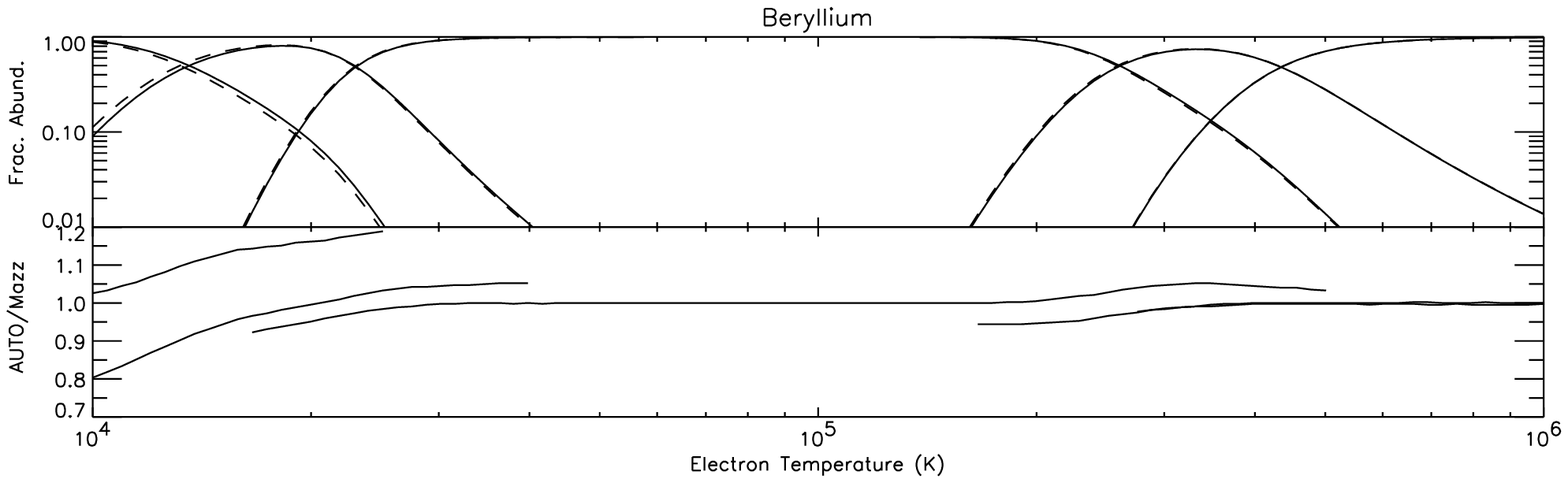}
  \caption[]{Same as Fig.~\protect\ref{fig:bad_maz_he} but for Be.}
  \label{fig:bad_maz_be}
\end{figure}

\begin{figure}
  \centering
  \includegraphics[angle=90]{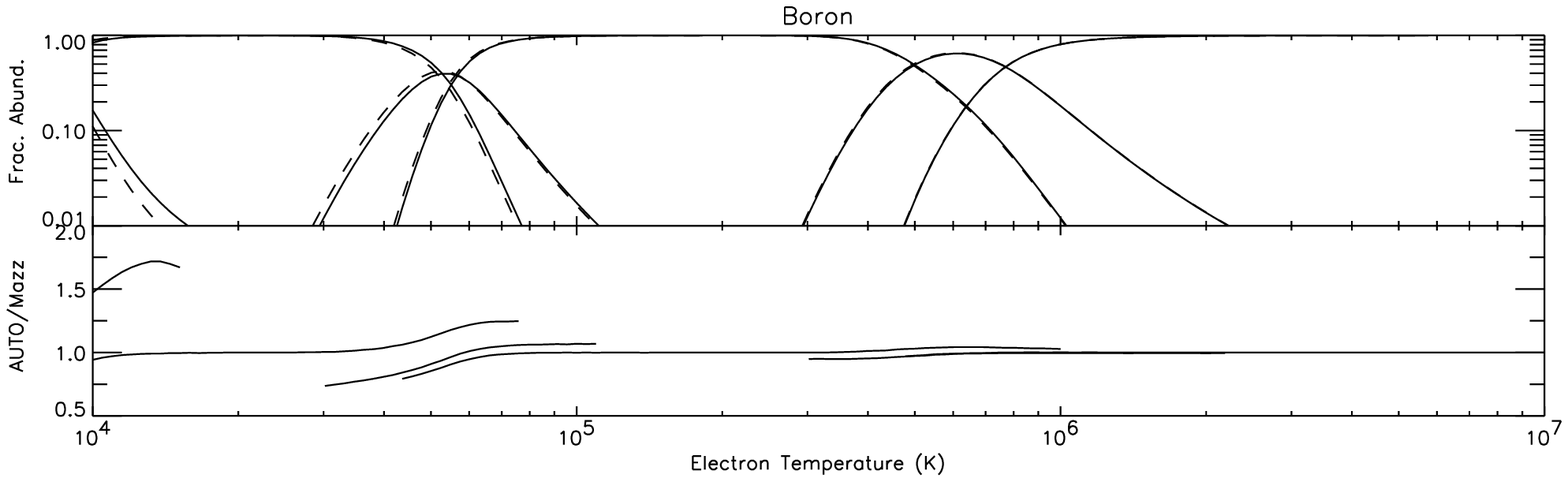}
  \caption[]{Same as Fig.~\protect\ref{fig:bad_maz_he} but for B.}
  \label{fig:bad_maz_b}
\end{figure}

\begin{figure}
  \centering
  \includegraphics[angle=90]{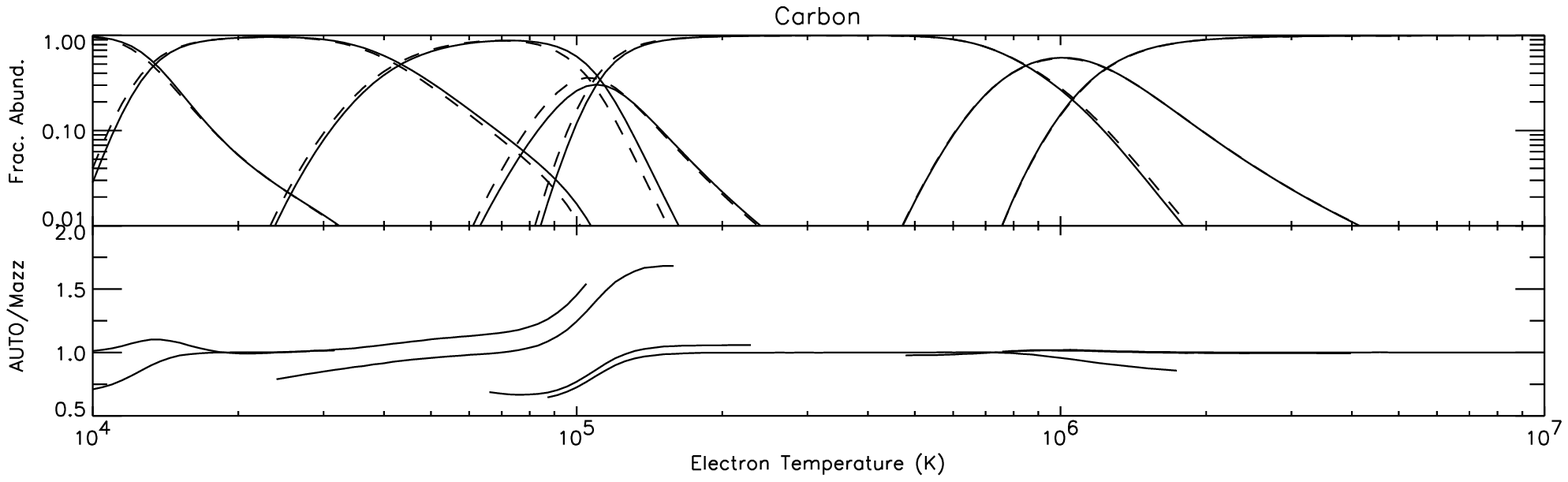}
  \caption[]{Same as Fig.~\protect\ref{fig:bad_maz_he} but for C.}
  \label{fig:bad_maz_c}
\end{figure}

\begin{figure}
  \centering
  \includegraphics[angle=90]{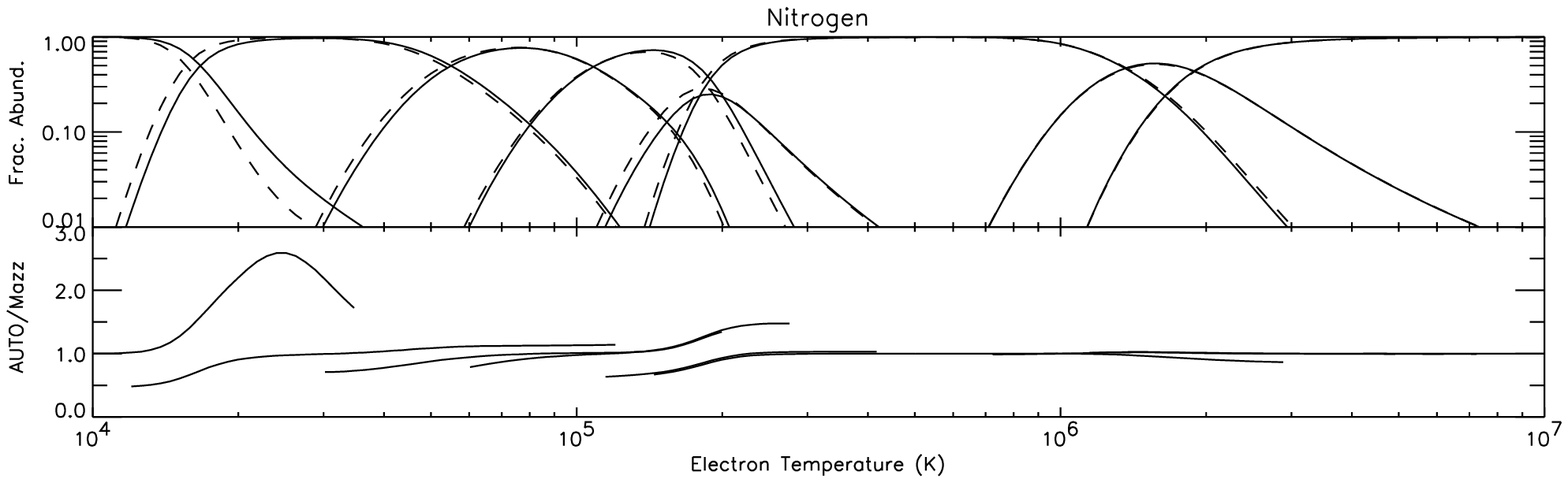}
  \caption[]{Same as Fig.~\protect\ref{fig:bad_maz_he} but for N.}
  \label{fig:bad_maz_n}
\end{figure}

\begin{figure}
  \centering
  \includegraphics[angle=90]{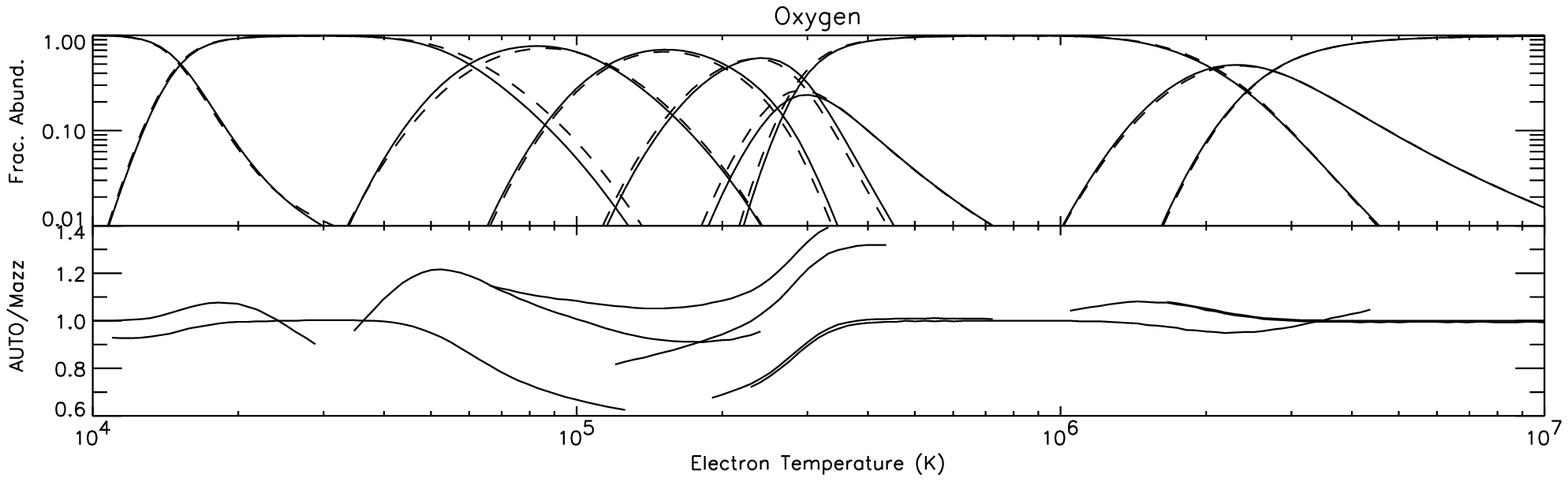}
  \caption[]{Same as Fig.~\protect\ref{fig:bad_maz_he} but for O.}
  \label{fig:bad_maz_o}
\end{figure}

\begin{figure}
  \centering
  \includegraphics[angle=90]{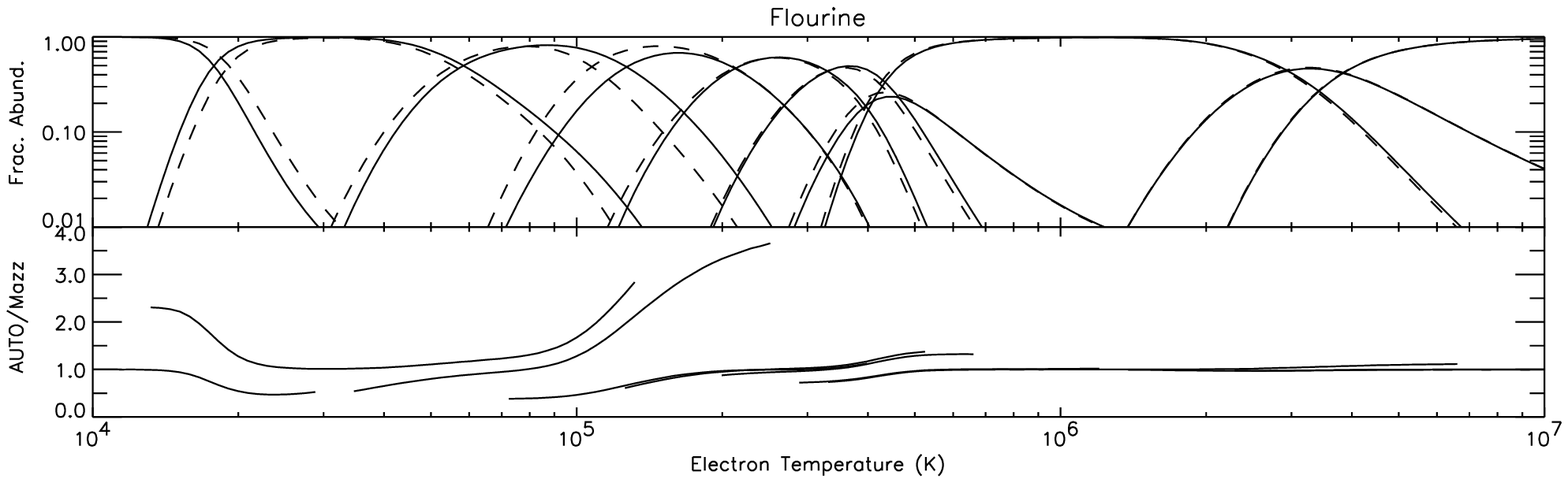}
  \caption[]{Same as Fig.~\protect\ref{fig:bad_maz_he} but for F.}
  \label{fig:bad_maz_f}
\end{figure}

\begin{figure}
  \centering
  \includegraphics[angle=90]{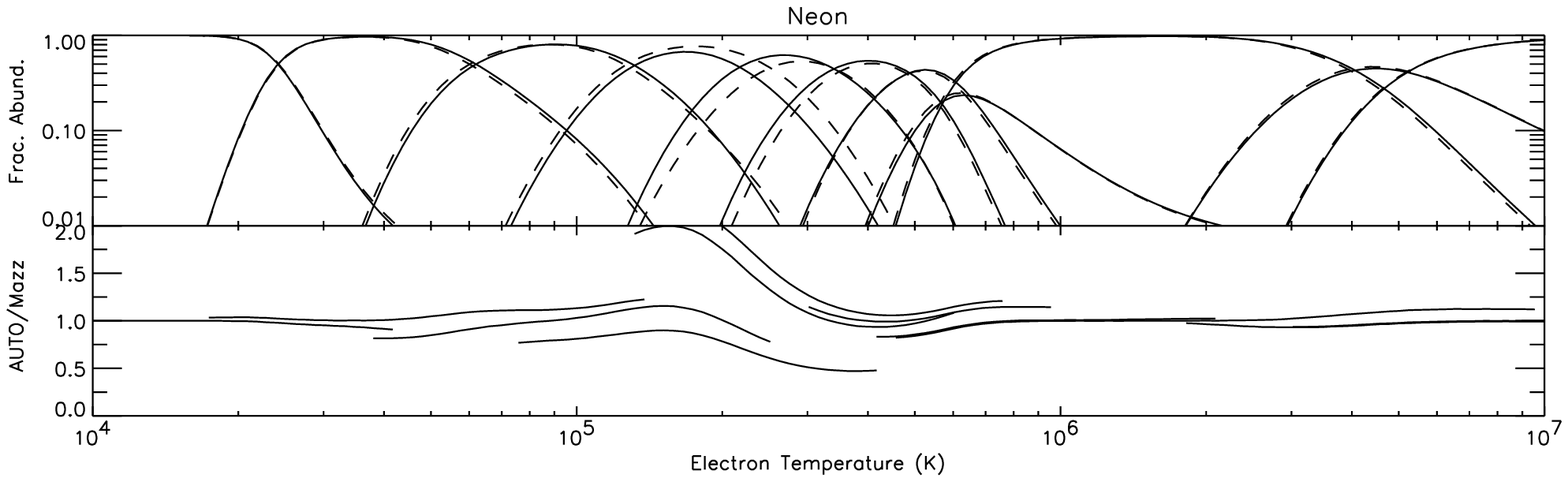}
  \caption[]{Same as Fig.~\protect\ref{fig:bad_maz_he} but for Ne.}
  \label{fig:bad_maz_ne}
\end{figure}

\begin{figure}
  \centering
  \includegraphics[angle=90]{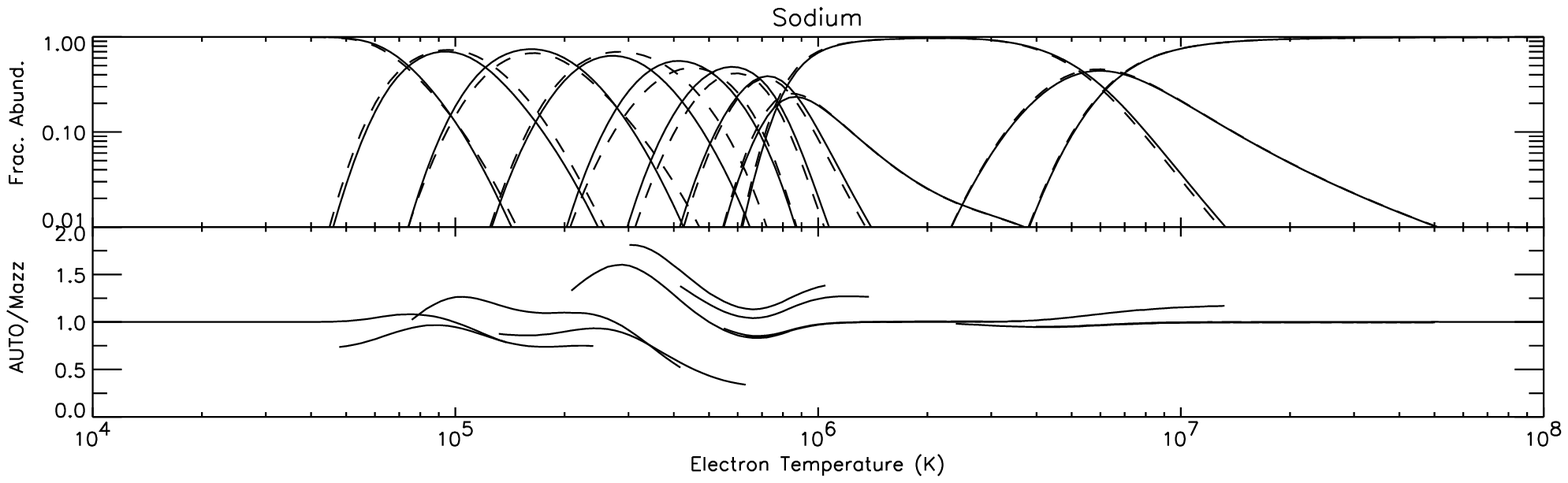}
  \caption[]{Same as Fig.~\protect\ref{fig:bad_maz_he} but for Na. The lowest ionization
             stage shown is Ne-like, the neutral ion forming at temperatures
	     below $10^4$~K.}
  \label{fig:bad_maz_na}
\end{figure}

\begin{figure}
  \centering
  \includegraphics[angle=90]{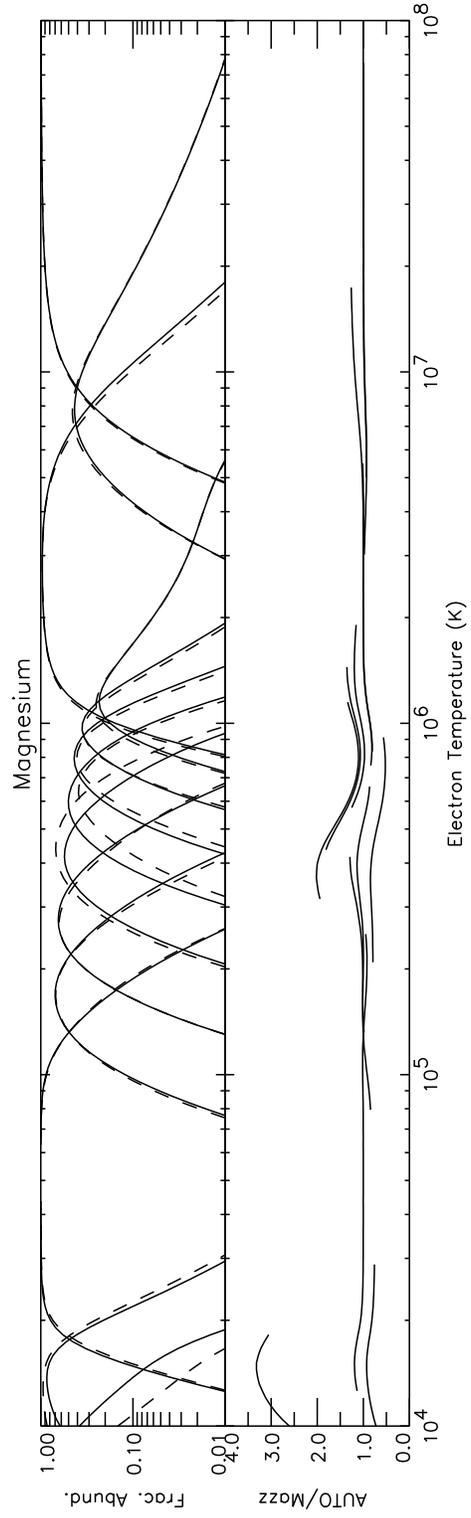}
  \caption[]{Same as Fig.~\protect\ref{fig:bad_maz_he} but for Mg.}
  \label{fig:bad_maz_mg}
\end{figure}

\begin{figure}
  \centering
  \includegraphics[angle=90]{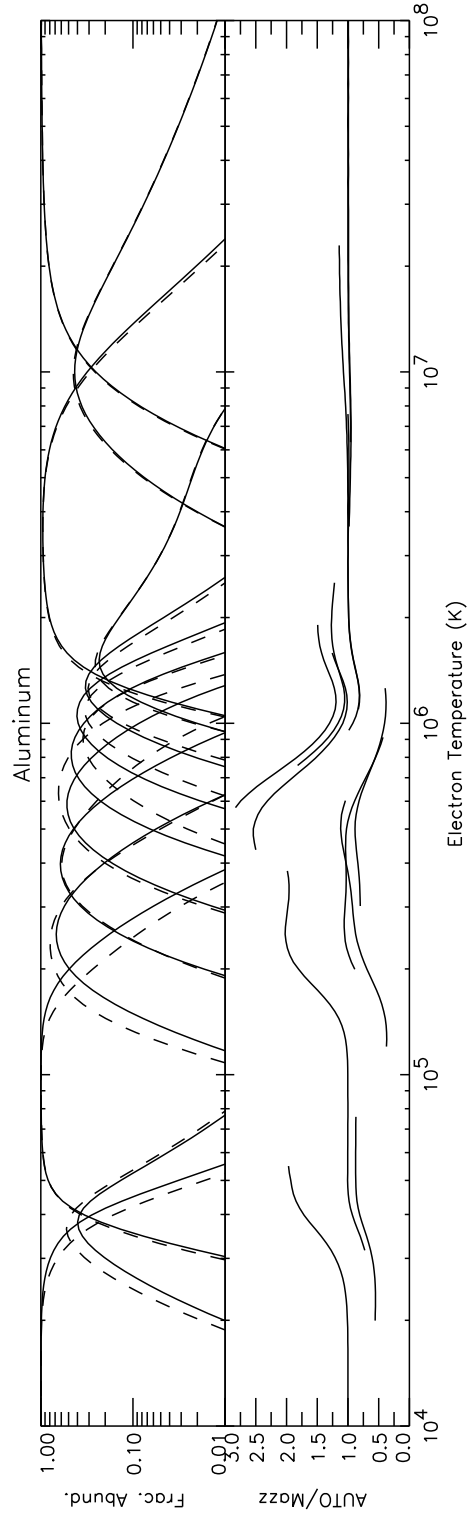}
  \caption[]{\footnotesize{Same as Fig.~\protect\ref{fig:bad_maz_he} but for Al, and using the
             DR and RR rate coefficients of \protect\citet{Mazz98a} for ions not
	     calculated by \protect\citet{Badn06a} and \protect\citet{Badn06b}, respectively. The lowest 
	     ionization stage shown is Mg-like, the neutral ion forming at 
	     temperatures below $10^4$~K.}}
  \label{fig:bad_maz_al}
\end{figure}

\begin{figure}
  \centering
  \includegraphics[angle=90]{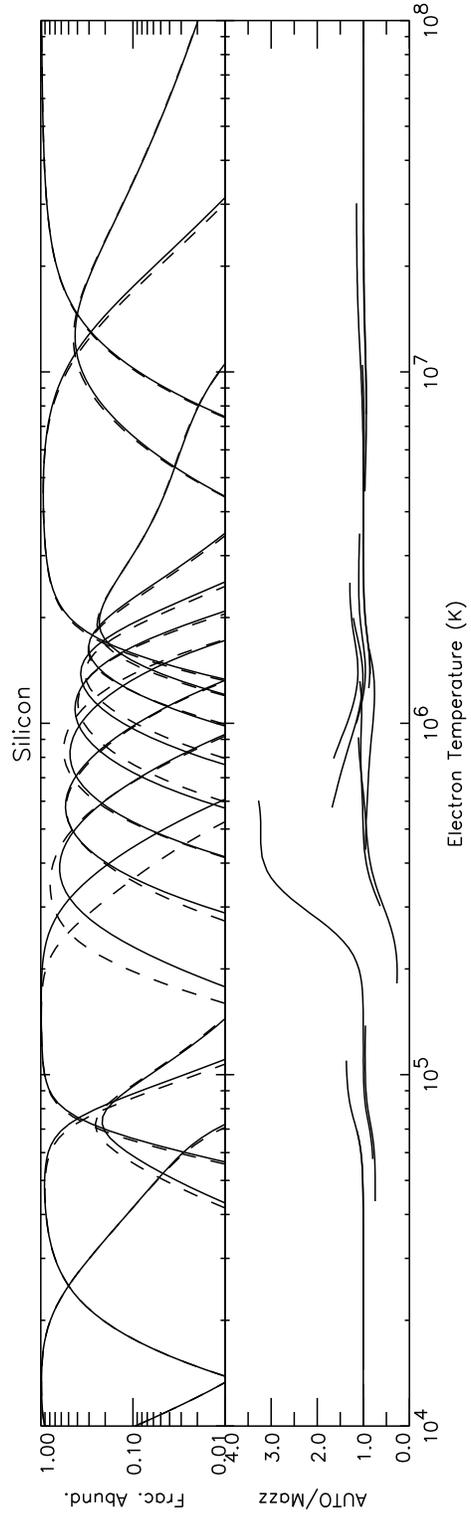}
  \caption[]{Same as Fig.~\protect\ref{fig:bad_maz_al} but for Si. All ionization stages are shown.}
  \label{fig:bad_maz_si}
\end{figure}

\begin{figure}
  \centering
  \includegraphics[angle=90]{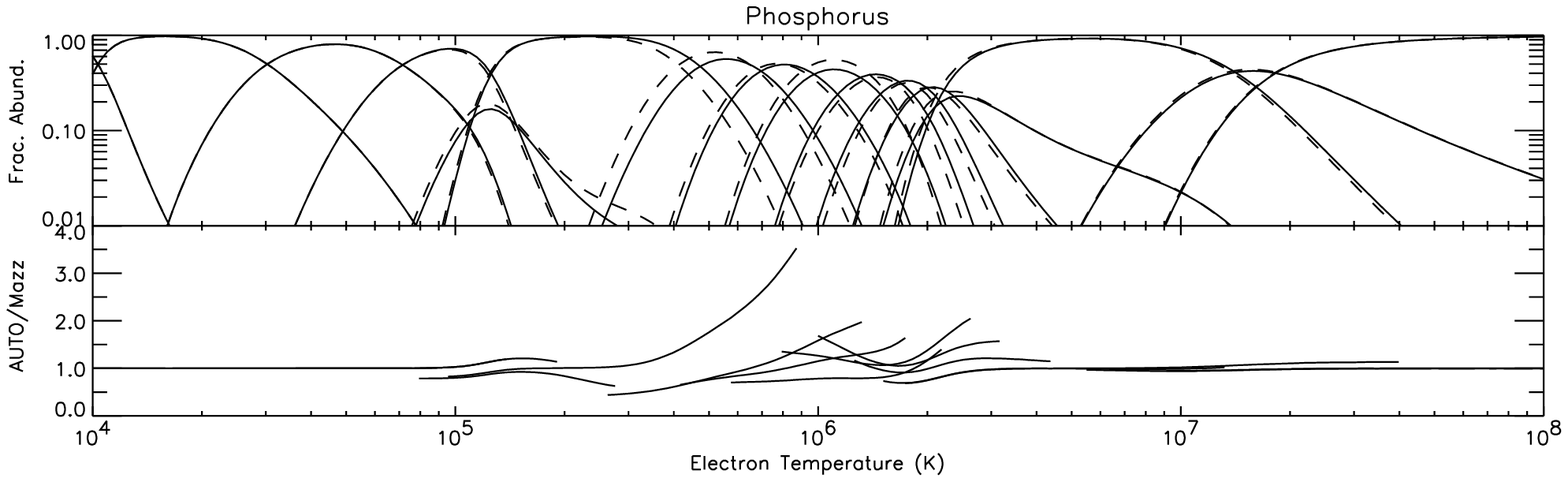}
  \caption[]{Same as Fig.~\protect\ref{fig:bad_maz_si} but for P.}
  \label{fig:bad_maz_p}
\end{figure}

\begin{figure}
  \centering
  \includegraphics[angle=90]{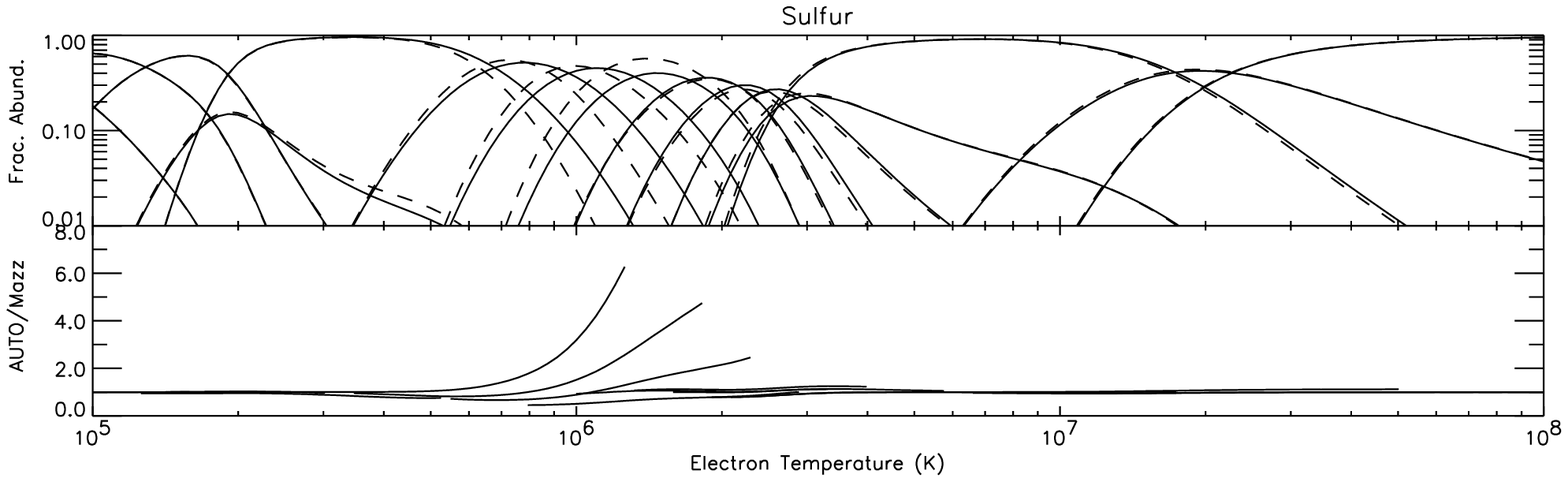}
  \caption[]{Same as Fig.~\protect\ref{fig:bad_maz_si} but for S. The lowest ionization
             stage shown is P-like.}
  \label{fig:bad_maz_s}
\end{figure}

\begin{figure}
  \centering
  \includegraphics[angle=90]{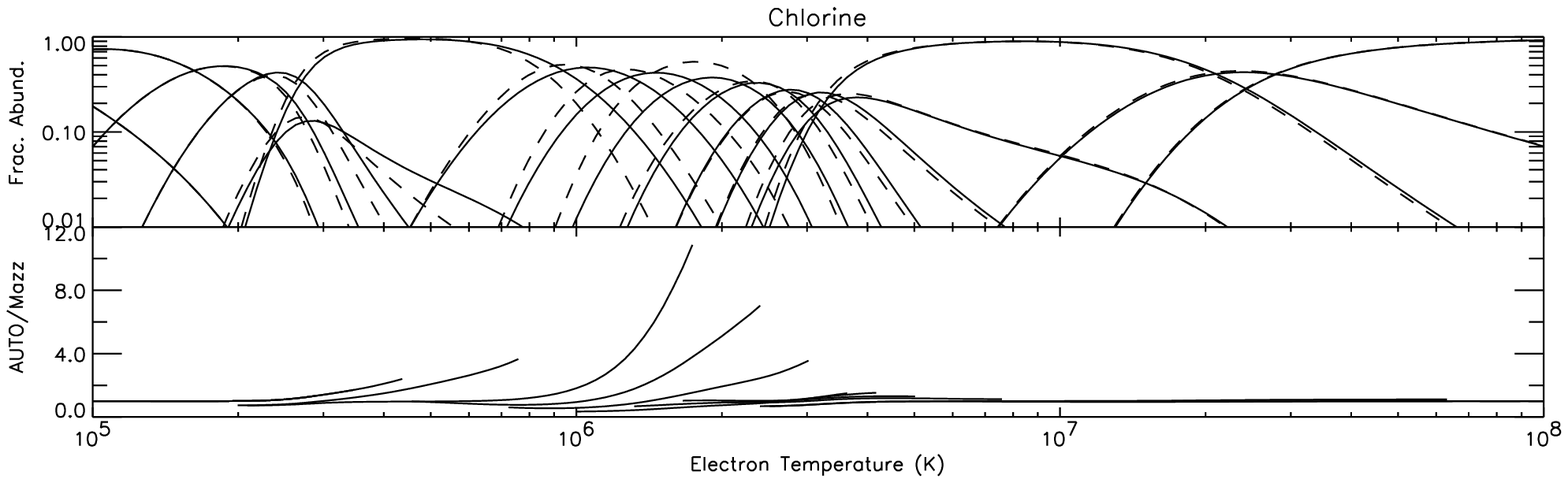}
  \caption[]{Same as Fig.~\protect\ref{fig:bad_maz_s} but for Cl.}
  \label{fig:bad_maz_cl}
\end{figure}

\begin{figure}
  \centering
  \includegraphics[angle=90]{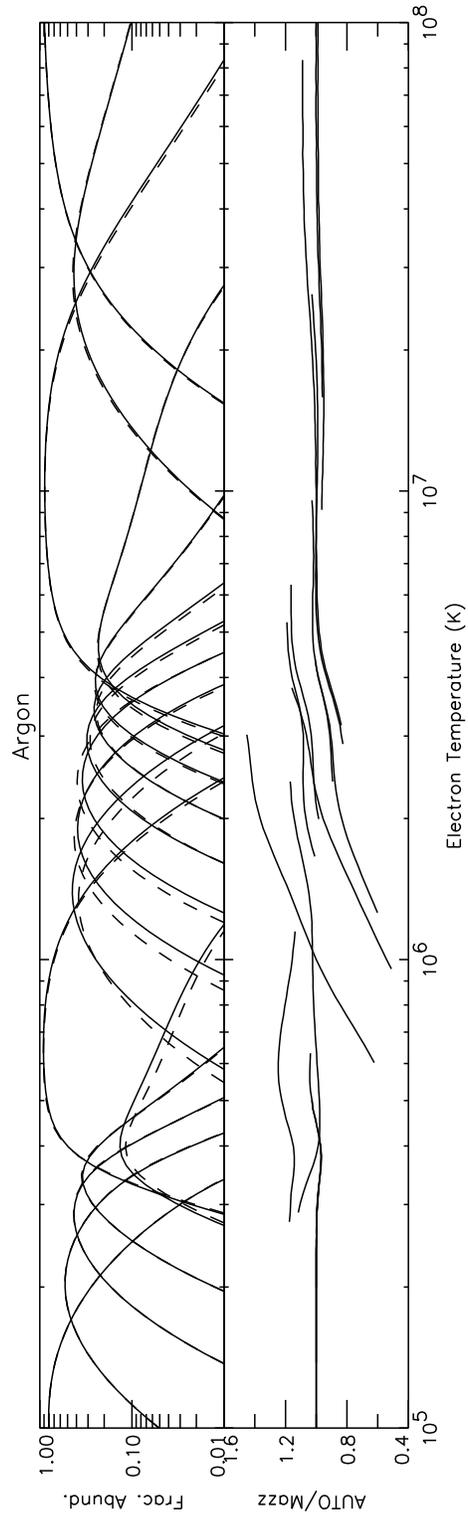}
  \caption[]{Same as Fig.~\protect\ref{fig:bad_maz_s} but for Ar.}
  \label{fig:bad_maz_ar}
\end{figure}

\begin{figure}
  \centering
  \includegraphics[angle=90]{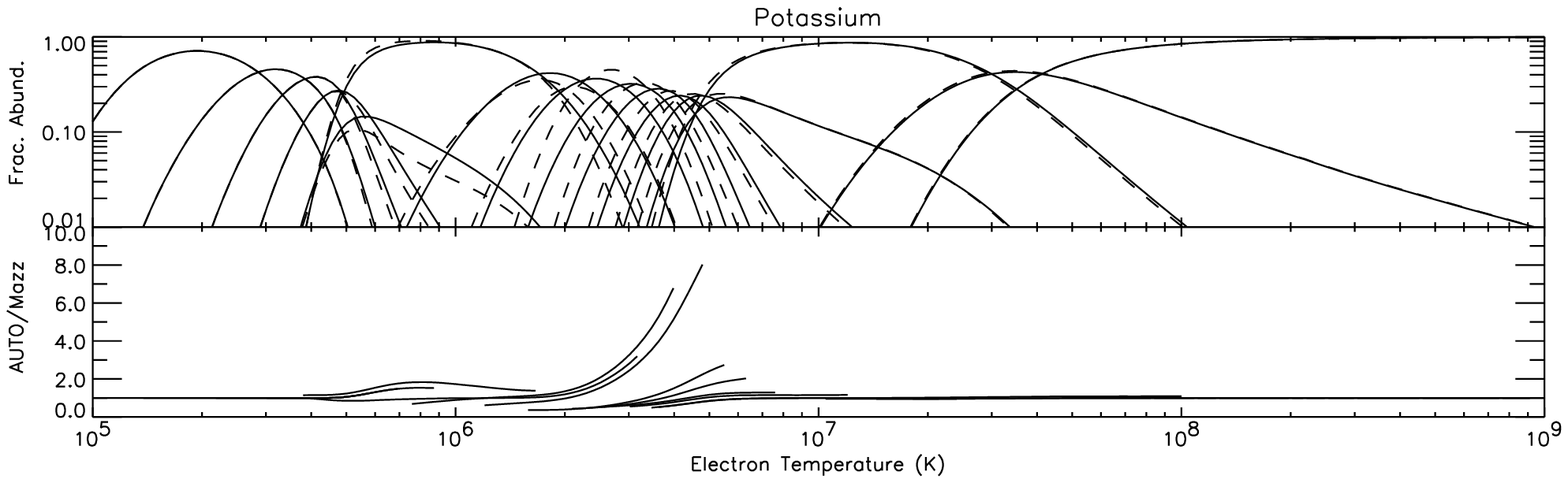}
  \caption[]{Same as Fig.~\protect\ref{fig:bad_maz_s} but for K.}
  \label{fig:bad_maz_k}
\end{figure}

\begin{figure}
  \centering
  \includegraphics[angle=90]{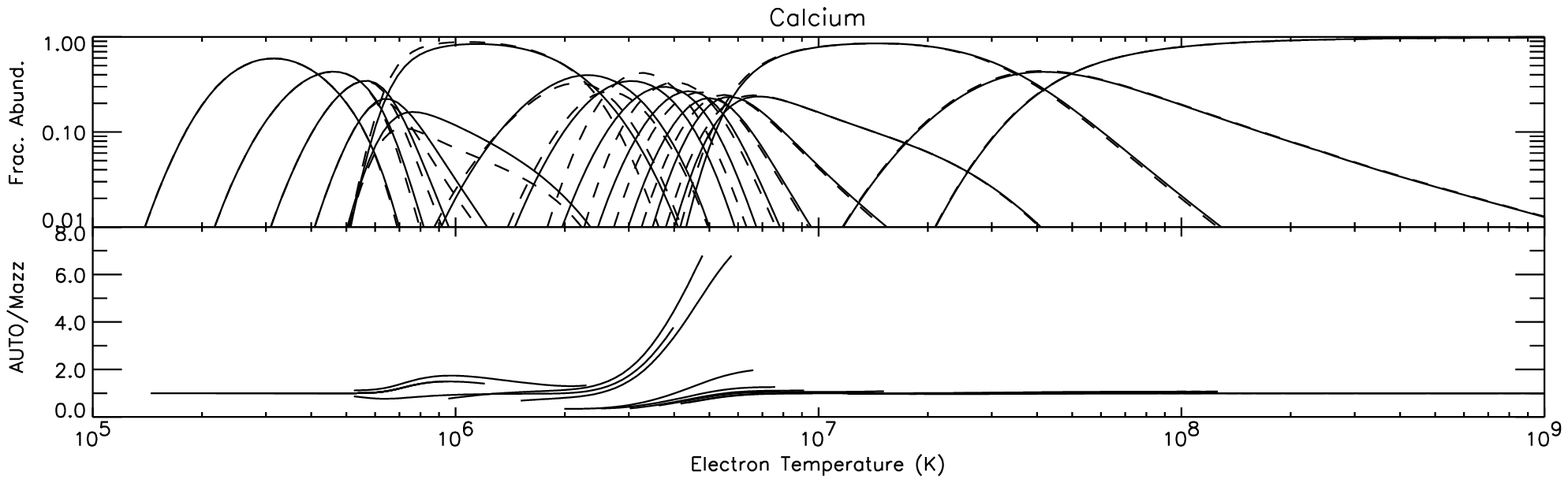}
  \caption[]{Same as Fig.~\protect\ref{fig:bad_maz_s} but for Ca.}
  \label{fig:bad_maz_ca}
\end{figure}

\begin{figure}
  \centering
  \includegraphics[angle=90]{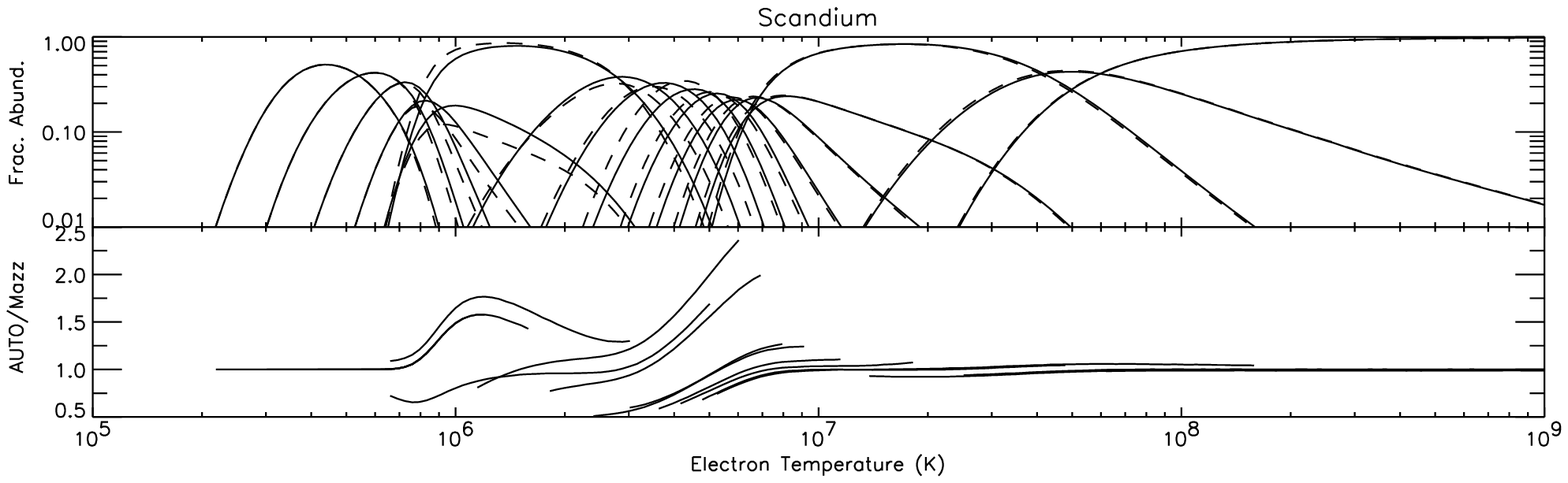}
  \caption[]{Same as Fig.~\protect\ref{fig:bad_maz_s} but for Sc.}
  \label{fig:bad_maz_sc}
\end{figure}

\begin{figure}
  \centering
  \includegraphics[angle=90]{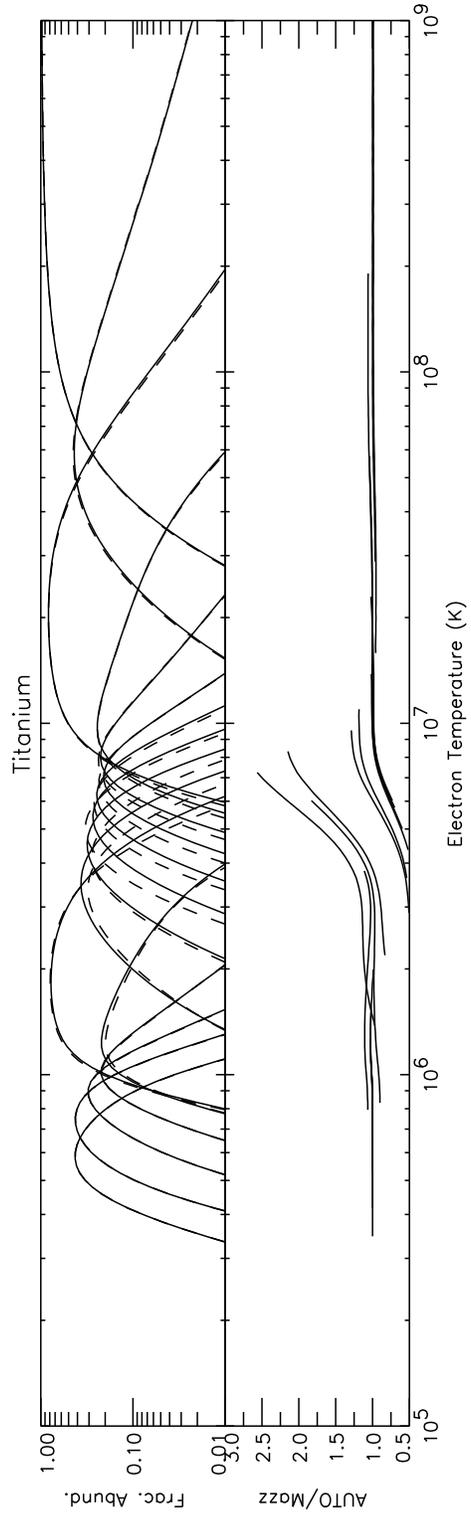}
  \caption[]{Same as Fig.~\protect\ref{fig:bad_maz_s} but for Ti.}
  \label{fig:bad_maz_ti}
\end{figure}

\begin{figure}
  \centering
  \includegraphics[angle=90]{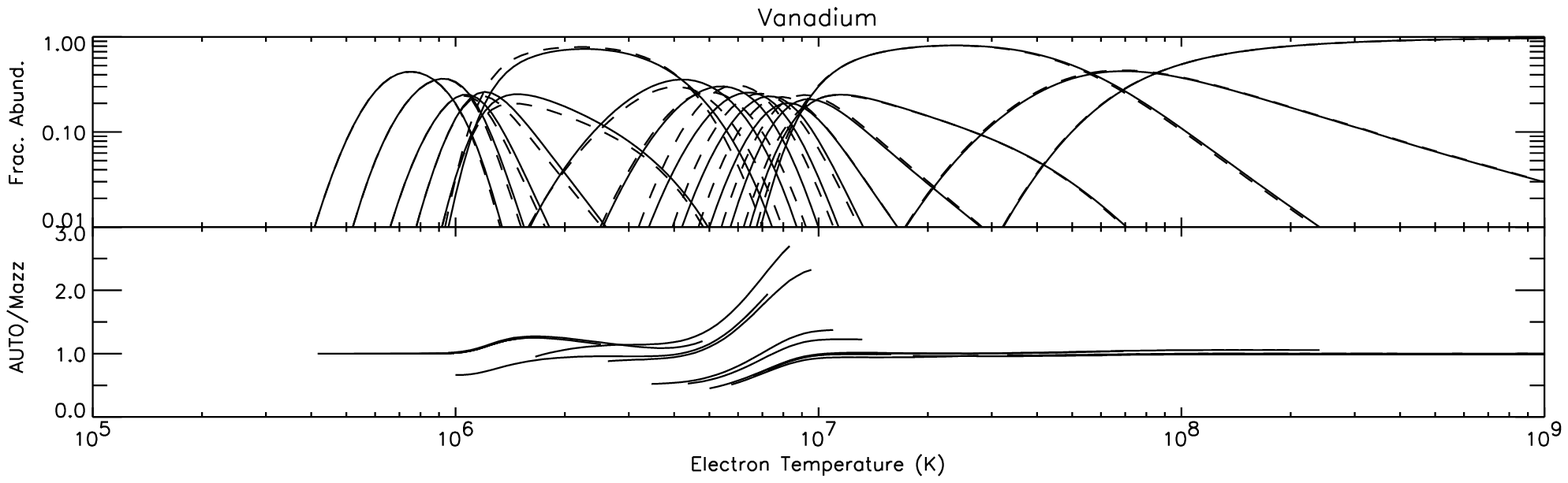}
  \caption[]{Same as Fig.~\protect\ref{fig:bad_maz_s} but for V.}
  \label{fig:bad_maz_v}
\end{figure}

\begin{figure}
  \centering
  \includegraphics[angle=90]{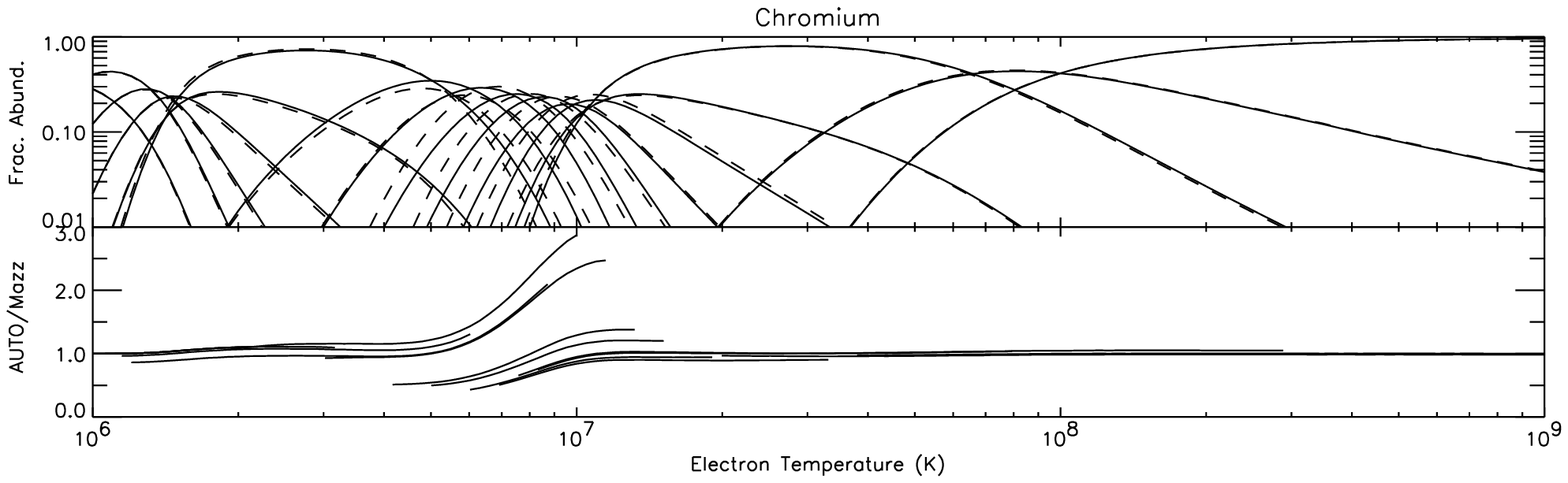}
  \caption[]{Same as Fig.~\protect\ref{fig:bad_maz_s} but for Cr.}
  \label{fig:bad_maz_cr}
\end{figure}

\begin{figure}
  \centering
  \includegraphics[angle=90]{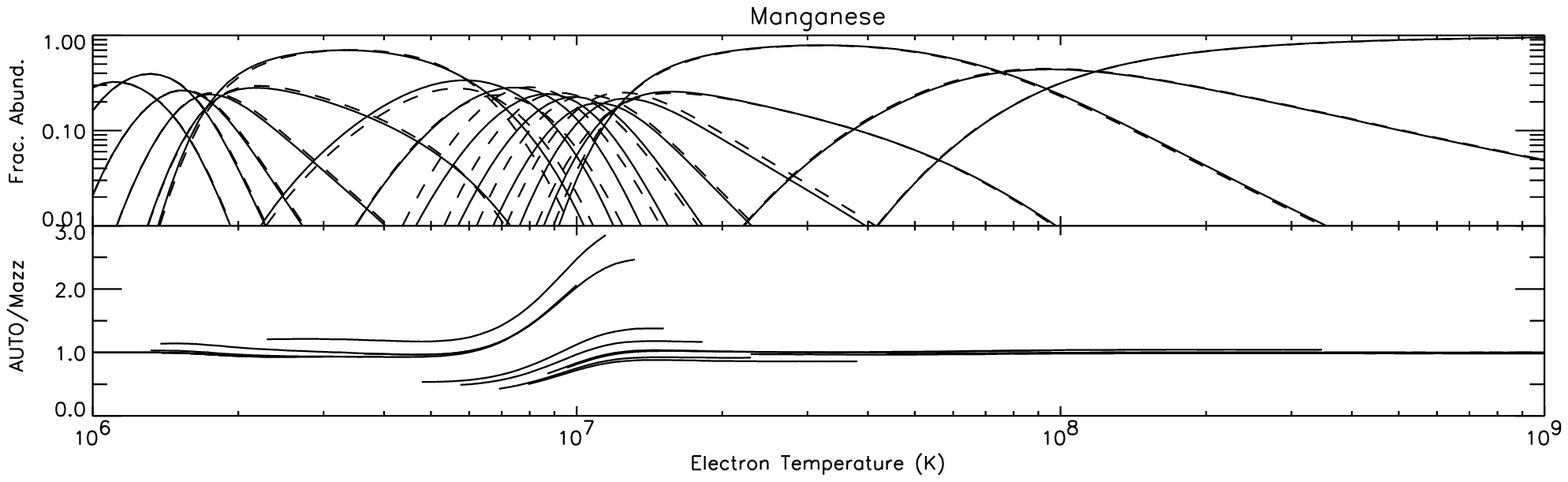}
  \caption[]{Same as Fig.~\protect\ref{fig:bad_maz_s} but for Mn.}
  \label{fig:bad_maz_mn}
\end{figure}

\begin{figure}
  \centering
  \includegraphics[angle=90]{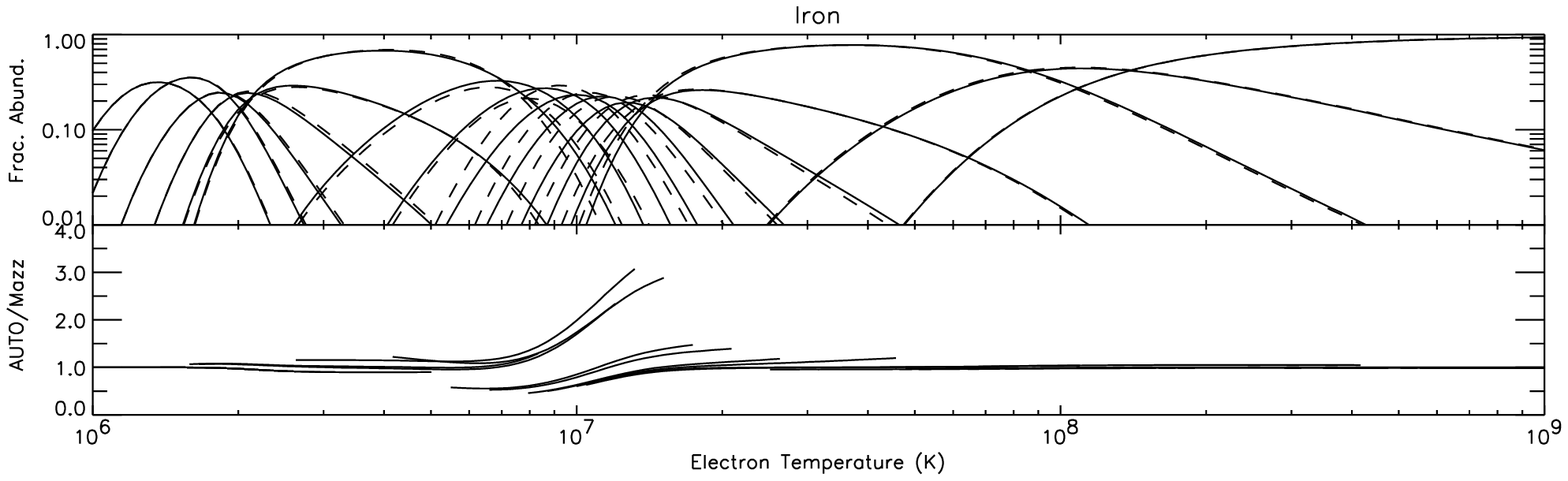}
  \caption[]{Same as Fig.~\protect\ref{fig:bad_maz_s} but for Fe.}
  \label{fig:bad_maz_fe}
\end{figure}

\begin{figure}
  \centering
  \includegraphics[angle=90]{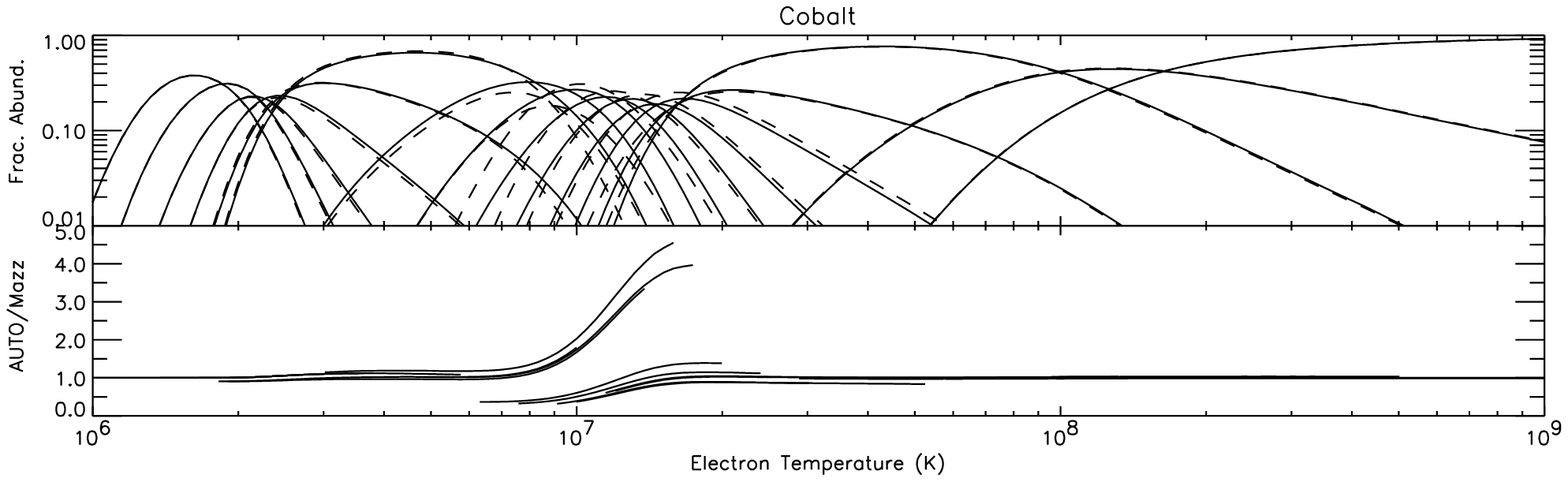}
  \caption[]{Same as Fig.~\protect\ref{fig:bad_maz_s} but for Co.}
  \label{fig:bad_maz_co}
\end{figure}

\begin{figure}
  \centering
  \includegraphics[angle=90]{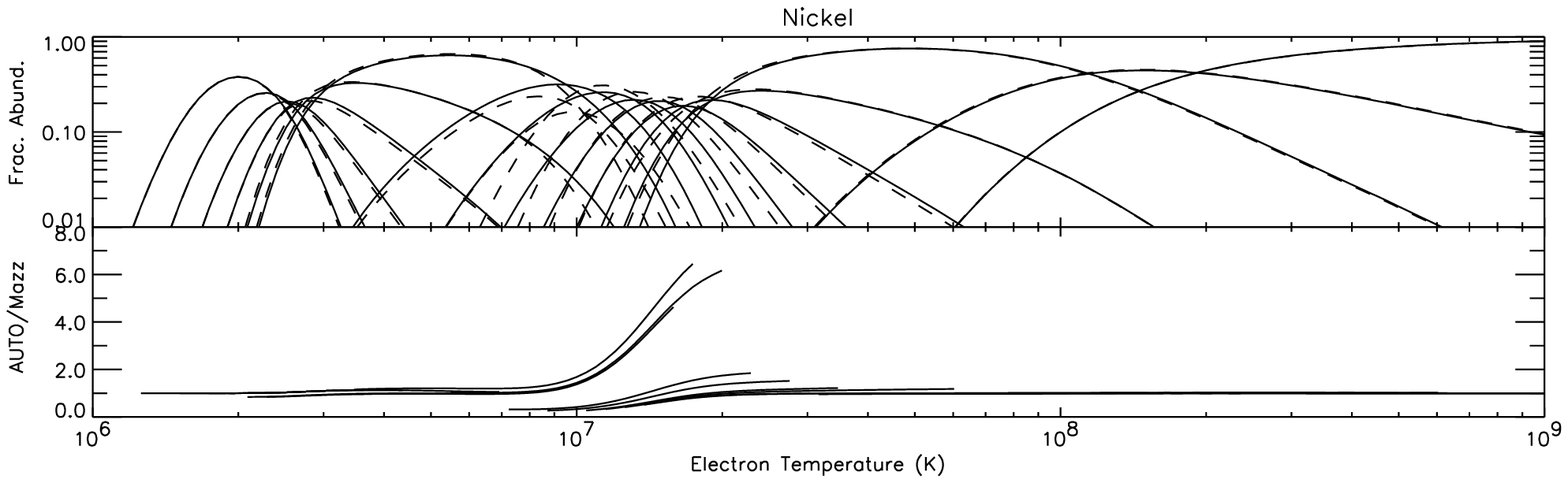}
  \caption[]{Same as Fig.~\protect\ref{fig:bad_maz_s} but for Ni.}
  \label{fig:bad_maz_ni}
\end{figure}

\begin{figure}
  \centering
  \includegraphics[angle=90]{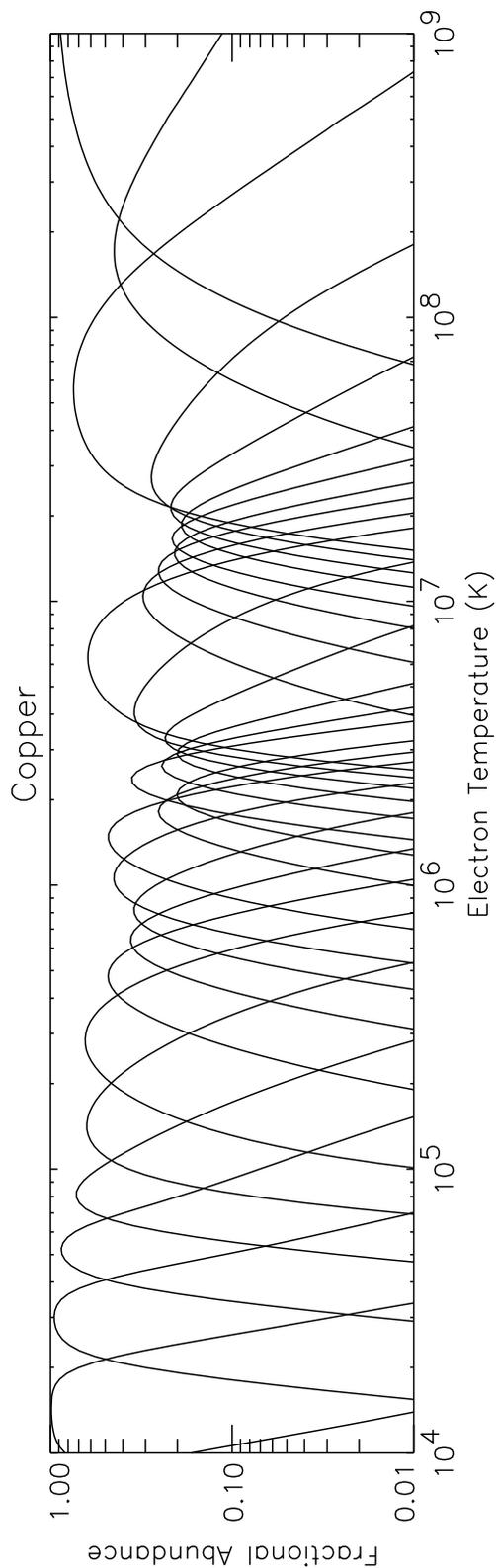}
  \caption[]{\scriptsize{Ionization fractional abundance versus electron temperature for 
             all ionization stages of Cu. The DR rate coefficients of 
	     \protect\citet{Badn06a} are used for H- through Na-like ions, the RR
	     rate coefficients of \protect\citet{Badn06b} are used for bare through
	     Na-like ions, and the recommended DR and RR data of Mazzotta 
	     (private communication) are used for all other ions. 
	     Also used are the EII data
	     of Mazzotta (private communication).}}
  \label{fig:cu}
\end{figure}

\begin{figure}
  \centering
  \includegraphics[angle=90]{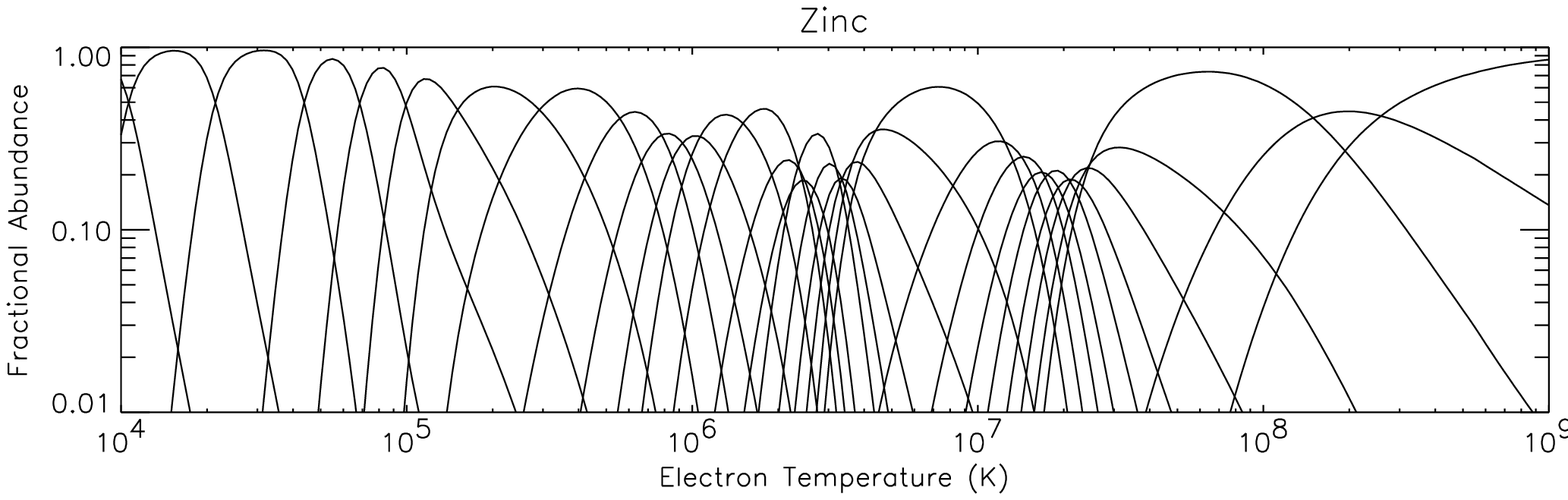}
  \caption[]{Same as Fig.~\protect\ref{fig:cu} but for Zn.}
  \label{fig:zn}
\end{figure}

\begin{figure}
  \centering
  \includegraphics[angle=90]{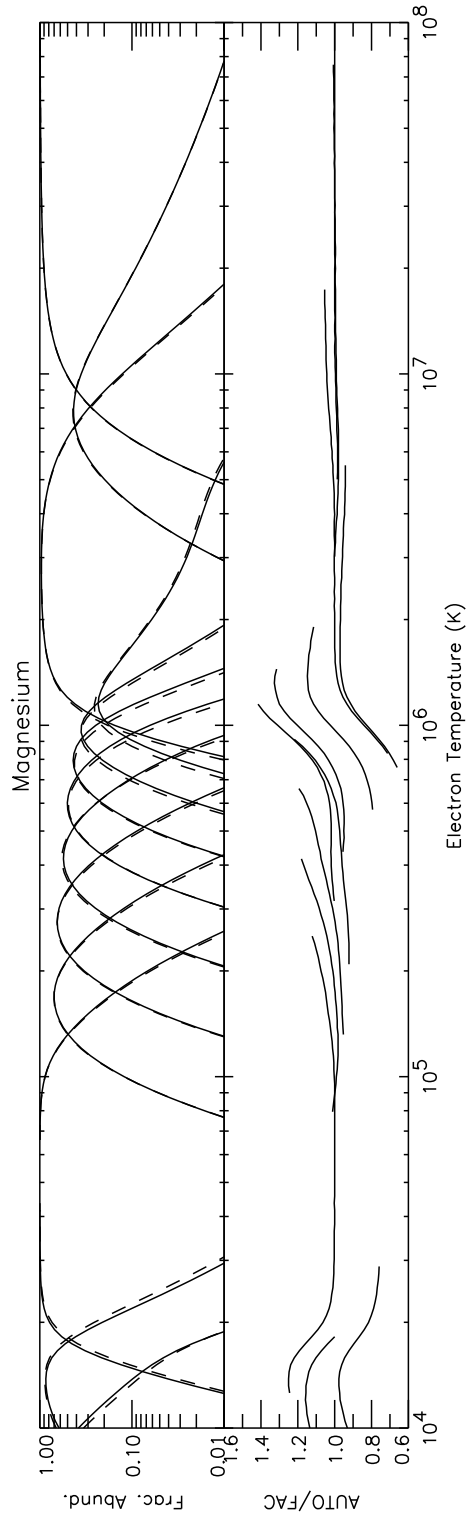}
  \caption[]{\tiny{Ionization fractional abundance versus electron temperature for 
             all ionization stages of Mg. 
	     The {\it solid curves} of the upper graph show
	     the ionization fractional abundance as calculated 
	     using the {\sc autostructure} DR  
	     and RR data of \protect\citet{Badn06a,Badn06b}.
	     The 
	     {\it dashed curves} show the abundances 
	     as calculated using the {\sc fac} DR rate coefficients of
	     \protect\citet{Gu03a,Gu04a} for H- through Na-like ions, 
	     the {\sc fac} RR rate coefficients of \protect\citet{Gu03b} for bare through F-like
	     ions, and the RR rate coefficient of \protect\citet{Mazz98a} for the
	     Na-like ion. 
	     The EII rate coefficients used are those of \protect\citet{Mazz98a}.
	     The lower graph shows the ratio of the calculated abundances.
	     Comparison is made only for fractional abundances greater than
	     $10^{-2}$.
	     We label the results using the data of 
	     \protect\citet{Badn06a} and \protect\citet{Badn06b} as `AUTO'
	     and \protect\citet{Gu03a,Gu03b,Gu04a} as `FAC'.}}
  \label{fig:bad_gu_mg}
\end{figure}

\begin{figure}
  \centering
  \includegraphics[angle=90]{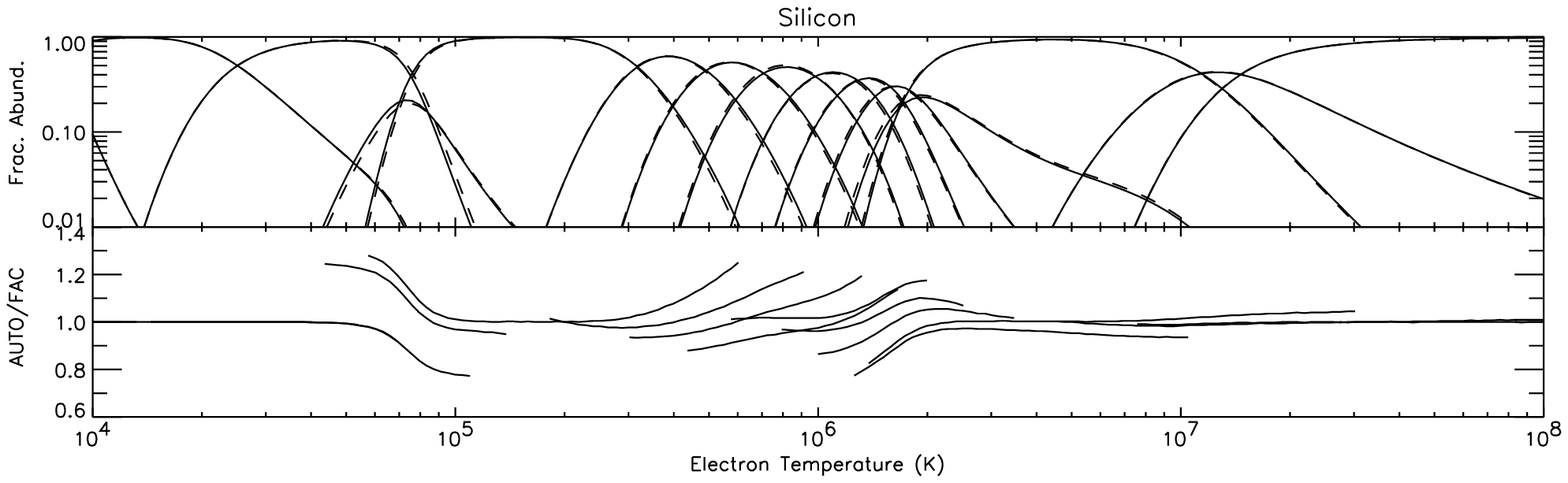}
  \caption[]{Same as Fig.~\protect\ref{fig:bad_gu_mg} but for Si, and using the DR and
             RR rate coefficients of \protect\citet{Mazz98a} for ions not calculated by
	     \protect\citet{Gu03a,Gu04a} or by \protect\citet{Badn06a,Badn06b}.}
  \label{fig:bad_gu_si}
\end{figure}

\begin{figure}
  \centering
  \includegraphics[angle=90]{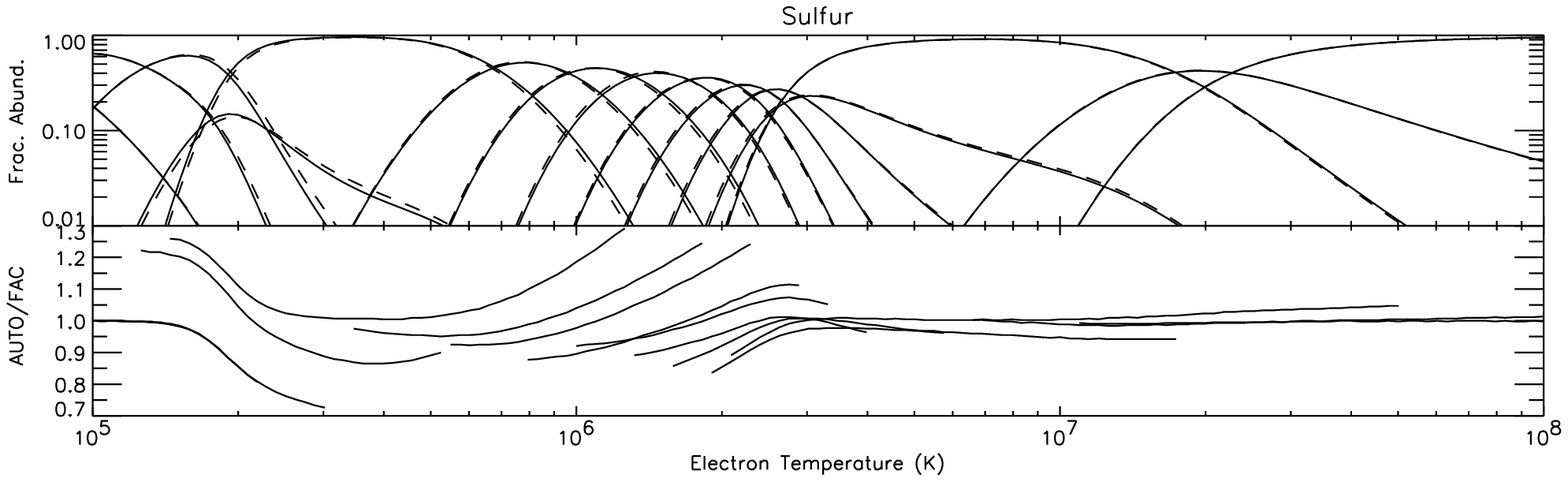}
  \caption[]{Same as Fig.~\protect\ref{fig:bad_gu_si} but for S. The lowest ionization
             stage shown is Si-like.}
  \label{fig:bad_gu_s}
\end{figure}

\begin{figure}
  \centering
  \includegraphics[angle=90]{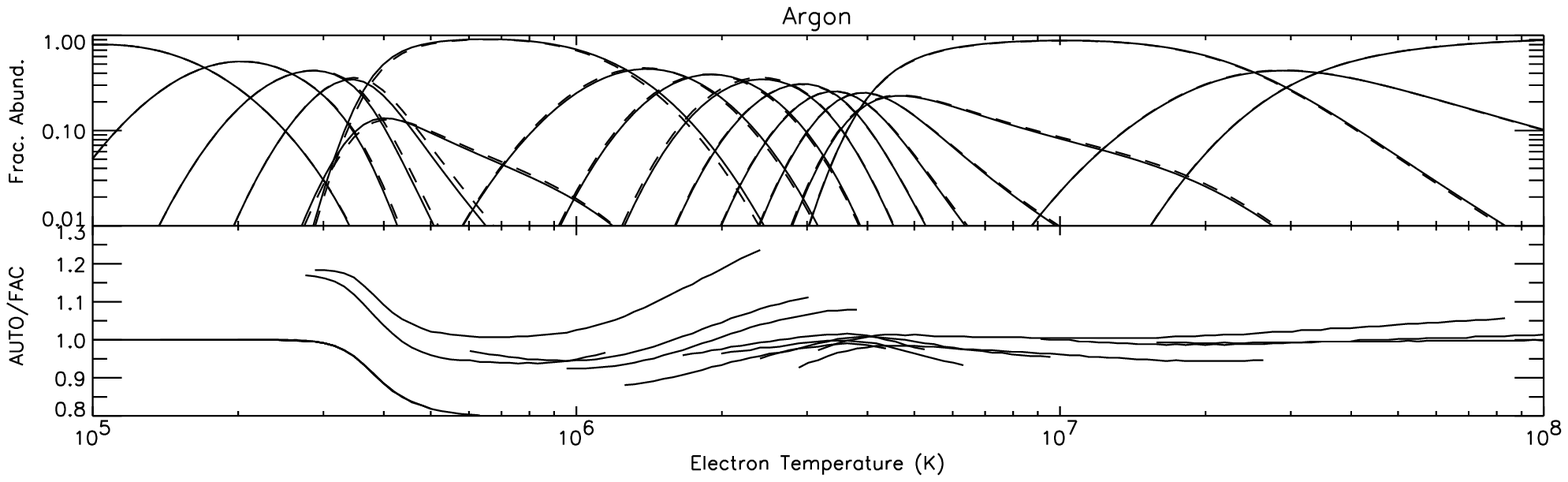}
  \caption[]{Same as Fig.~\protect\ref{fig:bad_gu_si} but for Ar. The lowest ionization
             stage shown is P-like.}
  \label{fig:bad_gu_ar}
\end{figure}

\begin{figure}
  \centering
  \includegraphics[angle=90]{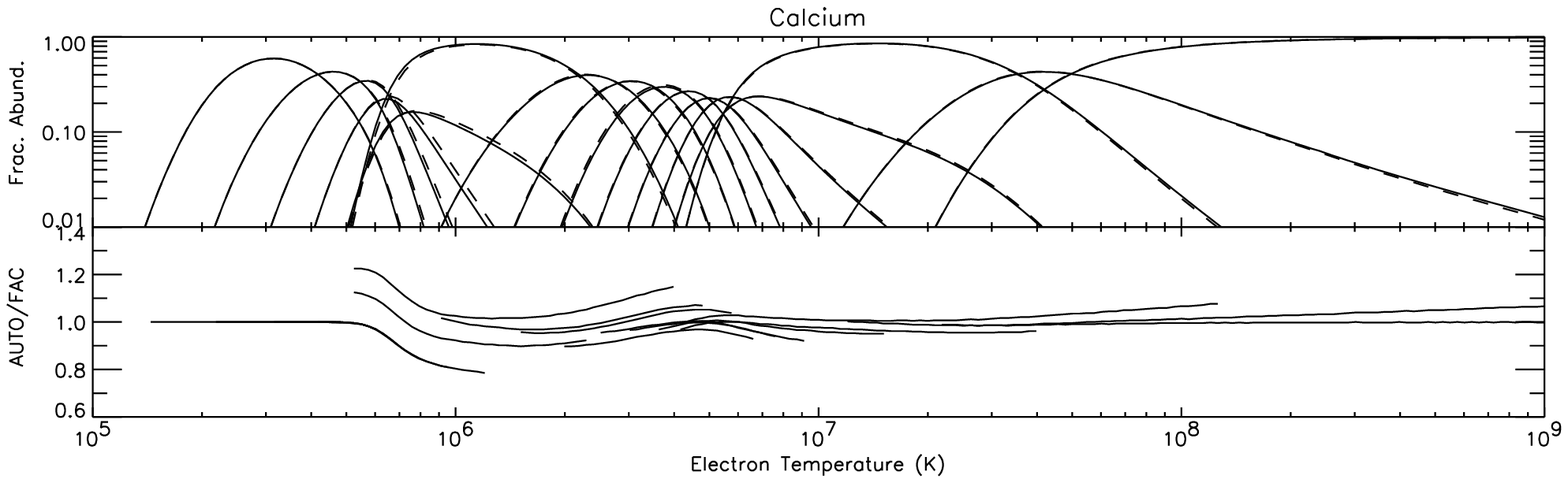}
  \caption[]{Same as Fig.~\protect\ref{fig:bad_gu_ar} but for Ca.}
  \label{fig:bad_gu_ca}
\end{figure}

\begin{figure}
  \centering
  \includegraphics[angle=90]{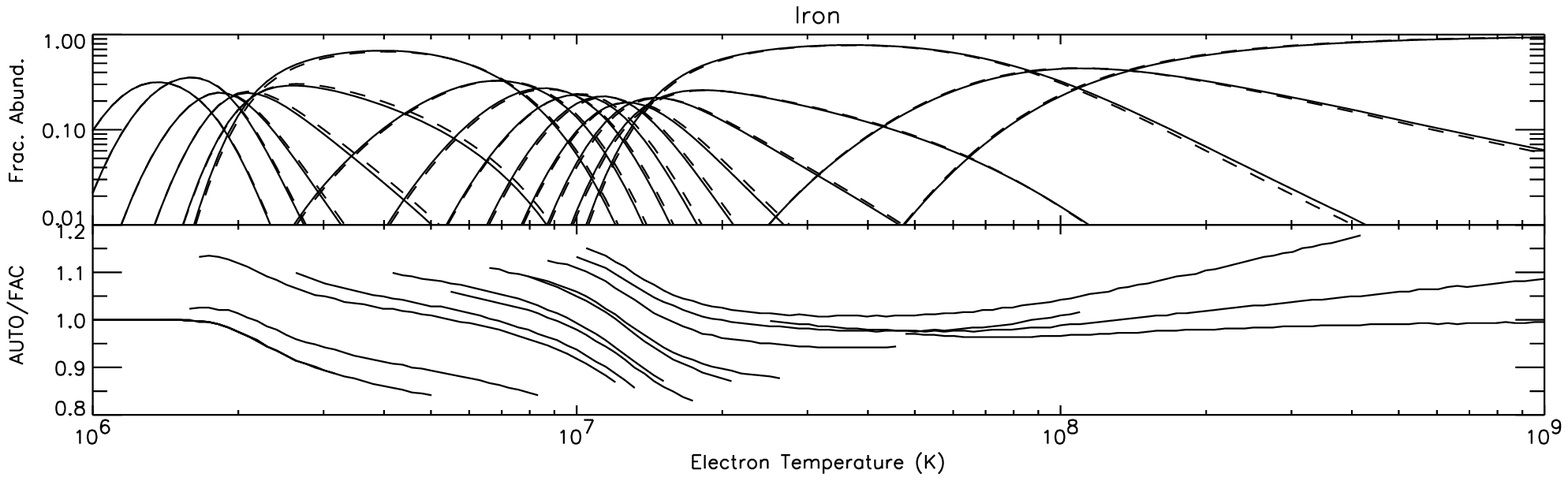}
  \caption[]{Same as Fig.~\protect\ref{fig:bad_gu_ar} but for Fe.}
  \label{fig:bad_gu_fe}
\end{figure}

\begin{figure}
  \centering
  \includegraphics[angle=90]{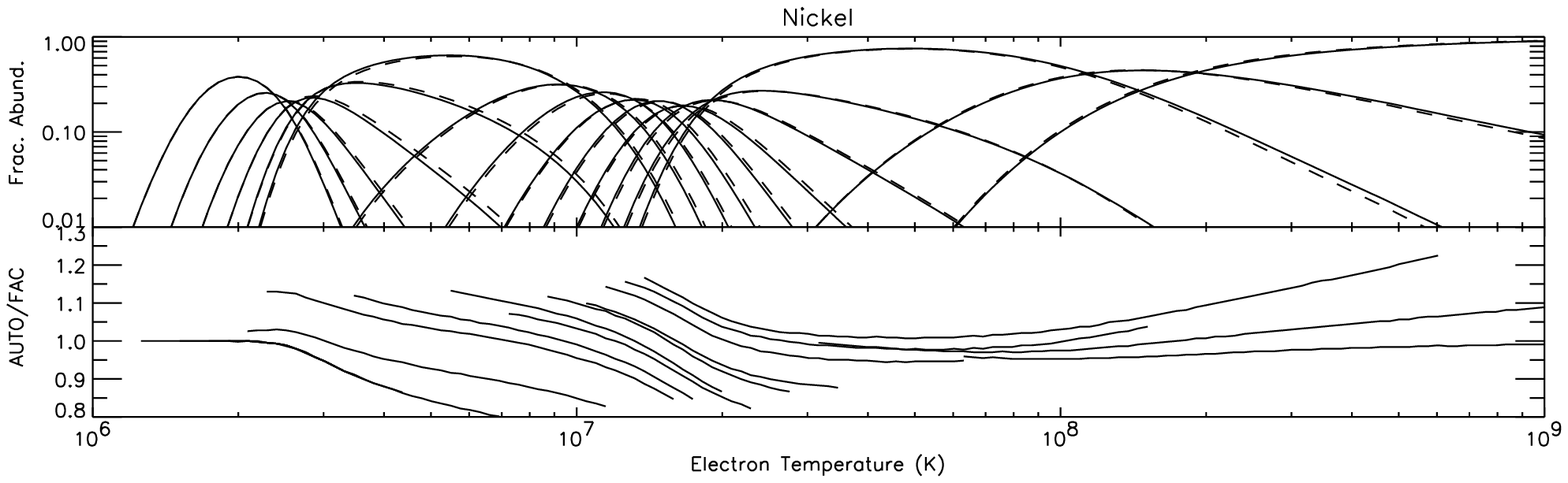}
  \caption[]{Same as Fig.~\protect\ref{fig:bad_gu_ar} but for Ni.}
  \label{fig:bad_gu_ni}
\end{figure}

Figs.~\ref{fig:bad_maz_h} to \ref{fig:bad_maz_ni} show our calculated
fractional abundances compared to those of \citet{Mazz98a}.  Mazzotta et al.\
did not publish results for the ionization balance of Cu and Zn, so we present
our results without comparison (Figs.~\ref{fig:cu} and \ref{fig:zn}). Our
calculated fractional abundances are given in tabular form in
Tables~\ref{tab:H} to \ref{tab:Zn}.  For the elements where DR and RR rate
coefficients are also provided by \citet{Gu03a,Gu03b,Gu04a}, we compare the
results using his data with those using the data of \citet{Badn06a,Badn06b}.
These comparisons are shown in Figs.~\ref{fig:bad_gu_mg} to
\ref{fig:bad_gu_ni}. The calculated fractional abundances based on the data of
Gu are given in Tables~\ref{tab:Mg_Gu} to \ref{tab:Ni_Gu}.
To make these tables easily machine readable,
we tabulate fractional abundances down
to $10^{-15}$ and fix fractional abundances below this value to $10^{-15}$.

We limit our studies to the temperature range $10^4$--$10^9$~K.  The recombination
data of \citet{Badn06a,Badn06b} and \citet{Gu03a,Gu03b,Gu04a} covers  ionization
stages from bare through Na-like. For ions with more electrons we use the data
recommended by \citet{Mazz98a} and Mazzotta (private communication). As the CIE
calculations move to ionization stages with more electrons than Na-like, which
has 11 electrons, the effects of the new DR and RR data decrease, as is expected.
These differences become insignificant typically  by the time one reaches the
Si- or P-like isoelectronic sequence, with 14 and 15 electrons, respectively.
Because of this, and to avoid figures becoming overly congested, we generally
plot our results only for ionization stages with 15 or fewer electrons. The
lower temperature limit shown is also increased to focus on these ionization
stages. Where all ionization stages are not shown, the figure caption indicates
such.

\section{Discussion}
\label{sec:Discussion}

In the discussion below we point out differences in the CIE ionic fractional
abundances we have calculated using the {\sc autostructure} data of
\citet{Badn06a,Badn06b} with calculations using other data. Firstly, we compare
to the recommended CIE results of \citet{Mazz98a}, and then to the CIE results
using the {\sc fac} data of \citet{Gu03a,Gu03b,Gu04a}.  We highlight ions and
temperatures where the differences are larger than 20\%. The differences quoted
are the percentage increase or decrease in our calculations relative to the
fractional abundances of \citet{Mazz98a} or those calculated using the  data of
\citet{Gu03a,Gu03b,Gu04a}.
All differences discussed below can be attributed to the use of different 
DR and RR data sets.

To simplify the comparison, we  point out where there are large differences at
peak fractional abundance and at fractional abundances of 0.1 and 0.01. 
Table~\ref{tab:peak} lists ions where our peak abundances differ from those of
\citet{Mazz98a} by more than 20\%, or where the difference in peak formation
temperature is $\ge0.05$ in the dex. This table gives the percentage change in
peak fractional abundance and the change in temperature relative to the results
of \citet{Mazz98a}.

It is interesting to note that the differences in our calculated CIE fractional
abundances relative to those of \citet{Mazz98a} are, in general, much larger
than the differences between our results and the results using the data of
\citet{Gu03a,Gu03b,Gu04a}. In the former case, peak abundance differences of
nearly 60\% are found (see Fig.~\ref{fig:bad_maz_sc}) and the differences can be
larger than a factor of 11 (i.e., 1000\%) at fractional abundances down to 0.01 (see
Fig.~\ref{fig:bad_maz_cl}). For the latter case, peak abundance differences are
within 10\% and differences for fractional abundances down to 0.01 are within
50\%. This reflects the fact that the modern DR and RR data are in better
agreement with one another than with the older data.

We have not
investigated the reliability of the DR and RR data at temperatures where the
fractional abundance is $<0.01$; so our calculated fractional abundances must be
used with caution outside this range.
Comparison of the fractional abundances using the data of
\citet{Badn06a,Badn06b} and \citet{Gu03a,Gu03b,Gu04a} can be used to give an
estimate of how the uncertainties in these modern DR and RR calculations
translate into uncertainties in the CIE calculations for fractional abundances
below 0.01.

\subsection{First Row Elements}

The differences between our calculated fractional abundances and  those of
\citet{Mazz98a} for H are negligible. There is no DR process for H, and for the
temperature range in Fig.~\ref{fig:bad_maz_h}, the difference between the RR
rate coefficients of \citet{Badn06c} and \citet{Mazz98a} is to within 0.2\% 
(which is better than the accuracy of the published RR rate coefficient fits).
For He (Fig.~\ref{fig:bad_maz_he}), we find differences between our calculated
fractional abundances and those of Mazzotta et al.\ to be within 20\% for the
neutral and singly-ionized ion. Differences in the bare ion are negligible.

\subsection{Second Row Elements}

For Li and Be we find differences between our calculated fractional abundances
relative to those of \citet{Mazz98a} to be within 20\% for all ionization stages
(Figs.~\ref{fig:bad_maz_li} and \ref{fig:bad_maz_be}, respectively). The
difference for B is also of this order except for the neutral atom, where 
our calculations  give an increase in abundance of $\sim50\%$ at a
fractional abundance of 0.1, rising to $\sim70\%$ at 0.01
(Fig.~\ref{fig:bad_maz_b}). We attribute this to the B$^{1+}$ DR rate
coefficient calculated by \citet{Col03} being almost an order of magnitude
larger than that recommended by Mazzotta et al.\ in the CIE formation zone.

Differences for C (Fig.~\ref{fig:bad_maz_c}) are found to be generally within
20\%. Exceptions are at temperatures of $1\times10^4$--$2\times10^4$~K where  there
are differences of up to 40\% (but only for fractional abundances less than 0.1)
and in the temperature range of $7\times10^4$--$2\times10^5$~K where the
differences are up to 60\% even at fractional abundances greater than 0.1. 

For N (Fig.~\ref{fig:bad_maz_n}) we find the largest differences in the neutral
and singly-charged fractional abundances. These differences are found at
temperatures of $1\times10^4$--$3\times 10^4$~K. They rise with decreasing
fractional abundance to an $\sim140\%$ increase in the neutral abundance and an
$\sim80\%$ decrease in singly-charged abundance at fractional abundances of
0.1. Outside this temperature range, other ionization stages have differences
within 50\%. As in the B case, the increase in the neutral abundance and
decrease in the singly-charged abundance seen in our calculations is due to the
N$^{1+}$ DR rate coefficient of \citet{Zat04a} being larger than that of
\citet{Mazz98a}.

For O (Fig.~\ref{fig:bad_maz_o}), the largest differences are found at
temperatures of $6\times10^4$--$4\times10^5$~K. For fractional abundances greater
than 0.1, the difference is as large as a 40\% decrease for the singly-charged
ion. The difference is within 30\% for the other ions. When fractional
abundances as low as 0.01 are considered, the decrease in the singly-charged
abundance is up to 60\%.

Of all the second row elements, F (Fig.~\ref{fig:bad_maz_f}) shows the largest
deviation from the  \citet{Mazz98a} results. This difference is most pronounced
for the first 4 ionization stages. In particular, at temperatures of
$1\times10^4$--$3\times10^4$~K we find differences up to 120\% for fractional
abundances from 0.1 to 0.01. Also for these 4 ions, at temperatures of
$7\times10^4$--$3\times10^5$~K the differences are $\sim200\%$ for fractional
abundances down to 0.1, and nearly 300\% for abundances down to 0.01.

Ne (Fig.~\ref{fig:bad_maz_ne}) shows differences of up to 100\% at
fractional abundances of 0.1. The largest of these is in the temperature range
$1\times10^5$--$4\times10^5$~K where the Ne$^{3+}$ abundance is decreased relative
to the \citet{Mazz98a} results and the Ne$^{4+}$ abundance is increased.

\subsection{Third Row Elements}

We find relatively small differences for Na (less than 30\%), except for
Na$^{3+}$, Na$^{4+}$, and Na$^{5+}$ at temperatures of
$2\times10^5$--$7\times10^5$~K (Fig.~\ref{fig:bad_maz_na}). In this temperature
range for Na$^{3+}$ we find a decrease of  $\sim120\%$ in the fractional abundance at
0.1 and $\sim200\%$ in the fractional abundance at 0.01. For Na$^{4+}$
and Na$^{5+}$, increases of $\sim60\%$ are seen at a fractional abundance of
0.1.

The largest differences in Mg peak abundance (Fig.~\ref{fig:bad_maz_mg}) are for
Mg$^{5+}$, which shows a 24\% decrease relative to \citet{Mazz98a}, and 
Mg$^{6+}$, which shows a 29\% increase. Off peak, the largest difference for Mg
is in the neutral atom. At $1\times10^4$--$2\times10^4$~K, for fractional
abundances between 0.01 and 0.4, our results are larger then those of
\citet{Mazz98a} by between 160 and 250\%.  This is due to the Mg$^{1+}$ DR and
RR rate coefficients of \citet{Alt06} and  \citet{Badn06b}, respectively, being
around a factor 2 larger than those recommended by \citet{Mazz98a}. Other
differences for Mg are concentrated around the $3\times 10^5$--$2\times 10^6$~K
temperature range, where they are up to 100\%.

Differences in our fractional abundance curves relative to those of
\citet{Mazz98a} for Al (Fig.~\ref{fig:bad_maz_al}) are seen across a wide range
of ionization stages.  Al$^{2+}$ shows a 32\% decrease in peak abundance,
Al$^{6+}$ shows a 23\% decrease, Al$^{7+}$ shows a 34\% increase, and Al$^{8+}$
shows a 22\% increase. At a fractional abundance of 0.1, the maximum difference
is $\sim150\%$ for all ions.

For Si (Fig.~\ref{fig:bad_maz_si}), the peak abundance of Si$^{5+}$ is decreased
by 27\%. The differences seen for Si are up to 70\% relative to \citet{Mazz98a}
for all but the F- and Ne-like ions. These two ions show differences of over
200\% at fractional abundances of 0.1 and temperatures between
$2\times10^5$--$6\times10^5$~K. 

The largest difference in peak abundance for P (Fig.~\ref{fig:bad_maz_p}) is
P$^{8+}$, which shows a 26\% decrease relative to the results of
\citet{Mazz98a}. Other differences in abundance between our results and
\citet{Mazz98a} for P are largest at $3\times10^5$--$3\times10^6$~K. In
particular, P$^{5+}$ shows an increase of $\sim140\%$ at a fractional abundance
of 0.1, and of $\sim250\%$ at an abundance of 0.01.

S, Cl, and Ar (Figs.~\ref{fig:bad_maz_s}--\ref{fig:bad_maz_ar}, respectively)
show the greatest peak abundance difference in the N-like ion. The decrease for
S$^{9+}$ is 42\% relative to the fractional abundance of \citet{Mazz98a}, for
Cl$^{10+}$ it is 47\%, and for Ar$^{11+}$ it is 14\%.  Also, the temperature of
peak formation of Cl$^{9+}$ is increased by 0.06 in the dex. These three
elements all show maximum discrepancy in the temperature range
$6\times10^5$--$3\times10^6$~K. For S these differences are up to a factor of 3
at fractional abundances of 0.1, and up to a factor of 6 at fractional
abundances of 0.01. For Cl these differences are up to a factor of 4.2 at
fractional abundances of 0.1, and up to a factor of 11 at fractional abundances
of 0.01.  For Ar these differences are no greater than a factor of 1.5 for
fractional abundances of 0.1 and up to a factor of 2 at fractional abundances
of 0.01.

Fig.~\ref{fig:bad_gu_mg} shows the differences in fractional abundances
calculated using the data of \citet{Badn06a,Badn06b} compared to the data of
\citet{Gu03a,Gu03b,Gu04a} for Mg. These differences are up to 30\% for
fractional abundances of 0.1 and greater, and up to 50\% for fractional
abundances of 0.01. They are concentrated in temperature regions of
$1\times10^4$--$2\times10^4$~K and $3\times10^5$--$2\times10^6$~K. Comparison
with the calculated abundances of Si (Fig.~\ref{fig:bad_gu_si}) shows
differences up to 25\% in fractional abundances of 0.1, rising to $\sim30\%$ at
abundances of 0.01. S and Ar abundances were also calculated using the data of
Gu. Agreement here for S (Fig.~\ref{fig:bad_gu_s}) is within 25\% at an
abundance of 0.1 and 40\% at an abundance of 0.01. For Ar
(Fig.~\ref{fig:bad_gu_ar}), the agreement is within 20\% at an abundance of 0.1
and 30\% at an abundance of 0.01. For S and Ar, the largest difference is seen
for the Ne- and Mg-like ions.

\subsection{Fourth Row Elements}

For K and Ca (Figs.~\ref{fig:bad_maz_k} and \ref{fig:bad_maz_ca},
respectively), the largest peak abundance differences  are for the Na-, O-, and
N-like ions. For Na-like K$^{8+}$ the increase is 36\% relative to
\citet{Mazz98a}, and for Ca$^{9+}$ it is 48\%. For O-like K$^{11+}$ the
increase is 22\%, and for Ca$^{12+}$ it is 29\%. For N-like K$^{12+}$ the
decrease is 41\% relative to \citet{Mazz98a}, and for Ca$^{13+}$ it is 40\%.
Large increases in the temperature of peak abundance are found for K$^{11+}$,
Ca$^{12+}$ and Ca$^{13+}$; they are 0.06, 0.07, and 0.06 in the dex,
respectively. For these elements the largest fractional abundance differences
are seen between  $2\times10^6$--$6\times10^6$~K and are  within a factor of 5
at fractional abundances of 0.1 and up to a factor of 8 at fractional
abundances of 0.01.

Differences between our results and those of \citet{Mazz98a} for Sc
(Fig.~\ref{fig:bad_maz_sc}) at peak abundance are a 57\% increase for Sc$^{10+}$
and a 21\% decrease for Sc$^{14+}$. Differences for all ions are up to 80\% at
fractional abundances of 0.1, and 150\% for fractional abundances of 0.01. The
largest differences are between $7\times10^5$--$1\times10^7$~K.

The largest Ti and V peak abundance differences (Figs.~\ref{fig:bad_maz_ti} and
\ref{fig:bad_maz_v}, respectively) are a 21\% decrease for Ti$^{15+}$ and a 25\%
increase for V$^{12+}$. These elements show differences in fractional abundance
from our calculations relative to those of \citet{Mazz98a} concentrated at
temperatures of $3\times10^6$--$1\times10^7$~K. These differences are less than a
factor of 2 for fractional abundances of 0.1 and above. They are up to a factor
of 3 at fractional abundances of 0.01.

Cr, Mn and Fe  (Figs.~\ref{fig:bad_maz_cr}--\ref{fig:bad_maz_fe}, respectively)
all show similar differences relative to the \citet{Mazz98a} data.  Of these
elements, the largest percentage difference in peak abundance is seen for the
Fe$^{18+}$ ion, which has an increase of 28\%. Other differences for these 3
elements are in the temperature range $4\times10^6$--$2\times10^7$~K and are up
to a factor of 2.2 at fractional abundances of 0.1 and a factor of 3 at
fractional abundances of 0.01.

Co and Ni (Figs.~\ref{fig:bad_maz_co} and \ref{fig:bad_maz_ni}, respectively)
have a number of ions with large differences in peak fractional abundances
relative to \citet{Mazz98a}. For Co, the largest of these are Co$^{19+}$, which
has a 48\% increase in peak abundance, and Co$^{20+}$, which has a 37\%
decrease. The temperature of peak formation is also increased by 0.05 in the
dex in both cases. The largest peak abundance differences for Ni are for
Ni$^{19+}$, Ni$^{20+}$, and  Ni$^{21+}$, with a 34\% increase, a 61\% increase,
and a 42\% decrease, respectively. Ni$^{20+}$ and Ni$^{21+}$ also have an
increase in peak abundance temperature of 0.06 and 0.07 in the dex,
respectively. Both Co and Ni show the largest discrepancies between our results
and those of \citet{Mazz98a} between $7\times10^6$--$2\times10^7$~K. For Co
these differences are up to a factor of 3.5 at fractional abundances of 0.1 and
up to a factor of 4 at fractional abundances of 0.01. For Ni they are up to a
factor of 4 at fractional abundances of 0.1 and up to a factor of 7 at
fractional abundances of 0.01.

In Figs.~\ref{fig:cu} and \ref{fig:zn} we present the ionization fractional
abundances of Cu and Zn without comparison. We note that \citet{Mm02} present
fractional abundance results for these elements using updated ionization rate
coefficients, but the purpose of our present work is to highlight the effect of
improved recombination data. Comparison to \citet{Mm02} will be considered in a
future work where we include updated EII  data.

Comparing with Ca abundances from the data of \citet{Gu03a,Gu03b,Gu04a} gives
differences not larger than 20\% at fractional abundances of 0.1 and not larger
than 30\% at fractional abundances of 0.01 for all temperatures covered here
(Fig.~\ref{fig:bad_gu_ca}).  Comparing for Fe gives differences in the
fractional abundance  that is within 20\% at all temperatures
(Fig.~\ref{fig:bad_gu_fe}). For Ni, differences in the fractional abundance  are
within 15\% at fractional abundances of 0.1 and 25\% at fractional abundances of
0.01 (Fig.~\ref{fig:bad_gu_ni}).

\section{Conclusion}
\label{sec:Conclusions}

This work has collected the most recent state-of-the-art theoretical DR and RR
rate coefficients and, based on these data, calculated new CIE ionic fractional
abundances of all elements from H to Zn. For these elements we have implemented
the data of \citet{Badn06a,Badn06b} for all charge states from bare through
Na-like.  DR data for  Mg, Si, S, Ar, Ca, Fe, and Ni has also been calculated
by \citet{Gu03a,Gu04a} for all charge states from H- through Na-like.
Additionally, RR data for  these 7 elements has been calculated by
\citet{Gu03b} for all charge states from bare through F-like. We have also
computed ionization balance results using these data of Gu. For ionization
stages not provided by the above calculations, we revert to the recombination
data recommended by \citet{Mazz98a} for all elements up to and  including Ni
and Mazzotta (private communication) for Cu and Zn.  We also use the EII data
from these two sources.

Our results represent a significant improvement over past CIE calculations. This
will impact directly on the plasma conditions inferred from spectral observations
and is thus of much importance for the astrophysics community. We will further
this study in subsequent work by the inclusion of experimentally-derived DR 
data for singly- and doubly-charged ions (where available), incorporating CT,
and updating the EII data to the extent possible.

We conclude by noting that further progress in CIE calculations will require a
concerted theoretical and experimental effort to generate the remaining needed
atomic data. Modern DR and RR data are urgently needed for ions with 12 or more
bound electrons.  There is also a need for improved EII and CT data.  There has
been no significant revision or laboratory benchmarking of the recommended EII
database since around 1990. Additionally, the latest compilation of recommended
CT rate coefficients dates back to \citet{King96}. This is in need of updating
to reflect advances in CT in the last decade.
We propose that all future data for DR, RR, CT, and EII
should be generated aiming for an accuracy of better than 35\%.  This will
match the accuracy of the modern electron-ion recombination measurements
and calculations and help to insure a uniformity of accuracy for future
CIE calculations.  Such an accurate and up-to-date database is crucial for
being able to produce reliable CIE calculations for the
astrophysics community.

\acknowledgements

We thank M. Bannister, M. Finkenthal, T. Kato, E. Landi, S. Loch, M. Mattioli,
P. Mazzotta, and R. Smith for stimulating conversations. 
P.B., T.W.G., W.M., and D.W.S.\ were supported in part by the NASA Solar and
Heliospheric Physics Supporting Research and Technology program and the NASA
Astronomy and Physics Research and Analysis Program.
N.R.B.\ was supported in part by PPARC
PPA$\backslash$G$\backslash$S2003$\backslash$00055.
J.M.L.\ was supported by NASA LWS Contract
NNH05AAOSI and by Basic Research Funds of the Office of Naval Research.

%\vfill
%\eject

%\clearpage

\clearpage

\clearpage

\clearpage

%\clearpage

%\clearpage

%\clearpage

%\clearpage

%\clearpage

%\clearpage

\clearpage

%\clearpage

%\clearpage

%\clearpage

%\clearpage

\clearpage

\clearpage

\clearpage

\clearpage

\clearpage

\clearpage

\begin{deluxetable}{cc}
\tabletypesize{\footnotesize}
\tablewidth{0pt}
\tablecaption{Published {\sc autostructure} DR Rate Coefficients
\label{tab:dr}}
\tablehead{\colhead{Isoelectronic Sequence} & \colhead{Publication}}
\startdata
H-like  & \protect\citet{Badn06c} \\
He-like & \protect\citet{Baut06} \\
Li-like & \protect\citet{Col04}  \\
Be-like & \protect\citet{Col03}  \\
B-like  & \protect\citet{Alt04}  \\
C-like  & \protect\citet{Zat04a} \\
N-like  & \protect\citet{Mit04}  \\
O-like  & \protect\citet{Zat03}  \\
F-like  & \protect\citet{Zat06}  \\
Ne-like & \protect\citet{Zat04b,Fu06} \\
Na-like & \protect\citet{Alt06}  
\enddata
\tablecomments{Data have been published for 
all elements from He through Zn. These data were calculated using 
the techniques described in \protect\citet{Badn03a}. The complete data set
is provided via the weblink of \protect\citet{Badn06a}.}
\end{deluxetable}

\pagebreak

\begin{deluxetable}{cccc}
\tabletypesize{\tiny}
\tablewidth{0pt}
\tablecaption{Sources of data for the {\sc autostructure}-based
CIE calculations
\label{tab:data_summary}}
\tablehead{\colhead{Isoelectronic Sequence} & \colhead{DR Data Source}& \colhead{RR Data Source}& \colhead{EII Data Source}}
\startdata
Bare    & --- & \protect\citet{Badn06d} & --- \\
H-like  & \protect\citet{Badn06c} & \protect\citet{Badn06d} & \protect\citet{Mazz98a}\tablenotemark{1} \\
He-like & \protect\citet{Baut06}  & \protect\citet{Badn06d} & \protect\citet{Mazz98a}\tablenotemark{1} \\
Li-like & \protect\citet{Col04}   & \protect\citet{Badn06d} & \protect\citet{Mazz98a}\tablenotemark{1} \\
Be-like & \protect\citet{Col03}   & \protect\citet{Badn06d} & \protect\citet{Mazz98a}\tablenotemark{1} \\
B-like  & \protect\citet{Alt04}   & \protect\citet{Badn06d} & \protect\citet{Mazz98a}\tablenotemark{1} \\
C-like  & \protect\citet{Zat04a}  & \protect\citet{Badn06d} & \protect\citet{Mazz98a}\tablenotemark{1} \\
N-like  & \protect\citet{Mit04}   & \protect\citet{Badn06d} & \protect\citet{Mazz98a}\tablenotemark{1} \\
\enddata	    
\tablecomments{These are the data used to produce the CIE results given in 
Figs.~\protect\ref{fig:bad_maz_h} to \protect\ref{fig:zn} and Tables~\protect\ref{tab:H} to 
\protect\ref{tab:Zn}. The {\sc autostructure} DR data \protect\citep{Badn06a} and RR data
\protect\citep{Badn06b} are also available online.}
\tablenotetext{1}{The data from \protect\citet{Mazz98a} is replaced by Mazzotta
(private communication) for the elements Cu and Zn.}
\end{deluxetable}	     				     

\pagebreak

\begin{deluxetable}{cccc}
\tabletypesize{\footnotesize}
\tablewidth{0pt}
\tablecaption{Sources of data used in the {\sc fac}-based
CIE calculations
\label{tab:data_summary_gu}}
\tablehead{\colhead{Isoelectronic Sequence} & \colhead{DR Data Source}& \colhead{RR Data Source}& \colhead{EII Data Source}}
\startdata
Bare    & --- & \protect\citet{Gu03b} & --- \\
H-like  & \protect\citet{Gu03a}   & \protect\citet{Gu03b} & \protect\citet{Mazz98a} \\
He-like & \protect\citet{Gu03a}   & \protect\citet{Gu03b} & \protect\citet{Mazz98a}\\
Li-like & \protect\citet{Gu03a}   & \protect\citet{Gu03b} & \protect\citet{Mazz98a} \\
Be-like & \protect\citet{Gu03a}   & \protect\citet{Gu03b} & \protect\citet{Mazz98a} \\
B-like  & \protect\citet{Gu03a}   & \protect\citet{Gu03b} & \protect\citet{Mazz98a} \\
C-like  & \protect\citet{Gu03a}   & \protect\citet{Gu03b} & \protect\citet{Mazz98a} \\
N-like  & \protect\citet{Gu03a}   & \protect\citet{Gu03b} & \protect\citet{Mazz98a} \\
\enddata
\tablecomments{These are the data used to produce the CIE results given in 
Figs.~\protect\ref{fig:bad_gu_mg} to \protect\ref{fig:bad_gu_ni} and Tables~\protect\ref{tab:Mg_Gu} to 
\protect\ref{tab:Ni_Gu}.}
\end{deluxetable}

\clearpage

\begin{deluxetable}{ccc}
\tabletypesize{\tiny}
\tablecaption{Hydrogen CIE Fractional Abundances ({\sc autostructure}-based
results)
\label{tab:H}}
\tablewidth{0pt}
\tablehead{
\colhead{log(T)} 
& \colhead{H{$^{0+}$}}
& \colhead{H{$^{1+}$}}}
\startdata
4.00 &  0.001 &  2.659 \\
4.10 &  0.031 &  1.161 \\
4.20 &  0.359 &  0.250 \\
4.30 &  1.151 &  0.032 \\
4.40 &  1.954 &  0.005 \\
4.50 &  2.636 &  0.001 \\
4.60 &  3.209 &  0.000 \\
4.70 &  3.692 &  0.000 \\
4.80 &  4.104 &  0.000 \\
4.90 &  4.459 &  0.000 \\
5.00 &  4.768 &  0.000 \\
5.10 &  5.041 &  0.000 \\
5.20 &  5.285 &  0.000 \\
5.30 &  5.503 &  0.000 \\
5.40 &  5.702 &  0.000 \\
5.50 &  5.883 &  0.000 \\
5.60 &  6.051 &  0.000 \\
5.70 &  6.206 &  0.000 \\
5.80 &  6.355 &  0.000 \\
5.90 &  6.508 &  0.000 \\
6.00 &  6.642 &  0.000 \\
6.10 &  6.766 &  0.000 \\
6.20 &  6.884 &  0.000 \\
6.30 &  6.998 &  0.000 \\
6.40 &  7.110 &  0.000 \\
6.50 &  7.219 &  0.000 \\
6.60 &  7.328 &  0.000 \\
6.70 &  7.435 &  0.000 \\
6.80 &  7.541 &  0.000 \\
6.90 &  7.646 &  0.000 \\
7.00 &  7.751 &  0.000 \\
7.10 &  7.855 &  0.000 \\
7.20 &  7.959 &  0.000 \\
7.30 &  8.063 &  0.000 \\
7.40 &  8.167 &  0.000 \\
7.50 &  8.271 &  0.000 \\
7.60 &  8.375 &  0.000 \\
7.70 &  8.479 &  0.000 \\
7.80 &  8.583 &  0.000 \\
7.90 &  8.687 &  0.000 \\
8.00 &  8.791 &  0.000 \\
8.10 &  8.896 &  0.000 \\
8.20 &  9.000 &  0.000 \\
8.30 &  9.105 &  0.000 \\
8.40 &  9.210 &  0.000 \\
8.50 &  9.315 &  0.000 \\
8.60 &  9.420 &  0.000 \\
8.70 &  9.525 &  0.000 \\
8.80 &  9.630 &  0.000 \\
8.90 &  9.735 &  0.000 \\
9.00 &  9.841 &  0.000 
\enddata
\tablecomments{Calculated $-\log_{10}$ of the fractional abundance for all
ionization stages. We use the RR rate 
coefficients of \protect\citet{Badn06b} and
EII rate coefficients of \protect\citet{Mazz98a}.}
\end{deluxetable}

\clearpage

\begin{deluxetable}{cccc}
\tabletypesize{\tiny}
\tablecaption{Helium CIE Fractional Abundances ({\sc autostructure}-based
results)
\label{tab:He}}
\tablewidth{0pt}
\tablehead{
\colhead{log(T)} 
& \colhead{He{$^{0+}$}}
& \colhead{He{$^{1+}$}}
& \colhead{He{$^{2+}$}}}
\startdata
4.00 &  0.000 &  8.774 & 15.000 \\
4.10 &  0.000 &  6.097 & 15.000 \\
4.20 &  0.000 &  3.941 & 15.000 \\
4.30 &  0.003 &  2.202 & 15.000 \\
4.40 &  0.066 &  0.852 &  8.848 \\
4.50 &  0.522 &  0.155 &  5.788 \\
4.60 &  1.331 &  0.021 &  3.751 \\
4.70 &  2.080 &  0.006 &  2.200 \\
\enddata
\tablecomments{Same as Table~\protect\ref{tab:H} but also using the DR rate 
coefficients of \protect\citet{Badn06a}. Fractional abundances are cut off at 
$10^{-15}$. For ease of machine readability, values less than
$10^{-15}$ are given $-\log_{10}$ values of 15.}
\end{deluxetable}

\clearpage

\begin{deluxetable}{ccccc}
\tabletypesize{\tiny}
\tablecaption{Lithium CIE Fractional Abundances ({\sc autostructure}-based
results)
\label{tab:Li}}
\tablewidth{0pt}
\tablehead{
\colhead{log(T)} 
& \colhead{Li{$^{0+}$}}
& \colhead{Li{$^{1+}$}}
& \colhead{Li{$^{2+}$}}
& \colhead{Li{$^{3+}$}}}
\startdata
4.00 &  2.646 &  0.001 & 15.000 & 15.000 \\
4.10 &  3.323 &  0.000 & 15.000 & 15.000 \\
4.20 &  3.885 &  0.000 & 15.000 & 15.000 \\
4.30 &  4.354 &  0.000 & 15.000 & 15.000 \\
4.40 &  4.749 &  0.000 & 15.000 & 15.000 \\
4.50 &  5.085 &  0.000 &  9.278 & 15.000 \\
4.60 &  5.372 &  0.000 &  6.675 & 15.000 \\
4.70 &  5.621 &  0.000 &  4.580 & 14.722 \\
\enddata
\tablecomments{Same as Table~\protect\ref{tab:He} but for lithium.}
\end{deluxetable}

\clearpage

\begin{deluxetable}{cccccc}
\tabletypesize{\tiny}
\tablecaption{Beryllium CIE Fractional Abundances ({\sc autostructure}-based
results)
\label{tab:Be}}
\tablewidth{0pt}
\tablehead{
\colhead{log(T)}
& \colhead{Be{$^{0+}$}} 
& \colhead{Be{$^{1+}$}}
& \colhead{Be{$^{2+}$}}
& \colhead{Be{$^{3+}$}}
& \colhead{Be{$^{4+}$}}}
\startdata
4.00 &  0.041 &  1.045 &  6.684 & 15.000 & 15.000 \\
4.10 &  0.203 &  0.429 &  4.059 & 15.000 & 15.000 \\
4.20 &  0.564 &  0.142 &  2.152 & 15.000 & 15.000 \\
4.30 &  1.088 &  0.116 &  0.815 & 15.000 & 15.000 \\
4.40 &  1.971 &  0.527 &  0.160 & 15.000 & 15.000 \\
4.50 &  3.118 &  1.263 &  0.025 & 15.000 & 15.000 \\
4.60 &  4.176 &  1.958 &  0.005 & 15.000 & 15.000 \\
4.70 &  5.086 &  2.545 &  0.001 & 15.000 & 15.000 \\
\enddata
\tablecomments{Same as Table~\protect\ref{tab:He} but for beryllium.}
\end{deluxetable}

\clearpage

\begin{deluxetable}{ccccccc}
\tabletypesize{\tiny}
\tablecaption{Boron CIE Fractional Abundances ({\sc autostructure}-based
results)
\label{tab:B}}
\tablewidth{0pt}
\tablehead{
\colhead{log(T)} 
& \colhead{B{$^{0+}$}}
& \colhead{B{$^{1+}$}}
& \colhead{B{$^{2+}$}}
& \colhead{B{$^{3+}$}}
& \colhead{B{$^{4+}$}}
& \colhead{B{$^{5+}$}}}
\startdata
4.00 &  0.788 &  0.077 &  9.849 & 15.000 & 15.000 & 15.000 \\
4.10 &  1.535 &  0.013 &  7.387 & 15.000 & 15.000 & 15.000 \\
4.20 &  2.013 &  0.004 &  5.487 & 14.839 & 15.000 & 15.000 \\
4.30 &  2.293 &  0.002 &  3.963 & 10.714 & 15.000 & 15.000 \\
4.40 &  2.493 &  0.002 &  2.722 &  7.382 & 15.000 & 15.000 \\
4.50 &  2.674 &  0.010 &  1.709 &  4.683 & 15.000 & 15.000 \\
4.60 &  2.890 &  0.059 &  0.911 &  2.522 & 15.000 & 15.000 \\
4.70 &  3.285 &  0.289 &  0.432 &  0.936 & 15.000 & 15.000 \\
\enddata
\tablecomments{Same as Table~\protect\ref{tab:He} but for boron.}
\end{deluxetable}

\clearpage

\begin{deluxetable}{cccccccc}
\tabletypesize{\tiny}
\tablecaption{Carbon CIE Fractional Abundances ({\sc autostructure}-based
results)
\label{tab:C}}
\tablewidth{0pt}
\tablehead{
\colhead{log(T)}
& \colhead{C{$^{0+}$}}
& \colhead{C{$^{1+}$}}
& \colhead{C{$^{2+}$}}
& \colhead{C{$^{3+}$}}
& \colhead{C{$^{4+}$}}
& \colhead{C{$^{5+}$}}
& \colhead{C{$^{6+}$}}}
\startdata
4.00 &  0.013 &  1.545 & 10.780 & 15.000 & 15.000 & 15.000 & 15.000 \\
4.10 &  0.175 &  0.480 &  7.098 & 15.000 & 15.000 & 15.000 & 15.000 \\
4.20 &  0.713 &  0.094 &  4.630 & 15.000 & 15.000 & 15.000 & 15.000 \\
4.30 &  1.256 &  0.025 &  2.951 & 13.067 & 15.000 & 15.000 & 15.000 \\
4.40 &  1.638 &  0.018 &  1.760 &  9.452 & 15.000 & 15.000 & 15.000 \\
4.50 &  1.967 &  0.058 &  0.938 &  6.710 & 14.681 & 15.000 & 15.000 \\
4.60 &  2.368 &  0.209 &  0.424 &  4.650 & 10.387 & 15.000 & 15.000 \\
4.70 &  2.892 &  0.497 &  0.168 &  3.137 &  7.074 & 15.000 & 15.000 \\
\enddata
\tablecomments{Same as Table~\protect\ref{tab:He} but for carbon.}
\end{deluxetable}

\clearpage

\begin{deluxetable}{ccccccccc}
\tabletypesize{\tiny}
\tablecaption{Nitrogen CIE Fractional Abundances ({\sc autostructure}-based
results)
\label{tab:N}}
\tablewidth{0pt}
\tablehead{
\colhead{log(T)} 
& \colhead{N{$^{0+}$}}
& \colhead{N{$^{1+}$}}
& \colhead{N{$^{2+}$}}
& \colhead{N{$^{3+}$}}
& \colhead{N{$^{4+}$}}
& \colhead{N{$^{5+}$}}
& \colhead{N{$^{6+}$}}
& \colhead{N{$^{7+}$}}}
\startdata
4.00 &  0.000 &  3.144 & 15.000 & 15.000 & 15.000 & 15.000 & 15.000 & 15.000 \\
4.10 &  0.013 &  1.539 & 10.328 & 15.000 & 15.000 & 15.000 & 15.000 & 15.000 \\
4.20 &  0.196 &  0.440 &  6.699 & 15.000 & 15.000 & 15.000 & 15.000 & 15.000 \\
4.30 &  0.788 &  0.077 &  4.373 & 14.157 & 15.000 & 15.000 & 15.000 & 15.000 \\
4.40 &  1.374 &  0.019 &  2.853 & 10.089 & 15.000 & 15.000 & 15.000 & 15.000 \\
4.50 &  1.796 &  0.015 &  1.764 &  7.010 & 15.000 & 15.000 & 15.000 & 15.000 \\
4.60 &  2.144 &  0.053 &  0.965 &  4.681 & 13.020 & 15.000 & 15.000 & 15.000 \\
4.70 &  2.542 &  0.200 &  0.438 &  2.972 &  9.302 & 15.000 & 15.000 & 15.000 \\
\enddata
\tablecomments{Same as Table~\protect\ref{tab:He} but for nitrogen.}
\end{deluxetable}

\clearpage

\begin{deluxetable}{cccccccccc}
\tabletypesize{\tiny}
\tablecaption{Oxygen CIE Fractional Abundances ({\sc autostructure}-based
results)
\label{tab:O}}
\tablewidth{0pt}
\tablehead{
\colhead{log(T)} 
& \colhead{O{$^{0+}$}}
& \colhead{O{$^{1+}$}}
& \colhead{O{$^{2+}$}}
& \colhead{O{$^{3+}$}}
& \colhead{O{$^{4+}$}}
& \colhead{O{$^{5+}$}}
& \colhead{O{$^{6+}$}}
& \colhead{O{$^{7+}$}}
& \colhead{O{$^{8+}$}}}
\startdata
4.00 &  0.001 &  2.540 & 15.000 & 15.000 & 15.000 & 15.000 & 15.000 & 15.000 & 15.000 \\
4.10 &  0.040 &  1.060 & 15.000 & 15.000 & 15.000 & 15.000 & 15.000 & 15.000 & 15.000 \\
4.20 &  0.412 &  0.213 &  8.299 & 15.000 & 15.000 & 15.000 & 15.000 & 15.000 & 15.000 \\
4.30 &  1.159 &  0.031 &  5.705 & 15.000 & 15.000 & 15.000 & 15.000 & 15.000 & 15.000 \\
4.40 &  1.733 &  0.008 &  3.778 & 12.163 & 15.000 & 15.000 & 15.000 & 15.000 & 15.000 \\
4.50 &  2.072 &  0.006 &  2.341 &  8.420 & 15.000 & 15.000 & 15.000 & 15.000 & 15.000 \\
4.60 &  2.325 &  0.024 &  1.315 &  5.641 & 13.834 & 15.000 & 15.000 & 15.000 & 15.000 \\
4.70 &  2.626 &  0.120 &  0.623 &  3.612 &  9.764 & 15.000 & 15.000 & 15.000 & 15.000 \\
\enddata
\tablecomments{Same as Table~\protect\ref{tab:He} but for oxygen.}
\end{deluxetable}

\clearpage

\begin{deluxetable}{ccccccccccc}
\tabletypesize{\tiny}
\tablecaption{Flourine CIE Fractional Abundances ({\sc autostructure}-based
results)
\label{tab:F}}
\tablewidth{0pt}
\tablehead{
\colhead{log(T)} 
& \colhead{F{$^{0+}$}}
& \colhead{F{$^{1+}$}}
& \colhead{F{$^{2+}$}}
& \colhead{F{$^{3+}$}}
& \colhead{F{$^{4+}$}}
& \colhead{F{$^{5+}$}}
& \colhead{F{$^{6+}$}}
& \colhead{F{$^{7+}$}}
& \colhead{F{$^{8+}$}}
& \colhead{F{$^{9+}$}}}
\startdata
4.00 &  0.000 &  4.142 & 15.000 & 15.000 & 15.000 & 15.000 & 15.000 & 15.000 & 15.000 & 15.000 \\
4.10 &  0.003 &  2.218 & 15.000 & 15.000 & 15.000 & 15.000 & 15.000 & 15.000 & 15.000 & 15.000 \\
4.20 &  0.085 &  0.748 &  8.757 & 15.000 & 15.000 & 15.000 & 15.000 & 15.000 & 15.000 & 15.000 \\
4.30 &  0.684 &  0.101 &  5.716 & 15.000 & 15.000 & 15.000 & 15.000 & 15.000 & 15.000 & 15.000 \\
4.40 &  1.554 &  0.012 &  3.718 & 13.684 & 15.000 & 15.000 & 15.000 & 15.000 & 15.000 & 15.000 \\
4.50 &  2.168 &  0.005 &  2.241 &  9.541 & 15.000 & 15.000 & 15.000 & 15.000 & 15.000 & 15.000 \\
4.60 &  2.522 &  0.030 &  1.202 &  6.430 & 15.000 & 15.000 & 15.000 & 15.000 & 15.000 & 15.000 \\
4.70 &  2.835 &  0.145 &  0.550 &  4.192 & 10.910 & 15.000 & 15.000 & 15.000 & 15.000 & 15.000 \\
\enddata
\tablecomments{Same as Table~\protect\ref{tab:He} but for flourine.}
\end{deluxetable}

\clearpage

\begin{deluxetable}{cccccccccccc}
\tabletypesize{\tiny}
\tablecaption{Neon CIE Fractional Abundances ({\sc autostructure}-based
results)
\label{tab:Ne}}
\tablewidth{0pt}
\tablehead{
\colhead{log(T)} 
& \colhead{Ne{$^{0+}$}}
& \colhead{Ne{$^{1+}$}}
& \colhead{Ne{$^{2+}$}}
& \colhead{Ne{$^{3+}$}}
& \colhead{Ne{$^{4+}$}}
& \colhead{Ne{$^{5+}$}}
& \colhead{Ne{$^{6+}$}}
& \colhead{Ne{$^{7+}$}}
& \colhead{Ne{$^{8+}$}}
& \colhead{Ne{$^{9+}$}}
& \colhead{Ne{$^{10+}$}}}
\startdata
4.00 &  0.000 &  6.854 & 15.000 & 15.000 & 15.000 & 15.000 & 15.000 & 15.000 & 15.000 & 15.000 & 15.000 \\
4.10 &  0.000 &  4.494 & 15.000 & 15.000 & 15.000 & 15.000 & 15.000 & 15.000 & 15.000 & 15.000 & 15.000 \\
4.20 &  0.001 &  2.593 & 12.315 & 15.000 & 15.000 & 15.000 & 15.000 & 15.000 & 15.000 & 15.000 & 15.000 \\
4.30 &  0.037 &  1.090 &  8.015 & 15.000 & 15.000 & 15.000 & 15.000 & 15.000 & 15.000 & 15.000 & 15.000 \\
4.40 &  0.410 &  0.214 &  4.895 & 15.000 & 15.000 & 15.000 & 15.000 & 15.000 & 15.000 & 15.000 & 15.000 \\
4.50 &  1.219 &  0.028 &  2.926 & 10.455 & 15.000 & 15.000 & 15.000 & 15.000 & 15.000 & 15.000 & 15.000 \\
4.60 &  1.897 &  0.018 &  1.570 &  6.950 & 15.000 & 15.000 & 15.000 & 15.000 & 15.000 & 15.000 & 15.000 \\
4.70 &  2.344 &  0.096 &  0.711 &  4.446 & 11.924 & 15.000 & 15.000 & 15.000 & 15.000 & 15.000 & 15.000 \\
\enddata
\tablecomments{Same as Table~\protect\ref{tab:He} but for neon.}
\end{deluxetable}

\clearpage

\begin{deluxetable}{ccccccccccccc}
\tabletypesize{\tiny}
\tablecaption{Sodium CIE Fractional Abundances ({\sc autostructure}-based
results)
\label{tab:Na}}
\tablewidth{0pt}
\tablehead{
\colhead{log(T)} 
& \colhead{Na{$^{0+}$}}
& \colhead{Na{$^{1+}$}}
& \colhead{Na{$^{2+}$}}
& \colhead{Na{$^{3+}$}}
& \colhead{Na{$^{4+}$}}
& \colhead{Na{$^{5+}$}}
& \colhead{Na{$^{6+}$}}
& \colhead{Na{$^{7+}$}}
& \colhead{Na{$^{8+}$}}
& \colhead{Na{$^{9+}$}}
& \colhead{Na{$^{10+}$}}
& \colhead{Na{$^{11+}$}}}
\startdata
4.00 &  3.126 &  0.000 & 15.000 & 15.000 & 15.000 & 15.000 & 15.000 & 15.000 & 15.000 & 15.000 & 15.000 & 15.000 \\
4.10 &  3.808 &  0.000 & 15.000 & 15.000 & 15.000 & 15.000 & 15.000 & 15.000 & 15.000 & 15.000 & 15.000 & 15.000 \\
4.20 &  4.379 &  0.000 & 15.000 & 15.000 & 15.000 & 15.000 & 15.000 & 15.000 & 15.000 & 15.000 & 15.000 & 15.000 \\
4.30 &  4.862 &  0.000 &  9.869 & 15.000 & 15.000 & 15.000 & 15.000 & 15.000 & 15.000 & 15.000 & 15.000 & 15.000 \\
4.40 &  5.273 &  0.000 &  7.014 & 15.000 & 15.000 & 15.000 & 15.000 & 15.000 & 15.000 & 15.000 & 15.000 & 15.000 \\
4.50 &  5.624 &  0.000 &  4.742 & 12.948 & 15.000 & 15.000 & 15.000 & 15.000 & 15.000 & 15.000 & 15.000 & 15.000 \\
4.60 &  5.909 &  0.001 &  2.927 &  8.694 & 15.000 & 15.000 & 15.000 & 15.000 & 15.000 & 15.000 & 15.000 & 15.000 \\
4.70 &  6.079 &  0.013 &  1.540 &  5.425 & 13.063 & 15.000 & 15.000 & 15.000 & 15.000 & 15.000 & 15.000 & 15.000 \\
\enddata
\tablecomments{Same as Table~\protect\ref{tab:He} but for sodium.}
\end{deluxetable}

\clearpage

\begin{deluxetable}{cccccccccccccc}
\tabletypesize{\tiny}
\tablecaption{Magnesium CIE Fractional Abundances ({\sc autostructure}-based
results)
\label{tab:Mg}}
\tablewidth{0pt}
\tablehead{
\colhead{log(T)} 
& \colhead{Mg{$^{0+}$}}
& \colhead{Mg{$^{1+}$}}
& \colhead{Mg{$^{2+}$}}
& \colhead{Mg{$^{3+}$}}
& \colhead{Mg{$^{4+}$}}
& \colhead{Mg{$^{5+}$}}
& \colhead{Mg{$^{6+}$}}
& \colhead{Mg{$^{7+}$}}
& \colhead{Mg{$^{8+}$}}
& \colhead{Mg{$^{9+}$}}
& \colhead{Mg{$^{10+}$}}
& \colhead{Mg{$^{11+}$}}
& \colhead{Mg{$^{12+}$}}}
\startdata
4.00 &  0.419 &  0.209 &  3.794 & 15.000 & 15.000 & 15.000 & 15.000 & 15.000 & 15.000 & 15.000 & 15.000 & 15.000 & 15.000 \\
4.10 &  0.826 &  0.076 &  1.977 & 15.000 & 15.000 & 15.000 & 15.000 & 15.000 & 15.000 & 15.000 & 15.000 & 15.000 & 15.000 \\
4.20 &  1.341 &  0.130 &  0.670 & 15.000 & 15.000 & 15.000 & 15.000 & 15.000 & 15.000 & 15.000 & 15.000 & 15.000 & 15.000 \\
4.30 &  2.281 &  0.673 &  0.107 & 15.000 & 15.000 & 15.000 & 15.000 & 15.000 & 15.000 & 15.000 & 15.000 & 15.000 & 15.000 \\
4.40 &  3.440 &  1.482 &  0.015 & 15.000 & 15.000 & 15.000 & 15.000 & 15.000 & 15.000 & 15.000 & 15.000 & 15.000 & 15.000 \\
4.50 &  4.476 &  2.209 &  0.003 &  9.771 & 15.000 & 15.000 & 15.000 & 15.000 & 15.000 & 15.000 & 15.000 & 15.000 & 15.000 \\
4.60 &  5.361 &  2.816 &  0.001 &  7.018 & 15.000 & 15.000 & 15.000 & 15.000 & 15.000 & 15.000 & 15.000 & 15.000 & 15.000 \\
4.70 &  6.115 &  3.319 &  0.000 &  4.823 & 13.180 & 15.000 & 15.000 & 15.000 & 15.000 & 15.000 & 15.000 & 15.000 & 15.000 \\
\enddata
\tablecomments{Same as Table~\protect\ref{tab:He} but for magnesium.}
\end{deluxetable}

\clearpage

\begin{deluxetable}{ccccccccccccccc}
\setlength{\tabcolsep}{0.06in}
\tabletypesize{\tiny}
\tablecaption{Aluminum CIE Fractional Abundances ({\sc autostructure}-based
results)
\label{tab:Al}}
\tablewidth{0pt}
\tablehead{
\colhead{log(T)}
& \colhead{Al{$^{0+}$}}
& \colhead{Al{$^{1+}$}}
& \colhead{Al{$^{2+}$}}
& \colhead{Al{$^{3+}$}}
& \colhead{Al{$^{4+}$}}
& \colhead{Al{$^{5+}$}}
& \colhead{Al{$^{6+}$}}
& \colhead{Al{$^{7+}$}}
& \colhead{Al{$^{8+}$}}
& \colhead{Al{$^{9+}$}}
& \colhead{Al{$^{10+}$}}
& \colhead{Al{$^{11+}$}}
& \colhead{Al{$^{12+}$}}
& \colhead{Al{$^{13+}$}}}
\startdata
4.00 &  1.997 &  0.004 &  6.195 & 15.000 & 15.000 & 15.000 & 15.000 & 15.000 & 15.000 & 15.000 & 15.000 & 15.000 & 15.000 & 15.000 \\
4.10 &  2.183 &  0.003 &  4.379 & 12.831 & 15.000 & 15.000 & 15.000 & 15.000 & 15.000 & 15.000 & 15.000 & 15.000 & 15.000 & 15.000 \\
4.20 &  2.278 &  0.003 &  3.037 &  9.025 & 15.000 & 15.000 & 15.000 & 15.000 & 15.000 & 15.000 & 15.000 & 15.000 & 15.000 & 15.000 \\
4.30 &  2.375 &  0.006 &  1.999 &  6.004 & 15.000 & 15.000 & 15.000 & 15.000 & 15.000 & 15.000 & 15.000 & 15.000 & 15.000 & 15.000 \\
4.40 &  2.508 &  0.031 &  1.181 &  3.587 & 15.000 & 15.000 & 15.000 & 15.000 & 15.000 & 15.000 & 15.000 & 15.000 & 15.000 & 15.000 \\
4.50 &  2.741 &  0.141 &  0.593 &  1.706 & 15.000 & 15.000 & 15.000 & 15.000 & 15.000 & 15.000 & 15.000 & 15.000 & 15.000 & 15.000 \\
4.60 &  3.284 &  0.548 &  0.416 &  0.478 & 15.000 & 15.000 & 15.000 & 15.000 & 15.000 & 15.000 & 15.000 & 15.000 & 15.000 & 15.000 \\
4.70 &  4.384 &  1.504 &  0.874 &  0.078 &  9.058 & 15.000 & 15.000 & 15.000 & 15.000 & 15.000 & 15.000 & 15.000 & 15.000 & 15.000 \\
\enddata
\tablecomments{Same as Table~\protect\ref{tab:He} but for aluminum and
also using the DR and RR rate coefficients of \protect\citet{Mazz98a} 
for ions not calculated by \protect\citet{Badn06a,Badn06b}.}
\end{deluxetable}

\clearpage

\begin{deluxetable}{cccccccccccccccc}
\setlength{\tabcolsep}{0.06in}
\tabletypesize{\tiny}
\tablecaption{Silicon CIE Fractional Abundances ({\sc autostructure}-based
results)
\label{tab:Si}}
\tablewidth{0pt}
\tablehead{
\colhead{log(T)} 
& \colhead{Si{$^{0+}$}}
& \colhead{Si{$^{1+}$}}
& \colhead{Si{$^{2+}$}}
& \colhead{Si{$^{3+}$}}
& \colhead{Si{$^{4+}$}}
& \colhead{Si{$^{5+}$}}
& \colhead{Si{$^{6+}$}}
& \colhead{Si{$^{7+}$}}
& \colhead{Si{$^{8+}$}}
& \colhead{Si{$^{9+}$}}
& \colhead{Si{$^{10+}$}}
& \colhead{Si{$^{11+}$}}
& \colhead{Si{$^{12+}$}}
& \colhead{Si{$^{13+}$}}
& \colhead{Si{$^{14+}$}}}
\startdata
4.00 &  1.024 &  0.043 &  4.192 & 15.000 & 15.000 & 15.000 & 15.000 & 15.000 & 15.000 & 15.000 & 15.000 & 15.000 & 15.000 & 15.000 & 15.000 \\
4.10 &  1.829 &  0.008 &  2.539 & 15.000 & 15.000 & 15.000 & 15.000 & 15.000 & 15.000 & 15.000 & 15.000 & 15.000 & 15.000 & 15.000 & 15.000 \\
4.20 &  2.463 &  0.020 &  1.391 &  9.534 & 15.000 & 15.000 & 15.000 & 15.000 & 15.000 & 15.000 & 15.000 & 15.000 & 15.000 & 15.000 & 15.000 \\
4.30 &  2.952 &  0.103 &  0.679 &  6.789 & 15.000 & 15.000 & 15.000 & 15.000 & 15.000 & 15.000 & 15.000 & 15.000 & 15.000 & 15.000 & 15.000 \\
4.40 &  3.400 &  0.304 &  0.299 &  4.836 & 11.175 & 15.000 & 15.000 & 15.000 & 15.000 & 15.000 & 15.000 & 15.000 & 15.000 & 15.000 & 15.000 \\
4.50 &  3.867 &  0.599 &  0.126 &  3.410 &  7.767 & 15.000 & 15.000 & 15.000 & 15.000 & 15.000 & 15.000 & 15.000 & 15.000 & 15.000 & 15.000 \\
4.60 &  4.339 &  0.928 &  0.057 &  2.322 &  5.079 & 15.000 & 15.000 & 15.000 & 15.000 & 15.000 & 15.000 & 15.000 & 15.000 & 15.000 & 15.000 \\
4.70 &  4.807 &  1.260 &  0.041 &  1.465 &  2.930 & 15.000 & 15.000 & 15.000 & 15.000 & 15.000 & 15.000 & 15.000 & 15.000 & 15.000 & 15.000 \\
\enddata
\tablecomments{Same as Table~\protect\ref{tab:Al} but for silicon.}
\end{deluxetable}

\clearpage

\begin{deluxetable}{ccccccccccccccccc}
\setlength{\tabcolsep}{0.05in}
\tabletypesize{\tiny}
\tablecaption{Phosphorus CIE Fractional Abundances ({\sc autostructure}-based
results)
\label{tab:P}}
\tablewidth{0pt}
\tablehead{
\colhead{log(T)} 
& \colhead{P{$^{0+}$}}
& \colhead{P{$^{1+}$}}
& \colhead{P{$^{2+}$}}
& \colhead{P{$^{3+}$}}
& \colhead{P{$^{4+}$}}
& \colhead{P{$^{5+}$}}
& \colhead{P{$^{6+}$}}
& \colhead{P{$^{7+}$}}
& \colhead{P{$^{8+}$}}
& \colhead{P{$^{9+}$}}
& \colhead{P{$^{10+}$}}
& \colhead{P{$^{11+}$}}
& \colhead{P{$^{12+}$}}
& \colhead{P{$^{13+}$}}
& \colhead{P{$^{14+}$}}
& \colhead{P{$^{15+}$}}}
\startdata
4.00 &  0.217 &  0.405 &  6.005 & 15.000 & 15.000 & 15.000 & 15.000 & 15.000 & 15.000 & 15.000 & 15.000 & 15.000 & 15.000 & 15.000 & 15.000 & 15.000 \\
4.10 &  1.027 &  0.043 &  3.556 & 12.051 & 15.000 & 15.000 & 15.000 & 15.000 & 15.000 & 15.000 & 15.000 & 15.000 & 15.000 & 15.000 & 15.000 & 15.000 \\
4.20 &  1.910 &  0.009 &  2.064 &  8.192 & 15.000 & 15.000 & 15.000 & 15.000 & 15.000 & 15.000 & 15.000 & 15.000 & 15.000 & 15.000 & 15.000 & 15.000 \\
4.30 &  2.538 &  0.033 &  1.155 &  5.652 & 15.000 & 15.000 & 15.000 & 15.000 & 15.000 & 15.000 & 15.000 & 15.000 & 15.000 & 15.000 & 15.000 & 15.000 \\
4.40 &  2.907 &  0.133 &  0.582 &  3.871 & 12.284 & 15.000 & 15.000 & 15.000 & 15.000 & 15.000 & 15.000 & 15.000 & 15.000 & 15.000 & 15.000 & 15.000 \\
4.50 &  3.230 &  0.352 &  0.258 &  2.579 &  9.020 & 15.000 & 15.000 & 15.000 & 15.000 & 15.000 & 15.000 & 15.000 & 15.000 & 15.000 & 15.000 & 15.000 \\
4.60 &  3.619 &  0.673 &  0.117 &  1.638 &  6.507 & 12.331 & 15.000 & 15.000 & 15.000 & 15.000 & 15.000 & 15.000 & 15.000 & 15.000 & 15.000 & 15.000 \\
4.70 &  4.073 &  1.053 &  0.098 &  0.946 &  4.542 &  8.556 & 15.000 & 15.000 & 15.000 & 15.000 & 15.000 & 15.000 & 15.000 & 15.000 & 15.000 & 15.000 \\
\enddata
\tablecomments{Same as Table~\protect\ref{tab:Al} but for phosphorus.}
\end{deluxetable}

\clearpage

\begin{deluxetable}{cccccccccccccccccc}
\setlength{\tabcolsep}{0.04in}
\tabletypesize{\tiny}
\tablecaption{Sulfur CIE Fractional Abundances ({\sc autostructure}-based
results)
\label{tab:S}}
\tablewidth{0pt}
\tablehead{
\colhead{log(T)} 
& \colhead{S{$^{0+}$}}
& \colhead{S{$^{1+}$}}
& \colhead{S{$^{2+}$}}
& \colhead{S{$^{3+}$}}
& \colhead{S{$^{4+}$}}
& \colhead{S{$^{5+}$}}
& \colhead{S{$^{6+}$}}
& \colhead{S{$^{7+}$}}
& \colhead{S{$^{8+}$}}
& \colhead{S{$^{9+}$}}
& \colhead{S{$^{10+}$}}
& \colhead{S{$^{11+}$}}
& \colhead{S{$^{12+}$}}
& \colhead{S{$^{13+}$}}
& \colhead{S{$^{14+}$}}
& \colhead{S{$^{15+}$}}
& \colhead{S{$^{16+}$}}}
\startdata
4.00 &  0.158 &  0.515 &  8.086 & 15.000 & 15.000 & 15.000 & 15.000 & 15.000 & 15.000 & 15.000 & 15.000 & 15.000 & 15.000 & 15.000 & 15.000 & 15.000 & 15.000 \\
4.10 &  0.881 &  0.061 &  5.139 & 15.000 & 15.000 & 15.000 & 15.000 & 15.000 & 15.000 & 15.000 & 15.000 & 15.000 & 15.000 & 15.000 & 15.000 & 15.000 & 15.000 \\
4.20 &  1.765 &  0.008 &  3.139 & 11.063 & 15.000 & 15.000 & 15.000 & 15.000 & 15.000 & 15.000 & 15.000 & 15.000 & 15.000 & 15.000 & 15.000 & 15.000 & 15.000 \\
4.30 &  2.430 &  0.011 &  1.662 &  7.285 & 15.000 & 15.000 & 15.000 & 15.000 & 15.000 & 15.000 & 15.000 & 15.000 & 15.000 & 15.000 & 15.000 & 15.000 & 15.000 \\
4.40 &  2.847 &  0.102 &  0.682 &  4.530 & 11.442 & 15.000 & 15.000 & 15.000 & 15.000 & 15.000 & 15.000 & 15.000 & 15.000 & 15.000 & 15.000 & 15.000 & 15.000 \\
4.50 &  3.257 &  0.385 &  0.232 &  2.798 &  8.061 & 15.000 & 15.000 & 15.000 & 15.000 & 15.000 & 15.000 & 15.000 & 15.000 & 15.000 & 15.000 & 15.000 & 15.000 \\
4.60 &  3.712 &  0.755 &  0.094 &  1.738 &  5.681 & 13.484 & 15.000 & 15.000 & 15.000 & 15.000 & 15.000 & 15.000 & 15.000 & 15.000 & 15.000 & 15.000 & 15.000 \\
4.70 &  4.181 &  1.131 &  0.081 &  1.016 &  3.879 &  9.877 & 15.000 & 15.000 & 15.000 & 15.000 & 15.000 & 15.000 & 15.000 & 15.000 & 15.000 & 15.000 & 15.000 \\
\enddata
\tablecomments{Same as Table~\protect\ref{tab:Al} but for sulfur.}
\end{deluxetable}

\clearpage

\begin{deluxetable}{ccccccccccccccccccc}
\setlength{\tabcolsep}{0.03in}
\tabletypesize{\tiny}
\tablecaption{Chlorine CIE Fractional Abundances ({\sc autostructure}-based
results)
\label{tab:Cl}}
\tablewidth{0pt}
\tablehead{
\colhead{log(T)}
& \colhead{Cl{$^{0+}$}}
& \colhead{Cl{$^{1+}$}}
& \colhead{Cl{$^{2+}$}}
& \colhead{Cl{$^{3+}$}}
& \colhead{Cl{$^{4+}$}}
& \colhead{Cl{$^{5+}$}}
& \colhead{Cl{$^{6+}$}}
& \colhead{Cl{$^{7+}$}}
& \colhead{Cl{$^{8+}$}}
& \colhead{Cl{$^{9+}$}}
& \colhead{Cl{$^{10+}$}}
& \colhead{Cl{$^{11+}$}}
& \colhead{Cl{$^{12+}$}}
& \colhead{Cl{$^{13+}$}}
& \colhead{Cl{$^{14+}$}}
& \colhead{Cl{$^{15+}$}}
& \colhead{Cl{$^{16+}$}}
& \colhead{Cl{$^{17+}$}}}
\startdata
%\label{tab:Cl}
4.00 &  0.015 &  1.457 &  9.251 & 15.000 & 15.000 & 15.000 & 15.000 & 15.000 & 15.000 & 15.000 & 15.000 &  15.000 & 15.000 & 15.000 & 15.000 & 15.000 & 15.000 & 15.000 \\
4.10 &  0.304 &  0.298 &  5.510 & 15.000 & 15.000 & 15.000 & 15.000 & 15.000 & 15.000 & 15.000 & 15.000 &  15.000 & 15.000 & 15.000 & 15.000 & 15.000 & 15.000 & 15.000 \\
4.20 &  1.203 &  0.028 &  3.179 & 12.049 & 15.000 & 15.000 & 15.000 & 15.000 & 15.000 & 15.000 & 15.000 &  15.000 & 15.000 & 15.000 & 15.000 & 15.000 & 15.000 & 15.000 \\
4.30 &  2.132 &  0.015 &  1.576 &  7.744 & 15.000 & 15.000 & 15.000 & 15.000 & 15.000 & 15.000 & 15.000 &  15.000 & 15.000 & 15.000 & 15.000 & 15.000 & 15.000 & 15.000 \\
4.40 &  2.990 &  0.128 &  0.595 &  4.613 & 12.177 & 15.000 & 15.000 & 15.000 & 15.000 & 15.000 & 15.000 &  15.000 & 15.000 & 15.000 & 15.000 & 15.000 & 15.000 & 15.000 \\
4.50 &  3.821 &  0.440 &  0.198 &  2.611 &  8.322 & 15.000 & 15.000 & 15.000 & 15.000 & 15.000 & 15.000 &  15.000 & 15.000 & 15.000 & 15.000 & 15.000 & 15.000 & 15.000 \\
4.60 &  4.482 &  0.842 &  0.085 &  1.470 &  5.825 & 12.599 & 15.000 & 15.000 & 15.000 & 15.000 & 15.000 &  15.000 & 15.000 & 15.000 & 15.000 & 15.000 & 15.000 & 15.000 \\
4.70 &  5.010 &  1.276 &  0.100 &  0.816 &  4.091 &  9.330 & 15.000 & 15.000 & 15.000 & 15.000 & 15.000 &  15.000 & 15.000 & 15.000 & 15.000 & 15.000 & 15.000 & 15.000 \\
\enddata
\tablecomments{Same as Table~\protect\ref{tab:Al} but for chlorine.}
\end{deluxetable}

\clearpage

\begin{deluxetable}{cccccccccccccccccccc}
\setlength{\tabcolsep}{0.02in}
\tabletypesize{\tiny}
\tablecaption{Argon CIE Fractional Abundances ({\sc autostructure}-based
results)
\label{tab:Ar}}
\tablewidth{0pt}
\tablehead{
\colhead{log(T)} 
& \colhead{Ar{$^{0+}$}}
& \colhead{Ar{$^{1+}$}}
& \colhead{Ar{$^{2+}$}}
& \colhead{Ar{$^{3+}$}}
& \colhead{Ar{$^{4+}$}}
& \colhead{Ar{$^{5+}$}}
& \colhead{Ar{$^{6+}$}}
& \colhead{Ar{$^{7+}$}}
& \colhead{Ar{$^{8+}$}}
& \colhead{Ar{$^{9+}$}}
& \colhead{Ar{$^{10+}$}}
& \colhead{Ar{$^{11+}$}}
& \colhead{Ar{$^{12+}$}}
& \colhead{Ar{$^{13+}$}}
& \colhead{Ar{$^{14+}$}}
& \colhead{Ar{$^{15+}$}}
& \colhead{Ar{$^{16+}$}}
& \colhead{Ar{$^{17+}$}}
& \colhead{Ar{$^{18+}$}}}
\startdata
4.00 &  0.000 &  3.318 & 15.000 & 15.000 & 15.000 & 15.000 & 15.000 & 15.000 & 15.000 & 15.000 & 15.000 & 15.000 & 15.000 & 15.000 & 15.000 & 15.000 & 15.000 & 15.000 & 15.000   \\
4.10 &  0.012 &  1.570 &  8.318 & 15.000 & 15.000 & 15.000 & 15.000 & 15.000 & 15.000 & 15.000 & 15.000 & 15.000 & 15.000 & 15.000 & 15.000 & 15.000 & 15.000 & 15.000 & 15.000   \\
4.20 &  0.239 &  0.373 &  4.734 & 13.856 & 15.000 & 15.000 & 15.000 & 15.000 & 15.000 & 15.000 & 15.000 & 15.000 & 15.000 & 15.000 & 15.000 & 15.000 & 15.000 & 15.000 & 15.000   \\
4.30 &  1.068 &  0.040 &  2.486 &  8.819 & 15.000 & 15.000 & 15.000 & 15.000 & 15.000 & 15.000 & 15.000 & 15.000 & 15.000 & 15.000 & 15.000 & 15.000 & 15.000 & 15.000 & 15.000   \\
4.40 &  2.032 &  0.054 &  0.966 &  5.068 & 13.870 & 15.000 & 15.000 & 15.000 & 15.000 & 15.000 & 15.000 & 15.000 & 15.000 & 15.000 & 15.000 & 15.000 & 15.000 & 15.000 & 15.000   \\
4.50 &  3.213 &  0.462 &  0.186 &  2.567 &  9.165 & 15.000 & 15.000 & 15.000 & 15.000 & 15.000 & 15.000 & 15.000 & 15.000 & 15.000 & 15.000 & 15.000 & 15.000 & 15.000 & 15.000   \\
4.60 &  4.467 &  1.129 &  0.058 &  1.296 &  6.280 & 13.524 & 15.000 & 15.000 & 15.000 & 15.000 & 15.000 & 15.000 & 15.000 & 15.000 & 15.000 & 15.000 & 15.000 & 15.000 & 15.000   \\
4.70 &  5.337 &  1.642 &  0.120 &  0.661 &  4.387 & 10.020 & 15.000 & 15.000 & 15.000 & 15.000 & 15.000 & 15.000 & 15.000 & 15.000 & 15.000 & 15.000 & 15.000 & 15.000 & 15.000   \\
\enddata
\tablecomments{Same as Table~\protect\ref{tab:Al} but for argon.}
\end{deluxetable}

\clearpage

\begin{deluxetable}{ccccccccccc}
\tabletypesize{\tiny}
\tablecaption{Potassium CIE Fractional Abundances ({\sc autostructure}-based
results)
\label{tab:K}}
\tablewidth{0pt}
\tablehead{
\colhead{log(T)} 
& \colhead{K{$^{0+}$}}
& \colhead{K{$^{1+}$}}
& \colhead{K{$^{2+}$}}
& \colhead{K{$^{3+}$}}
& \colhead{K{$^{4+}$}}
& \colhead{K{$^{5+}$}}
& \colhead{K{$^{6+}$}}
& \colhead{K{$^{7+}$}}
& \colhead{K{$^{8+}$}}
& \colhead{K{$^{9+}$}}}
\startdata
4.00 &  3.150 &  0.000 & 15.000 & 15.000 & 15.000 & 15.000 & 15.000 & 15.000 & 15.000 & 15.000  \\
4.10 &  3.717 &  0.000 &  8.788 & 15.000 & 15.000 & 15.000 & 15.000 & 15.000 & 15.000 & 15.000  \\
4.20 &  4.189 &  0.000 &  6.067 & 15.000 & 15.000 & 15.000 & 15.000 & 15.000 & 15.000 & 15.000  \\
4.30 &  4.585 &  0.000 &  3.881 & 11.743 & 15.000 & 15.000 & 15.000 & 15.000 & 15.000 & 15.000  \\
4.40 &  4.923 &  0.003 &  2.124 &  7.491 & 15.000 & 15.000 & 15.000 & 15.000 & 15.000 & 15.000  \\
4.50 &  5.283 &  0.078 &  0.783 &  4.165 & 10.459 & 15.000 & 15.000 & 15.000 & 15.000 & 15.000  \\
4.60 &  5.956 &  0.537 &  0.154 &  2.055 &  6.440 & 14.492 & 15.000 & 15.000 & 15.000 & 15.000  \\
4.70 &  6.526 &  1.152 &  0.078 &  1.029 &  4.106 & 10.369 & 15.000 & 15.000 & 15.000 & 15.000  \\
\enddata
\tablecomments{Same as Table~\protect\ref{tab:Al} but for potassium.}
\end{deluxetable}

\clearpage

\begin{deluxetable}{cccccccccccc}
\tabletypesize{\tiny}
\tablecaption{Calcium CIE Fractional Abundances ({\sc autostructure}-based
results)
\label{tab:Ca}}
\tablewidth{0pt}
\tablehead{
\colhead{log(T)} 
& \colhead{Ca{$^{0+}$}}
& \colhead{Ca{$^{1+}$}}
& \colhead{Ca{$^{2+}$}}
& \colhead{Ca{$^{3+}$}}
& \colhead{Ca{$^{4+}$}}
& \colhead{Ca{$^{5+}$}}
& \colhead{Ca{$^{6+}$}}
& \colhead{Ca{$^{7+}$}}
& \colhead{Ca{$^{8+}$}}
& \colhead{Ca{$^{9+}$}}
& \colhead{Ca{$^{10+}$}}}
\startdata
4.00 &  1.113 &  0.041 &  1.914 & 15.000 & 15.000 & 15.000 & 15.000 & 15.000 & 15.000 & 15.000 & 15.000 \\
4.10 &  1.707 &  0.123 &  0.642 & 15.000 & 15.000 & 15.000 & 15.000 & 15.000 & 15.000 & 15.000 & 15.000 \\
4.20 &  2.711 &  0.683 &  0.102 & 15.000 & 15.000 & 15.000 & 15.000 & 15.000 & 15.000 & 15.000 & 15.000 \\
4.30 &  3.912 &  1.493 &  0.014 &  9.480 & 15.000 & 15.000 & 15.000 & 15.000 & 15.000 & 15.000 & 15.000 \\
4.40 &  4.986 &  2.219 &  0.003 &  6.707 & 15.000 & 15.000 & 15.000 & 15.000 & 15.000 & 15.000 & 15.000 \\
4.50 &  5.897 &  2.817 &  0.001 &  4.496 & 12.067 & 15.000 & 15.000 & 15.000 & 15.000 & 15.000 & 15.000 \\
4.60 &  6.566 &  3.200 &  0.001 &  2.798 &  8.130 & 15.000 & 15.000 & 15.000 & 15.000 & 15.000 & 15.000 \\
4.70 &  6.841 &  3.212 &  0.009 &  1.701 &  5.483 & 11.169 & 15.000 & 15.000 & 15.000 & 15.000 & 15.000 \\
\enddata
\tablecomments{Same as Table~\protect\ref{tab:Al} but for calcium.}
\end{deluxetable}

\clearpage

\begin{deluxetable}{cccccccccccc}
\tabletypesize{\tiny}
\tablecaption{Scandium CIE Fractional Abundances ({\sc autostructure}-based
results)
\label{tab:Sc}}
\tablewidth{0pt}
\tablehead{
\colhead{log(T)} 
& \colhead{Sc{$^{0+}$}}
& \colhead{Sc{$^{1+}$}}
& \colhead{Sc{$^{2+}$}}
& \colhead{Sc{$^{3+}$}}
& \colhead{Sc{$^{4+}$}}
& \colhead{Sc{$^{5+}$}}
& \colhead{Sc{$^{6+}$}}
& \colhead{Sc{$^{7+}$}}
& \colhead{Sc{$^{8+}$}}
& \colhead{Sc{$^{9+}$}}
& \colhead{Sc{$^{10+}$}}}
\startdata
4.00 &  0.895 &  0.060 &  3.006 & 12.390 & 15.000 & 15.000 & 15.000 & 15.000 & 15.000 & 15.000 & 15.000 \\
4.10 &  1.269 &  0.033 &  1.725 &  8.416 & 15.000 & 15.000 & 15.000 & 15.000 & 15.000 & 15.000 & 15.000 \\
4.20 &  1.652 &  0.053 &  1.029 &  5.557 & 15.000 & 15.000 & 15.000 & 15.000 & 15.000 & 15.000 & 15.000 \\
4.30 &  2.064 &  0.134 &  0.590 &  3.374 & 15.000 & 15.000 & 15.000 & 15.000 & 15.000 & 15.000 & 15.000 \\
4.40 &  2.534 &  0.301 &  0.321 &  1.696 & 15.000 & 15.000 & 15.000 & 15.000 & 15.000 & 15.000 & 15.000 \\
4.50 &  3.180 &  0.667 &  0.306 &  0.538 &  9.347 & 15.000 & 15.000 & 15.000 & 15.000 & 15.000 & 15.000 \\
4.60 &  4.256 &  1.482 &  0.781 &  0.096 &  6.384 & 15.000 & 15.000 & 15.000 & 15.000 & 15.000 & 15.000 \\
4.70 &  5.374 &  2.358 &  1.349 &  0.022 &  4.340 & 11.252 & 15.000 & 15.000 & 15.000 & 15.000 & 15.000 \\
\enddata
\tablecomments{Same as Table~\protect\ref{tab:Al} but for scandium.}
\end{deluxetable}

\clearpage

\begin{deluxetable}{ccccccccccccc}
\tabletypesize{\tiny}
\tablecaption{Titanium CIE Fractional Abundances ({\sc autostructure}-based
results)
\label{tab:Ti}}
\tablewidth{0pt}
\tablehead{
\colhead{log(T)} 
& \colhead{Ti{$^{0+}$}}
& \colhead{Ti{$^{1+}$}}
& \colhead{Ti{$^{2+}$}}
& \colhead{Ti{$^{3+}$}}
& \colhead{Ti{$^{4+}$}}
& \colhead{Ti{$^{5+}$}}
& \colhead{Ti{$^{6+}$}}
& \colhead{Ti{$^{7+}$}}
& \colhead{Ti{$^{8+}$}}
& \colhead{Ti{$^{9+}$}}
& \colhead{Ti{$^{10+}$}}
& \colhead{Ti{$^{11+}$}}}
\startdata
4.00 &  2.138 &  0.003 &  3.310 & 15.000 & 15.000 & 15.000 & 15.000 & 15.000 & 15.000 & 15.000 & 15.000 & 15.000  \\
4.10 &  2.465 &  0.010 &  1.739 &  9.011 & 15.000 & 15.000 & 15.000 & 15.000 & 15.000 & 15.000 & 15.000 & 15.000  \\
4.20 &  2.654 &  0.136 &  0.575 &  5.670 & 15.000 & 15.000 & 15.000 & 15.000 & 15.000 & 15.000 & 15.000 & 15.000  \\
4.30 &  3.251 &  0.678 &  0.103 &  3.693 & 11.759 & 15.000 & 15.000 & 15.000 & 15.000 & 15.000 & 15.000 & 15.000  \\
4.40 &  3.911 &  1.254 &  0.026 &  2.508 &  8.201 & 15.000 & 15.000 & 15.000 & 15.000 & 15.000 & 15.000 & 15.000  \\
4.50 &  4.334 &  1.564 &  0.023 &  1.618 &  5.401 & 15.000 & 15.000 & 15.000 & 15.000 & 15.000 & 15.000 & 15.000  \\
4.60 &  4.687 &  1.786 &  0.064 &  0.924 &  3.170 & 13.768 & 15.000 & 15.000 & 15.000 & 15.000 & 15.000 & 15.000  \\
4.70 &  5.138 &  2.091 &  0.215 &  0.455 &  1.496 &  9.898 & 15.000 & 15.000 & 15.000 & 15.000 & 15.000 & 15.000  \\
\enddata
\tablecomments{Same as Table~\protect\ref{tab:Al} but for titanium.}
\end{deluxetable}

\clearpage

\begin{deluxetable}{ccccccccccccc}
\tabletypesize{\tiny}
\tablecaption{Vanadium CIE Fractional Abundances ({\sc autostructure}-based
results)
\label{tab:V}}
\tablewidth{0pt}
\tablehead{
\colhead{log(T)} 
& \colhead{V{$^{0+}$}}
& \colhead{V{$^{1+}$}}
& \colhead{V{$^{2+}$}}
& \colhead{V{$^{3+}$}}
& \colhead{V{$^{4+}$}}
& \colhead{V{$^{5+}$}}
& \colhead{V{$^{6+}$}}
& \colhead{V{$^{7+}$}}
& \colhead{V{$^{8+}$}}
& \colhead{V{$^{9+}$}}
& \colhead{V{$^{10+}$}}
& \colhead{V{$^{11+}$}}}
\startdata
4.00 &  2.305 &  0.002 &  3.327 & 12.183 & 15.000 & 15.000 & 15.000 & 15.000 & 15.000 & 15.000 & 15.000 & 15.000  \\
4.10 &  2.987 &  0.010 &  1.685 &  7.933 & 15.000 & 15.000 & 15.000 & 15.000 & 15.000 & 15.000 & 15.000 & 15.000  \\
4.20 &  3.305 &  0.156 &  0.522 &  4.672 & 15.000 & 15.000 & 15.000 & 15.000 & 15.000 & 15.000 & 15.000 & 15.000  \\
4.30 &  3.770 &  0.665 &  0.107 &  2.565 & 11.377 & 15.000 & 15.000 & 15.000 & 15.000 & 15.000 & 15.000 & 15.000  \\
4.40 &  4.209 &  1.139 &  0.066 &  1.166 &  7.424 & 15.000 & 15.000 & 15.000 & 15.000 & 15.000 & 15.000 & 15.000  \\
4.50 &  4.732 &  1.652 &  0.280 &  0.345 &  4.548 & 12.556 & 15.000 & 15.000 & 15.000 & 15.000 & 15.000 & 15.000  \\
4.60 &  5.440 &  2.308 &  0.703 &  0.100 &  2.647 &  8.393 & 15.000 & 15.000 & 15.000 & 15.000 & 15.000 & 15.000  \\
4.70 &  6.087 &  2.869 &  1.054 &  0.066 &  1.286 &  5.231 & 15.000 & 15.000 & 15.000 & 15.000 & 15.000 & 15.000  \\
\enddata
\tablecomments{Same as Table~\protect\ref{tab:Al} but for vanadium.}
\end{deluxetable}

\clearpage

\begin{deluxetable}{cccccccccccccc}
\tabletypesize{\tiny}
\tablecaption{Chromium CIE Fractional Abundances ({\sc autostructure}-based
results)
\label{tab:Cr}}
\tablewidth{0pt}
\tablehead{
\colhead{log(T)} 
& \colhead{Cr{$^{0+}$}}
& \colhead{Cr{$^{1+}$}}
& \colhead{Cr{$^{2+}$}}
& \colhead{Cr{$^{3+}$}}
& \colhead{Cr{$^{4+}$}}
& \colhead{Cr{$^{5+}$}}
& \colhead{Cr{$^{6+}$}}
& \colhead{Cr{$^{7+}$}}
& \colhead{Cr{$^{8+}$}}
& \colhead{Cr{$^{9+}$}}
& \colhead{Cr{$^{10+}$}}
& \colhead{Cr{$^{11+}$}}
& \colhead{Cr{$^{12+}$}}}
\startdata
4.00 &  2.377 &  0.002 &  4.229 & 15.000 & 15.000 & 15.000 & 15.000 & 15.000 & 15.000 & 15.000 & 15.000 & 15.000 & 15.000  \\
4.10 &  3.255 &  0.002 &  2.414 & 11.554 & 15.000 & 15.000 & 15.000 & 15.000 & 15.000 & 15.000 & 15.000 & 15.000 & 15.000  \\
4.20 &  4.004 &  0.035 &  1.109 &  7.571 & 15.000 & 15.000 & 15.000 & 15.000 & 15.000 & 15.000 & 15.000 & 15.000 & 15.000  \\
4.30 &  4.618 &  0.200 &  0.433 &  4.744 & 14.377 & 15.000 & 15.000 & 15.000 & 15.000 & 15.000 & 15.000 & 15.000 & 15.000  \\
4.40 &  4.931 &  0.484 &  0.174 &  2.759 &  9.712 & 15.000 & 15.000 & 15.000 & 15.000 & 15.000 & 15.000 & 15.000 & 15.000  \\
4.50 &  5.131 &  0.825 &  0.094 &  1.352 &  6.161 & 14.553 & 15.000 & 15.000 & 15.000 & 15.000 & 15.000 & 15.000 & 15.000  \\
4.60 &  5.454 &  1.267 &  0.175 &  0.557 &  3.696 &  9.739 & 15.000 & 15.000 & 15.000 & 15.000 & 15.000 & 15.000 & 15.000  \\
4.70 &  5.925 &  1.805 &  0.390 &  0.244 &  2.198 &  6.527 & 13.424 & 15.000 & 15.000 & 15.000 & 15.000 & 15.000 & 15.000  \\
\enddata
\tablecomments{Same as Table~\protect\ref{tab:Al} but for chromium.}
\end{deluxetable}

\clearpage

\begin{deluxetable}{cccccccccccccc}
\setlength{\tabcolsep}{0.06in}
\tabletypesize{\tiny}
\tablecaption{Manganese CIE Fractional Abundances ({\sc autostructure}-based
results)
\label{tab:Mn}}
\tablewidth{0pt}
\tablehead{
\colhead{log(T)} 
& \colhead{Mn{$^{0+}$}}
& \colhead{Mn{$^{1+}$}}
& \colhead{Mn{$^{2+}$}}
& \colhead{Mn{$^{3+}$}}
& \colhead{Mn{$^{4+}$}}
& \colhead{Mn{$^{5+}$}}
& \colhead{Mn{$^{6+}$}}
& \colhead{Mn{$^{7+}$}}
& \colhead{Mn{$^{8+}$}}
& \colhead{Mn{$^{9+}$}}
& \colhead{Mn{$^{10+}$}}
& \colhead{Mn{$^{11+}$}}
& \colhead{Mn{$^{12+}$}}}
\startdata
4.00 &  1.974 &  0.005 &  4.026 & 15.000 & 15.000 & 15.000 & 15.000 & 15.000 & 15.000 & 15.000 & 15.000 & 15.000 & 15.000  \\
4.10 &  2.801 &  0.003 &  2.266 & 15.000 & 15.000 & 15.000 & 15.000 & 15.000 & 15.000 & 15.000 & 15.000 & 15.000 & 15.000  \\
4.20 &  3.285 &  0.059 &  0.895 &  8.168 & 15.000 & 15.000 & 15.000 & 15.000 & 15.000 & 15.000 & 15.000 & 15.000 & 15.000  \\
4.30 &  3.792 &  0.473 &  0.178 &  5.134 & 15.000 & 15.000 & 15.000 & 15.000 & 15.000 & 15.000 & 15.000 & 15.000 & 15.000  \\
4.40 &  4.425 &  1.094 &  0.037 &  3.180 & 10.449 & 15.000 & 15.000 & 15.000 & 15.000 & 15.000 & 15.000 & 15.000 & 15.000  \\
4.50 &  4.838 &  1.484 &  0.020 &  1.886 &  6.968 & 15.000 & 15.000 & 15.000 & 15.000 & 15.000 & 15.000 & 15.000 & 15.000  \\
4.60 &  5.161 &  1.760 &  0.046 &  1.085 &  4.573 & 11.057 & 15.000 & 15.000 & 15.000 & 15.000 & 15.000 & 15.000 & 15.000  \\
4.70 &  5.533 &  2.062 &  0.137 &  0.582 &  2.992 &  7.514 & 14.820 & 15.000 & 15.000 & 15.000 & 15.000 & 15.000 & 15.000  \\
\enddata
\tablecomments{Same as Table~\protect\ref{tab:Al} but for manganese.}
\end{deluxetable}

\clearpage

\begin{deluxetable}{ccccccccccccccc}
\setlength{\tabcolsep}{0.06in}
\tabletypesize{\tiny}
\tablecaption{Iron CIE Fractional Abundances ({\sc autostructure}-based
results)
\label{tab:Fe}}
\tablewidth{0pt}
\tablehead{
\colhead{log(T)} 
& \colhead{Fe{$^{0+}$}}
& \colhead{Fe{$^{1+}$}}
& \colhead{Fe{$^{2+}$}}
& \colhead{Fe{$^{3+}$}}
& \colhead{Fe{$^{4+}$}}
& \colhead{Fe{$^{5+}$}}
& \colhead{Fe{$^{6+}$}}
& \colhead{Fe{$^{7+}$}}
& \colhead{Fe{$^{8+}$}}
& \colhead{Fe{$^{9+}$}}
& \colhead{Fe{$^{10+}$}}
& \colhead{Fe{$^{11+}$}}
& \colhead{Fe{$^{12+}$}}
& \colhead{Fe{$^{13+}$}}}
\startdata
4.00 &  0.916 &  0.056 &  4.006 & 15.000 & 15.000 & 15.000 & 15.000 & 15.000 & 15.000 & 15.000 & 15.000 & 15.000 & 15.000 & 15.000  \\
4.10 &  1.450 &  0.019 &  2.165 & 15.000 & 15.000 & 15.000 & 15.000 & 15.000 & 15.000 & 15.000 & 15.000 & 15.000 & 15.000 & 15.000  \\
4.20 &  1.981 &  0.085 &  0.776 &  8.101 & 15.000 & 15.000 & 15.000 & 15.000 & 15.000 & 15.000 & 15.000 & 15.000 & 15.000 & 15.000  \\
4.30 &  2.883 &  0.589 &  0.130 &  5.069 & 15.000 & 15.000 & 15.000 & 15.000 & 15.000 & 15.000 & 15.000 & 15.000 & 15.000 & 15.000  \\
4.40 &  3.933 &  1.291 &  0.023 &  3.046 & 10.640 & 15.000 & 15.000 & 15.000 & 15.000 & 15.000 & 15.000 & 15.000 & 15.000 & 15.000  \\
4.50 &  4.698 &  1.747 &  0.021 &  1.520 &  6.727 & 15.000 & 15.000 & 15.000 & 15.000 & 15.000 & 15.000 & 15.000 & 15.000 & 15.000  \\
4.60 &  5.371 &  2.145 &  0.158 &  0.527 &  3.813 & 10.373 & 15.000 & 15.000 & 15.000 & 15.000 & 15.000 & 15.000 & 15.000 & 15.000  \\
4.70 &  6.163 &  2.688 &  0.507 &  0.170 &  1.969 &  6.405 & 13.158 & 15.000 & 15.000 & 15.000 & 15.000 & 15.000 & 15.000 & 15.000  \\
\enddata
\tablecomments{Same as Table~\protect\ref{tab:Al} but for iron.}
\end{deluxetable}

\clearpage

\begin{deluxetable}{ccccccccccccccc}
\setlength{\tabcolsep}{0.05in}
\tabletypesize{\tiny}
\tablecaption{Cobalt CIE Fractional Abundances ({\sc autostructure}-based
results)
\label{tab:Co}}
\tablewidth{0pt}
\tablehead{
\colhead{log(T)} 
& \colhead{Co{$^{0+}$}}
& \colhead{Co{$^{1+}$}}
& \colhead{Co{$^{2+}$}}
& \colhead{Co{$^{3+}$}}
& \colhead{Co{$^{4+}$}}
& \colhead{Co{$^{5+}$}}
& \colhead{Co{$^{6+}$}}
& \colhead{Co{$^{7+}$}}
& \colhead{Co{$^{8+}$}}
& \colhead{Co{$^{9+}$}}
& \colhead{Co{$^{10+}$}}
& \colhead{Co{$^{11+}$}}
& \colhead{Co{$^{12+}$}}
& \colhead{Co{$^{13+}$}}}
\startdata
4.00 &  0.410 &  0.214 &  3.969 & 15.000 & 15.000 & 15.000 & 15.000 & 15.000 & 15.000 & 15.000 & 15.000 & 15.000 & 15.000 & 15.000  \\
4.10 &  0.821 &  0.077 &  1.965 & 15.000 & 15.000 & 15.000 & 15.000 & 15.000 & 15.000 & 15.000 & 15.000 & 15.000 & 15.000 & 15.000  \\
4.20 &  1.323 &  0.136 &  0.656 &  7.275 & 15.000 & 15.000 & 15.000 & 15.000 & 15.000 & 15.000 & 15.000 & 15.000 & 15.000 & 15.000  \\
4.30 &  2.055 &  0.485 &  0.178 &  4.488 & 13.889 & 15.000 & 15.000 & 15.000 & 15.000 & 15.000 & 15.000 & 15.000 & 15.000 & 15.000  \\
4.40 &  2.801 &  0.886 &  0.063 &  2.521 &  9.142 & 15.000 & 15.000 & 15.000 & 15.000 & 15.000 & 15.000 & 15.000 & 15.000 & 15.000  \\
4.50 &  3.520 &  1.283 &  0.065 &  1.064 &  5.455 & 15.000 & 15.000 & 15.000 & 15.000 & 15.000 & 15.000 & 15.000 & 15.000 & 15.000  \\
4.60 &  4.428 &  1.885 &  0.332 &  0.284 &  2.888 &  9.874 & 15.000 & 15.000 & 15.000 & 15.000 & 15.000 & 15.000 & 15.000 & 15.000  \\
4.70 &  5.488 &  2.654 &  0.803 &  0.102 &  1.308 &  6.118 & 13.700 & 15.000 & 15.000 & 15.000 & 15.000 & 15.000 & 15.000 & 15.000  \\
\enddata
\tablecomments{Same as Table~\protect\ref{tab:Al} but for cobalt.}
\end{deluxetable}

\clearpage

\begin{deluxetable}{cccccccccccccccc}
\setlength{\tabcolsep}{0.05in}
\tabletypesize{\tiny}
\tablecaption{Nickel CIE Fractional Abundances ({\sc autostructure}-based
results)
\label{tab:Ni}}
\tablewidth{0pt}
\tablehead{
\colhead{log(T)} 
& \colhead{Ni{$^{0+}$}}
& \colhead{Ni{$^{1+}$}}
& \colhead{Ni{$^{2+}$}}
& \colhead{Ni{$^{3+}$}}
& \colhead{Ni{$^{4+}$}}
& \colhead{Ni{$^{5+}$}}
& \colhead{Ni{$^{6+}$}}
& \colhead{Ni{$^{7+}$}}
& \colhead{Ni{$^{8+}$}}
& \colhead{Ni{$^{9+}$}}
& \colhead{Ni{$^{10+}$}}
& \colhead{Ni{$^{11+}$}}
& \colhead{Ni{$^{12+}$}}
& \colhead{Ni{$^{13+}$}}}
\startdata
4.00 &  0.790 &  0.077 &  4.755 & 15.000 & 15.000 & 15.000 & 15.000 & 15.000 & 15.000 & 15.000 & 15.000 & 15.000 & 15.000 & 15.000  \\
4.10 &  1.615 &  0.012 &  2.696 & 15.000 & 15.000 & 15.000 & 15.000 & 15.000 & 15.000 & 15.000 & 15.000 & 15.000 & 15.000 & 15.000  \\
4.20 &  2.323 &  0.037 &  1.116 &  8.205 & 15.000 & 15.000 & 15.000 & 15.000 & 15.000 & 15.000 & 15.000 & 15.000 & 15.000 & 15.000  \\
4.30 &  3.094 &  0.420 &  0.209 &  4.876 & 15.000 & 15.000 & 15.000 & 15.000 & 15.000 & 15.000 & 15.000 & 15.000 & 15.000 & 15.000  \\
4.40 &  4.008 &  1.235 &  0.027 &  2.749 &  9.768 & 15.000 & 15.000 & 15.000 & 15.000 & 15.000 & 15.000 & 15.000 & 15.000 & 15.000  \\
4.50 &  4.627 &  1.856 &  0.035 &  1.198 &  5.840 & 14.463 & 15.000 & 15.000 & 15.000 & 15.000 & 15.000 & 15.000 & 15.000 & 15.000  \\
4.60 &  5.173 &  2.397 &  0.319 &  0.288 &  3.027 &  9.039 & 15.000 & 15.000 & 15.000 & 15.000 & 15.000 & 15.000 & 15.000 & 15.000  \\
4.70 &  5.866 &  3.053 &  0.852 &  0.089 &  1.356 &  5.283 & 13.145 & 15.000 & 15.000 & 15.000 & 15.000 & 15.000 & 15.000 & 15.000  \\
\enddata
\tablecomments{Same as Table~\protect\ref{tab:Al} but for nickel.}
\end{deluxetable}

\clearpage

\begin{deluxetable}{cccccccccccccccc}
\setlength{\tabcolsep}{0.04in}
\tabletypesize{\tiny}
\tablecaption{Copper CIE Fractional Abundances ({\sc autostructure}-based
results)
\label{tab:Cu}}
\tablewidth{0pt}
\tablehead{
\colhead{log(T)} 
& \colhead{Cu{$^{0+}$}}
& \colhead{Cu{$^{1+}$}}
& \colhead{Cu{$^{2+}$}}
& \colhead{Cu{$^{3+}$}}
& \colhead{Cu{$^{4+}$}}
& \colhead{Cu{$^{5+}$}}
& \colhead{Cu{$^{6+}$}}
& \colhead{Cu{$^{7+}$}}
& \colhead{Cu{$^{8+}$}}
& \colhead{Cu{$^{9+}$}}
& \colhead{Cu{$^{10+}$}}
& \colhead{Cu{$^{11+}$}}
& \colhead{Cu{$^{12+}$}}
& \colhead{Cu{$^{13+}$}}
& \colhead{Cu{$^{14+}$}}}
\startdata
4.00 &  0.056 &  0.920 &  7.465 & 15.000 & 15.000 & 15.000 & 15.000 & 15.000 & 15.000 & 15.000 & 15.000 & 15.000 & 15.000 & 15.000 & 15.000\\
4.10 &  0.232 &  0.383 &  5.188 & 15.000 & 15.000 & 15.000 & 15.000 & 15.000 & 15.000 & 15.000 & 15.000 & 15.000 & 15.000 & 15.000 & 15.000\\
4.20 &  0.595 &  0.127 &  3.543 & 11.331 & 15.000 & 15.000 & 15.000 & 15.000 & 15.000 & 15.000 & 15.000 & 15.000 & 15.000 & 15.000 & 15.000\\
4.30 &  1.052 &  0.043 &  2.326 &  7.960 & 15.000 & 15.000 & 15.000 & 15.000 & 15.000 & 15.000 & 15.000 & 15.000 & 15.000 & 15.000 & 15.000\\
4.40 &  1.520 &  0.032 &  1.381 &  5.424 & 13.345 & 15.000 & 15.000 & 15.000 & 15.000 & 15.000 & 15.000 & 15.000 & 15.000 & 15.000 & 15.000\\
4.50 &  2.016 &  0.107 &  0.679 &  3.467 &  9.323 & 15.000 & 15.000 & 15.000 & 15.000 & 15.000 & 15.000 & 15.000 & 15.000 & 15.000 & 15.000\\
4.60 &  2.629 &  0.348 &  0.268 &  2.035 &  6.304 & 13.060 & 15.000 & 15.000 & 15.000 & 15.000 & 15.000 & 15.000 & 15.000 & 15.000 & 15.000\\
4.70 &  3.375 &  0.762 &  0.131 &  1.056 &  4.051 &  9.039 & 15.000 & 15.000 & 15.000 & 15.000 & 15.000 & 15.000 & 15.000 & 15.000 & 15.000\\
\enddata
\tablecomments{Calculated $-\log_{10}$ of the fractional abundance for all 
ionization stages. We use the 
DR rate coefficients of \protect\citet{Badn06a}  for H-like through 
Na-like ions, RR rate coefficients of \protect\citet{Badn06b}  for bare through 
Na-like ions,
DR and RR rate coefficients of Mazzotta (private communication)
for all other ions, and 
the EII rate coefficients of Mazzotta (private communication).
Fractional abundances are cut off at 
$10^{-15}$. For ease of machine readability, values less than
$10^{-15}$ are given $-\log_{10}$ values of 15.}
\end{deluxetable}%

\end{document}